\documentclass[useAMS,usenatbib,usegraphicx]{mn2e}
\usepackage{rotating}
\bibpunct {(} {)} {;} {a} {} {,}
\title[Recovering galaxy properties from SED-fitting]{Recovering galaxy stellar population properties from broad-band spectral energy distribution fitting}%{Recovering galaxy stellar population properties from broad-band spectral energy distribution fitting I: a comprehensive study of parameters at given redshift}
\author[Pforr, Maraston \& Tonini]{Janine~Pforr$^1$\thanks{currently at: NOAO, 950 N. Cherry Ave., Tucson, AZ, 85719, USA, pforr@noao.edu}, Claudia~Maraston$^1$, Chiara~Tonini$^{1,2}$\\
$^1$ Institute of Cosmology and Gravitation, University of Portsmouth, Dennis Sciama Building, Burnaby Road, Portsmouth, PO1 3FX, UK\\
$^2$ Centre for Astrophysics and Supercomputing, Swinburne University of Technology, Hawthorn, Victoria 3122, Australia}
\begin{document}

\newcommand{\hs}{\emph{HyperZspec}}
\newcommand{\q}{\emph{Q}}
%\newcommand{\ch2}{$chi^2$}

%\begin{document}
\maketitle
\begin{abstract}
We explore the dependence of galaxy stellar population properties that are derived from broad-band spectral energy distribution (SED) fitting - such as age, stellar mass, dust reddening, etc. - on a variety of parameters, such as star formation histories, age grid, metallicity, initial mass function (IMF), dust reddening and reddening law, filter setup and wavelength coverage. Mock galaxies are used as test particles. We confirm our earlier results based on real $z=2$ galaxies, that usually adopted $\tau$-models lead to overestimate the star formation rate and to underestimate the stellar mass. Here, we show that - for star-forming galaxies - galaxy ages, masses and reddening, can be well determined {\it simultaneously} only when the correct star formation history is identified. This is the case for {\it inverted}-$\tau$ models at high-$z$, for which we find that the mass recovery (at fixed IMF) is as good as $\sim0.04$ dex. However, since the right star formation history is usually unknown, we quantify the offsets generated by adopting standard fitting setups. Stellar masses are generally underestimated, which results from underestimating ages. For mixed fitting setups with a variety of star formation histories the median mass recovery at $z\sim 2-3$ is as decent as $\sim0.1$ dex (at fixed IMF), albeit with large scatter. The situation worsens towards lower redshifts, because of the variety of possible star formation histories and ages. At $z\sim0.5$ the stellar mass can be underestimated by as much as $\sim0.6$ dex (at fixed IMF). A practical trick to improve upon this figure is to exclude reddening from the fitting parameters, as this helps to avoid unrealistically young and dusty solutions. Stellar masses are underestimated by a smaller amount ($\sim0.3$ dex at $z\sim0.5$). Reddening and the star formation rate should then be determined via a separate fitting. As expected, the recovery of properties is better for passive galaxies, for which e.g. the mass can be fully recovered (within $\sim0.01$ dex at fixed IMF) when using a fitting setup including metallicity effects. In both cases of star-forming as well as passive galaxies, the recovery of physical parameters is dependent on the spectral range involved in the fitting. We find that a coverage from the rest-frame UV to the rest-frame near-IR appears to be optimal. We also quantify the effect of narrowing the wavelength coverage or adding and removing filter bands, which can be useful for planning observational surveys. Finally, we provide scaling relations that allow the transformation of stellar masses obtained using different template fitting setups and stellar population models.
\end{abstract}
\begin{keywords}
galaxies: general -- galaxies: evolution -- galaxies: formation -- galaxies: fundamental parameters
\end{keywords}

%%%%%%%%%%%%% %%%%%%%%%%%%%%%%%%%%%%%%%%%%%%%%%%%%%
%%% INTRODUCTION 
%%%%%%%%%%%%%%%%%%%%%%%%%%%%%%%%%%%%%%%%%%%%%%%%%%

\section{Introduction}
Over the last two decades numerous galaxy surveys were designed and performed to study galaxy formation and evolution, e.g. CANDELS \citep{Grogin2011,Koekemoer2011}, COMBO-17 \citep{Wolf2001}, COSMOS \citep{Scoville2007}, DEEP2 \citep{Davis2003}, GMASS \citep{Kurk2009}, GOODS \citep{GOODS}, SDSS \citep{SDSS}, SERVS \citep{SERVS}. To this end, the determination of the physical properties of galaxies, such as ages, stellar masses, star formation rates (SFRs), star formation mode, dust reddening, metallicity and - when necessary - photometric redshifts possibly as a function of look-back time and environment is crucial. Even for galaxy surveys devoted primarily to Cosmology, such as BOSS \citep{Schlegel2009,SDSSBOSS}, DES (http://www.darkenergysurvey.org/), LSST \citep{LSST} and WiggleZ \citep{WiggleZ}, galaxy properties need to be accurately determined as galaxies bias the read-out of the cosmological signal \citep[e.g.][]{White2011}.\\ 
Galaxy properties are derived by fitting stellar population models to data, and several approaches are taken, which have pros and cons \citep[for recent reviews see][]{Walcher2011,M11BF}. For our purpose let us briefly distinguish between spectroscopic and photometric methods. While the availability of spectra is mandatory for the determination of detailed chemical abundances, such as chemical abundance ratios, which is accomplished through using selected absorption line indices and comparing to element-ratio sensitive models \citep[e.g.][]{Worthey1992,Trager2000a,Kuntschner2000,Thomas2005,Schiavon2007,Thomas2010}, the high S/N that is required made this method practical only for a limited number of galaxies at limited redshifts so far. Also popular is the so-called full spectral fitting approach in which population parameters are extracted by fitting the full available spectrum \citep{Heavens2000,Cid2004,Ocvirk2006,Panter2007,Tojeiro2007,Tojeiro2009,Chen2012} and which also requires high quality and high S/N spectra in order to succeed. These are often limited in wavelength coverage, notoriously time consuming and thus been limited so far in redshift and to small samples. \\
Fitting of broad-band photometry is the favourite approach at high redshift, because it is relatively cheap and yet - by covering a wide wavelength range as in the modern approach \citep[e.g. The Great Observatories Origins Deep Survey (GOODS)][]{GOODS} although at a coarser resolution compared to spectroscopy - can be effective in breaking the age-metallicity degeneracy and in providing sensible physical properties for galaxies. It is common to exploit such data in spectral energy distribution (SED) fitting and many conclusions on galaxy evolution are based on these results. \\
Hence, the interesting question is to assess how well galaxy properties can effectively be recovered with this method, and many papers have been devoted to this scope over recent years. The first papers that addressed the dependency of the derived galaxy properties at high redshift on the input stellar population models are \citet[][hereafter M05]{M05}, \citet{vdW06} and \citet[][hereafter M06]{M06}. These works consistently found that stellar ages and masses, of observed objects are - not surprisingly - a function of the stellar evolution pattern adopted in the models used to derive them \citep[for a review see][]{M11a}. This information is useful, but does not yet inform us on which fitting method provides the correct galaxy parameters. On the other hand, using mock galaxies for which the properties are known a priori allows one to understand the effectiveness of the assumptions made in the fitting, even though there is no guarantee that simulated galaxies have star formation histories similar to the real ones in the Universe. In addition, different parameters can be recovered with different accuracies. For example it is well known that stellar masses are generally better recovered than ages and SFRs \citep{Papovich2001,Pozzetti2007,Bol2009,Lee2009,M10}.\\
In recent years there have been several works in this direction. \citet{Longhetti08}, \citet{Wuyts2009} and \citet{Lee2009} address the dependency of the results on fitting parameters using simulated galaxies. \citet{Longhetti08} study the dependence of stellar mass estimates on a variety of models and their parameters like age, metallicity, IMF and star formation history (SFH) for early-type galaxies at $1<z<2$. They assume different stellar population synthesis codes for stellar evolution modelling \citep[][hereafter BC03; M05; and various others]{BC03}. They simulate early-type galaxies using BC03 stellar population models assuming exponentially declining star formation rates (with timescale $\tau$). Furthermore, they superimpose secondary star bursts with a smaller value of $\tau$ at later epochs. They find that stellar masses can not be recovered better than a factor of 2-3, at fixed known IMF.\\
\citet{Wuyts2009} use mock galaxies from a hydrodynamical merger simulation based on BC03 models. They simulate all types of galaxies, but focus on a small redshift interval between 1.5 and 3 and also do not explore metallicity effects. They conclude that properties of spheroidal galaxies (which means galaxies with little star-formation) are generally well reproduced, whereas those of star-forming galaxies suffer from large uncertainties.\\
\citet{Lee2009} focus on mock Lyman break galaxies at $z\sim3.4$, 4 and 5 based on BC03 stellar population models. They find that masses and SFRs tend to be underestimated and mean ages overestimated. They attribute these effects to the difference in SFH between the mocks and the $\tau$-model template used in the fitting. Furthermore, \citet{Lee2009} show that a large wavelength coverage from optical to rest-frame near-IR is required to best recover the input physical parameters.\\
In this paper we follow the same approach as these works. We perform our investigation by using both mock star-forming as well as passive galaxies, for which the physical properties - age, metallicity, reddening, mass and star formation rate - are well defined. We then treat these mocks as observed data and SED-fit them with several different setups. The comparison between input properties and those derived from the fitting allows us to understand and quantify the robustness of the derived properties as a function of the fitting parameters.\\
Our work is complementary to these earlier works on several aspects. Firstly, it relies on the M05 stellar population models both as ingredient of the fitting and of the mock galaxies, but also explores the case of using different stellar population models in the mock and in the fitting. Also, a wider redshift span is considered, down to redshift 0.5, and a larger set of parameters in terms of metallicity and reddening prescriptions. Moreover, a detailed exploration of the effects of wavelength coverage and photometric filters is performed. This latter analysis aims at providing a practical guide for planning observational surveys and proposals. 
Finally, we provide scaling relations between properties obtained with different fitting setups and population models, which will help unifying results obtained from different data sets and using different modelling.\\
This work complements a recent paper \citet[][hereafter M10]{M10} in which we focused on real $z\sim2$ star-forming galaxies from the GOODS survey and studied how well - or how bad - star formation histories that are typically adopted for the fitting templates are able to recover the galaxy star formation rates, stellar masses and reddening. One of the main conclusions of this work is that - because of the overshining effect from massive stars even if present in small proportions - essentially no star formation history is able to provide an accurate determination of star formation rates, reddening and masses if the age is left as a free parameter in the fit. This happens because the fit tries to match the light from the most massive stars, which is best accomplished with a low age. As a main conclusion a template was found that was able to recover the galaxy properties very well, which is an exponentially increasing star formation - that we named {\it inverted}-$\tau$ models as opposite to the usually assumed declining star formation or $\tau$-model but with two important priors, namely: i) that the starting of star formation was at a much earlier epoch with respect to the time of observation ($z\sim5$); ii) that the typical timescale for star formation ($\tau$) could not be too short, namely $\tau >0.3$ Gyr. This conclusion is able to provide information on the actual star formation mode of high redshift star-forming galaxies, which triggers a lot of astrophysical consequences.\\
In the present paper we do not pursue the aim at unveiling the galaxy star formation mode. Rather, we quantify the galaxy property offsets that are obtained with usually adopted templates expanding upon the investigation of  M10 by exploring a wider parameter space in terms of reddening, metallicity, etc..\\
For real passive galaxies at $z\sim2$, M06 found that a fitting setup including a wide range in metallicities, star formation histories and reddening laws was optimal as the derived best-fitting star formation histories were able to match the observed strength of the Mg$_{2800}$ line-strength in the galaxy spectra, which was regarded as a test independent of the photometric SED-fitting. In the present work we test the same fitting setup on mock passive galaxies and we reach similar conclusions.\\
%Paper organisation
The paper is organised as follows. We introduce the mock galaxy samples in sections \ref{mocks}. The method of SED-fitting and the different fitting setups are described in section \ref{sed}. Results and discussion are presented in section \ref{results}. The comparison of our results with the literature is carried out in section \ref{litcomp}. In section \ref{scalerel} we provide the scaling relations. We summarize our work in section \ref{summ}.\\
%General remarks like Cosmoslogy, magnitudes...
Throughout the paper we use a standard cosmology of $H_0=71.9$ km/s/Mpc, $\Omega_{\Lambda}=0.742$ and $\Omega_{M}=0.258$ as used in GalICS \citep[see][]{Galics} and magnitudes in the Vega system. %\footnote{Note that we run \hs \space with slightly different cosmological parameters for consistency with works on observed samples. This has however no effect on the final results.}

%%%%%%%%%%%%%%%%%%%%%%%%%%%%%%%%%%%%%%%%%%%%%%%%%%
%%% MOCK GALAXY SAMPLE
%%%%%%%%%%%%%%%%%%%%%%%%%%%%%%%%%%%%%%%%%%%%%%%%%%

\section{Mock galaxies}\label{mocks}
In the following we describe the samples of mock galaxies used to carry out our analysis. A summary of their properties is given in Table \ref{mockprops}.

\subsection{Star-forming galaxies from a semi-analytic model}
\begin{figure*}\includegraphics[width=84mm]{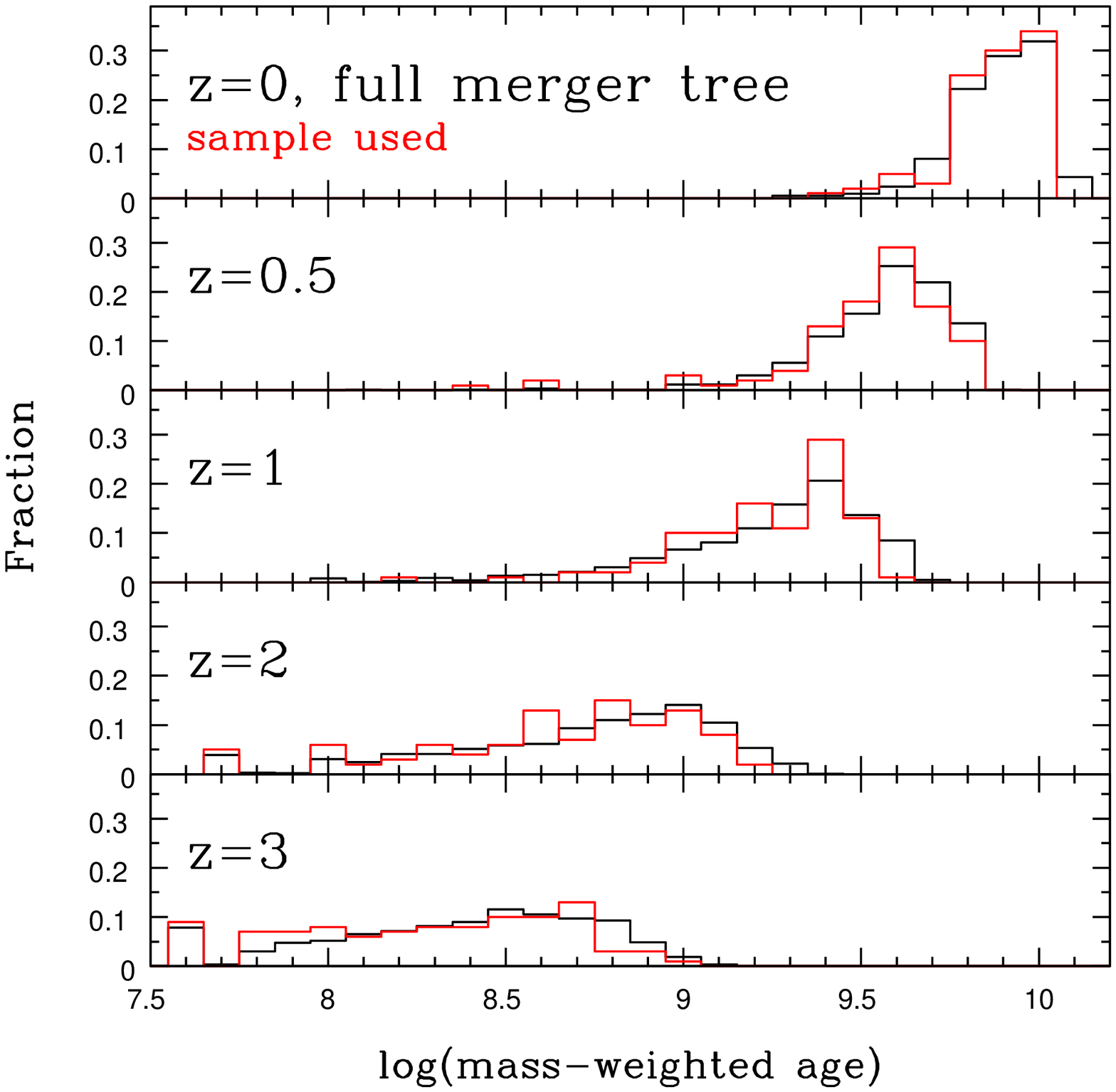}
\includegraphics[width=84mm]{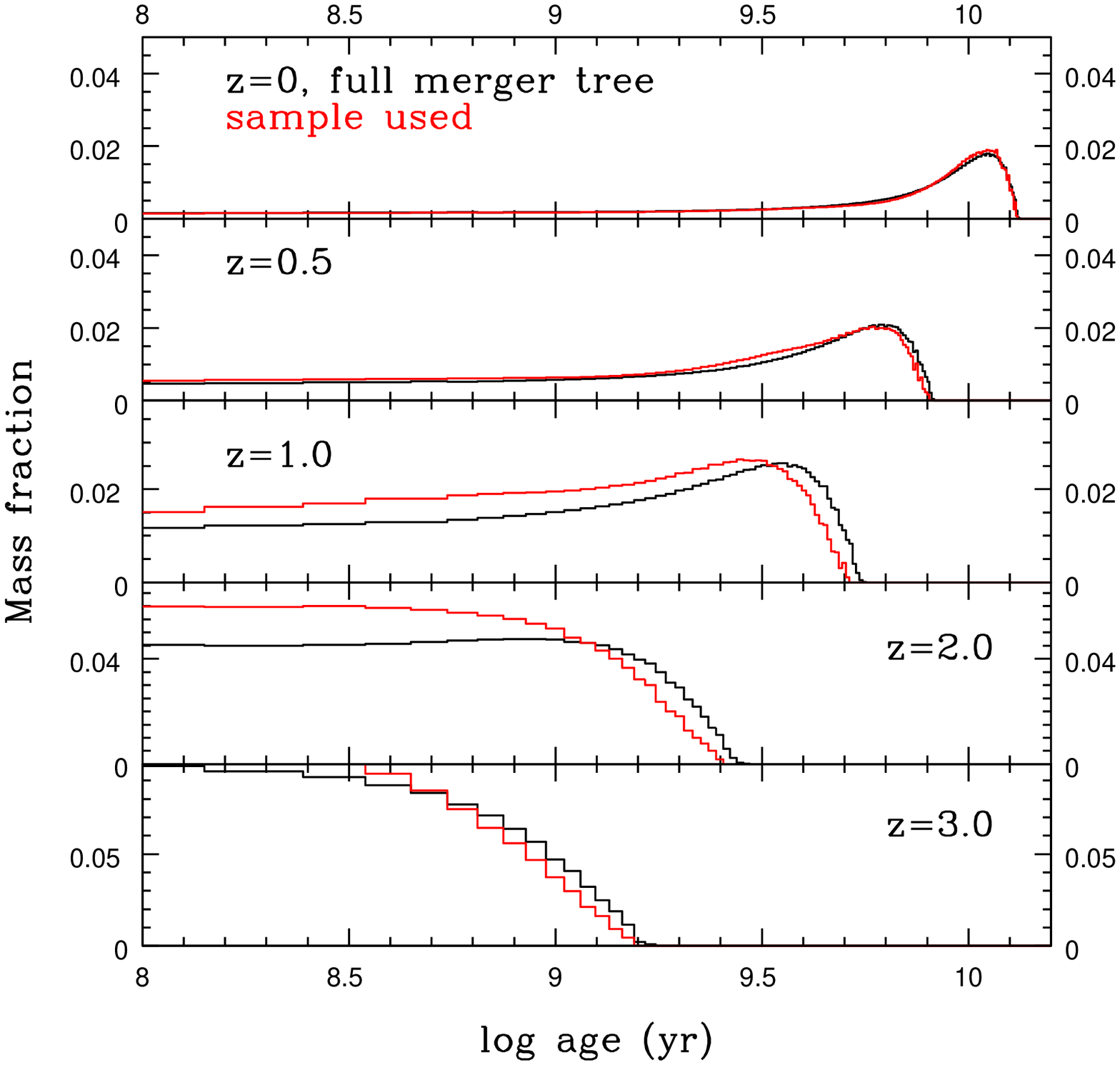}
\caption{\label{inputage} Left: Mass-weighted ages of mock star-forming galaxies at redshifts 0, 0.5, 1, 2 and 3 for the entire merger tree (black) and the sub-sample used in this paper (red). Right: Age distribution.}
\end{figure*}
\begin{figure}\includegraphics[width=84mm]{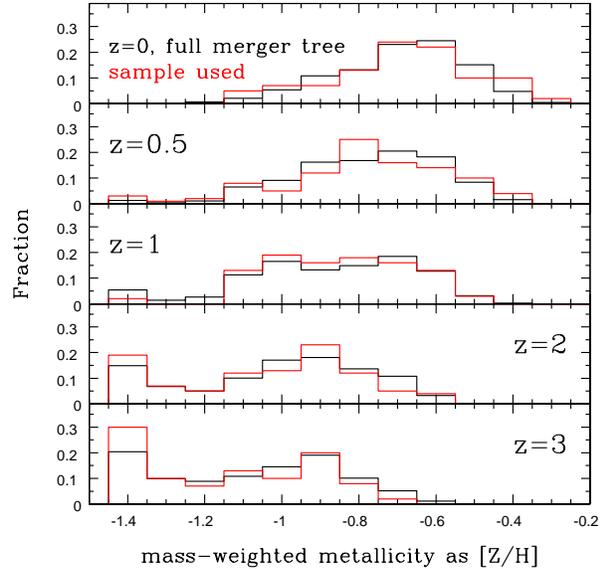}
\caption{\label{inputmetal} Mass-weighted metallicities of the mock star-forming galaxies at redshifts 0, 0.5, 1, 2 and 3 for the entire merger tree  (black) and the sub-sample used in this paper (red). }
\end{figure}
\begin{figure*}
\includegraphics[width=84mm]{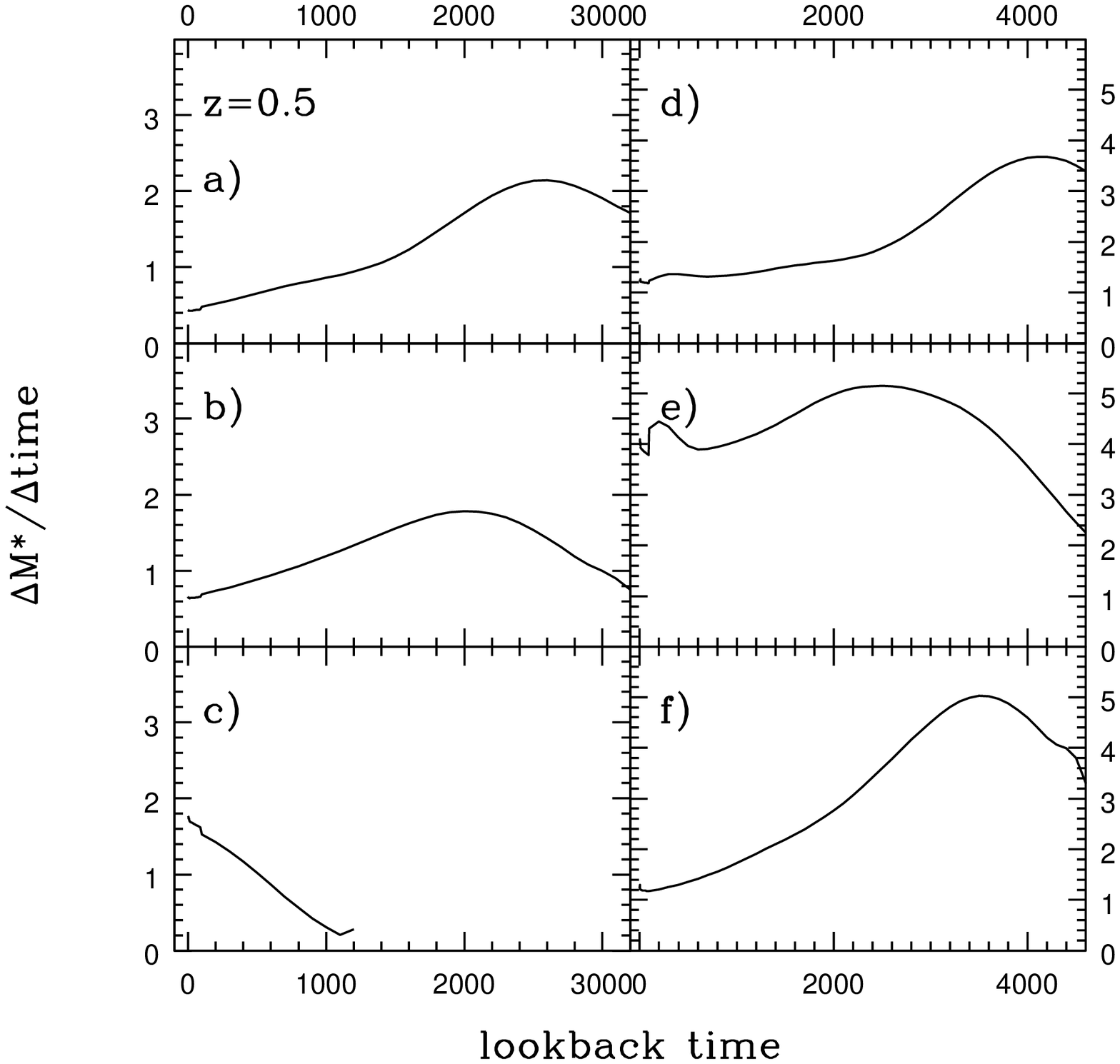}
\includegraphics[width=84mm]{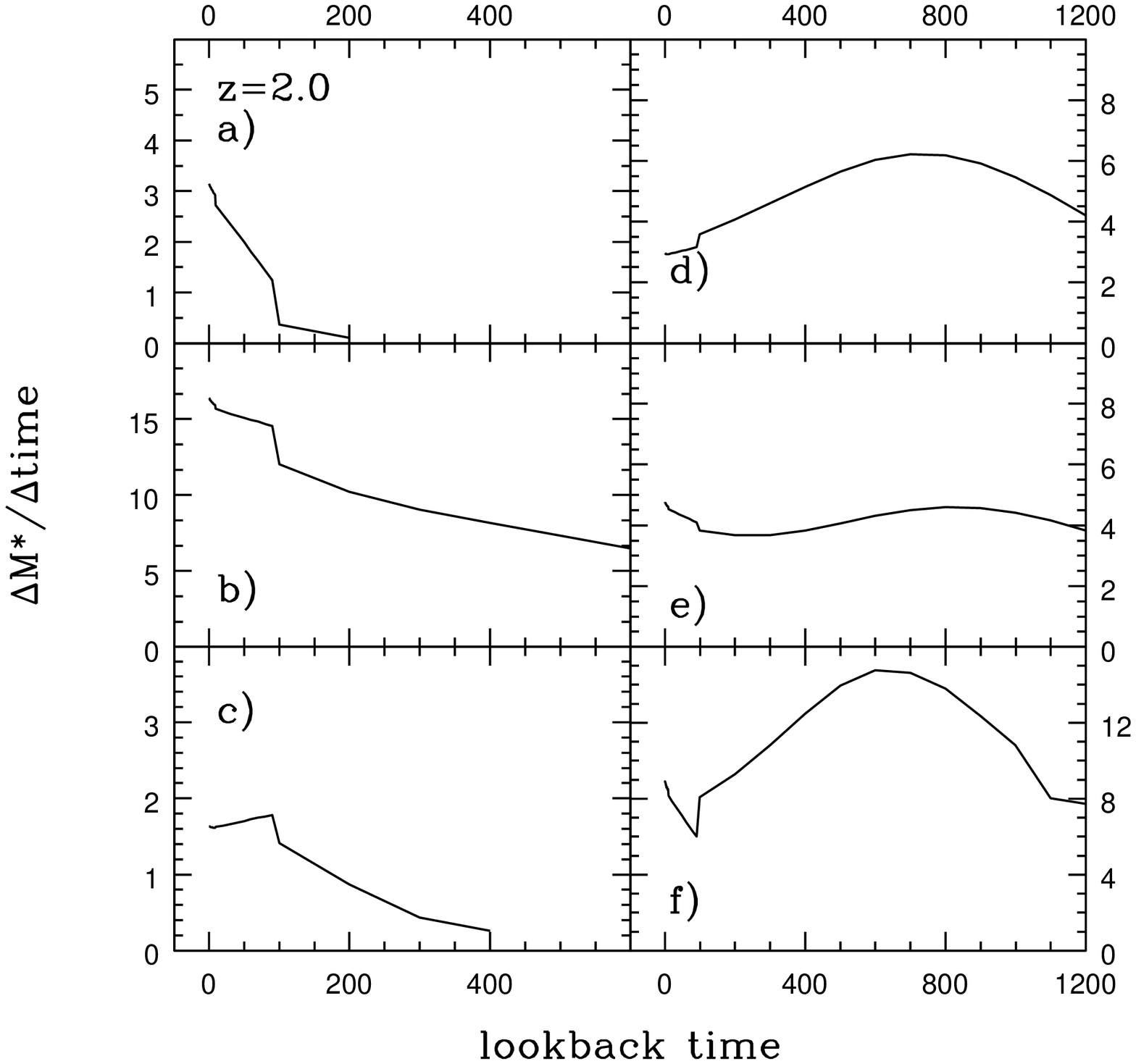}
\caption{\label{inputsfh} Selection of star formation histories of semi-analytic galaxies at z=0.5 and 2. The zeropoint of the lookback time lies at the given redshift. Galaxy star formation rates at $z=2$ increase, while those at $z=0.5$ predominantly decline. Galaxies at redshift 2 are younger than those at redshift 0.5 (see Fig. \ref{inputage}).}
\end{figure*}
\begin{figure}\includegraphics[width=84mm]{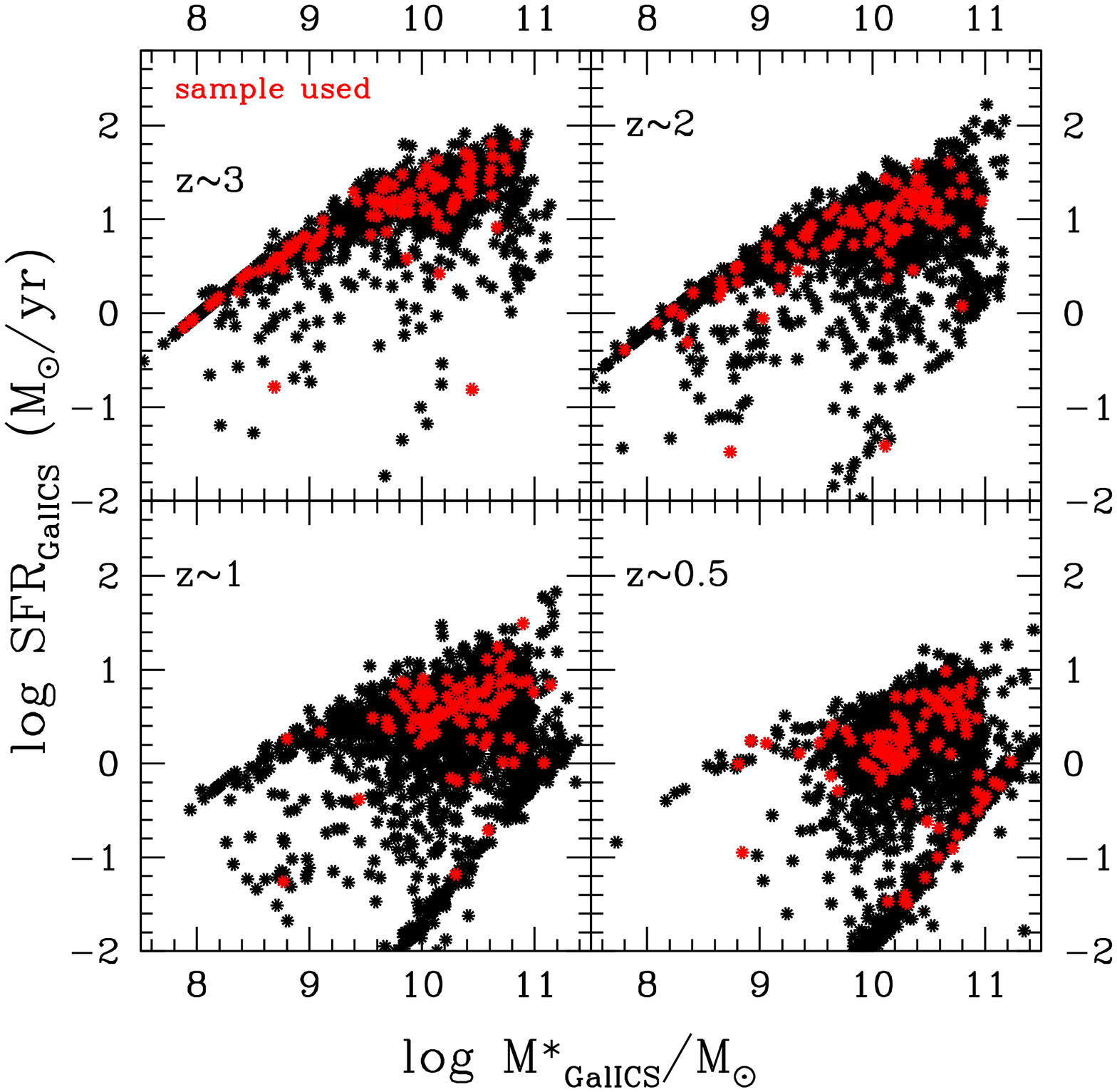}
\caption{\label{input} Stellar masses and star formation rates of mock star-forming galaxies at redshifts 0.5, 1, 2 and 3 for the entire merger tree (black dots). The sub-sample used in this paper is marked in red. In the bottom right of each panel the build-up of the red sequence towards lower redshift is visible.}
\end{figure}

%\textbf{Chiara's part} \\
We extract model galaxies from the hybrid hierarchical model GalICS \citep{Galics}. This uses dark matter merger trees obtained with an N-body simulation covering the redshift range $\sim 35 >z=0$. Baryonic physics is implemented semi-analytically following standard recipes \citep[for details about GalICS see Hatton et al. 2003 and for a review see][]{Baugh2006}.
The latter includes gas cooling, star formation, chemical enrichment, feedback from supernovae and AGN and treats the dynamical interactions of galaxies in clusters and groups, such as tidal stripping, dynamical friction and mergers.\\
The mass distribution inside each single galaxy is governed by its evolution.
Cooling of hot gas into the dark matter potential wells forms disks where star formation
occurs depending on the mass of the available cold gas and on the galaxy dynamical
time with the star formation efficiency as a free parameter.
Secular evolution in disks and mergers produces bulges at the centres of disks.\\ 
For each galaxy and its components (disk, bulge, burst) separately the code produces rest-frame spectral energy distributions. We use the GalICS version in which these are based on the Salpeter IMF \citep{salp} M05 stellar population models \citep{Tonini2009, Tonini2010}. We then apply dust reddening to the total spectra. Following \citet{Tonini20092}, we adopt a Calzetti extinction curve and a colour-excess $E(B-V)$ proportional to the galaxy star formation rate, parameterized as $E(B-V) = 0.33 \cdot (Log(SFR)-2) + 1/3$  and $E(B-V) = 0$ for SFRs less than $10\,M_{\odot}/yr$. This parameterization is based on the reddening derived for a sample of real star-forming galaxies at $z\sim2$ in the GOODS fields \citep[][M10]{Daddi2007a}.\\
We do not randomize the inclination of disk galaxies (which reduces the dust effect in face-on objects) but we redden the spectrum by the total amount of extinction calculated for each galaxy, therefore considering the maximal reddening for each object. As in \citet{Tonini2010} the mock catalogues are built by redshifting the (unreddened and reddened) spectra in each simulation timestep to the corresponding redshift value, obtaining the observer's frame spectra. We then filter the spectra with a chosen set of filters to obtain the observer's frame broad-band magnitudes. Finally, we scatter the magnitudes with Gaussian errors with 3 $\sigma$ errorbars where $\sigma=0.1$. In the fitting we adopt errorbars typical of those of the COSMOS survey \citep{Capak2007} for mock galaxies at redshifts $\leq1$ ($\sim$ 0.3 mag for near-IR bands and $\sim$ 0.05 mag otherwise) and those of GOODS star-forming objects \citep{Gia2004} for mock galaxies at redshift $\geq2$ ($\sim$ 0.1 mag). A minimum photometric error of 0.05 mag is applied in the fitting when the given photometric error in this band is smaller than 0.05 mag. We use a sample of 100 mock galaxies in both flavours - with dust reddening added as described above and without dust reddening - at each of the five redshifts z=0, 0.5, 1, 2 and 3, respectively.\\
The fitting method is described in section \ref{sed}. \\
The stellar population properties of the mock galaxies of the full merger tree are shown in conjunction with those of the 100 galaxy sample used in this paper in Figs. \ref{inputage}-\ref{input}. The mass-weighted ages of galaxies at z$\sim$0, 0.5 and 1 are confined around a few Gyr (Fig. \ref{inputage}, left). However, star formation is never completely shut off in semi-analytic models such that low redshift objects still contain a small fraction of populations that are much younger than the mass-weighted age. This is shown in Fig. \ref{inputage}, right hand panel. Therefore, the intrinsic stellar populations of these galaxies cover a much broader age range than those at higher redshift. This will have consequences on the capability of simple composite models to recover the actual properties of these galaxies (see section \ref{results}). Mass-weighted metallicities increase with redshift, but hardly reach half-solar metallicity by z$\sim$0 (Fig. \ref{inputmetal}). Note, that the latest formed populations have indeed solar and super-solar metallicities but their contribution to the total stellar mass is very small. 
At high redshift, the mass-weighted ages of most galaxies are younger than 1 Gyr, and much more similar one to each other with respect to those at low redshift. As expected, metallicities are lower.\\
As a further illustration we show the typical star formation histories of galaxies at redshift 0.5 and 2 in Fig. \ref{inputsfh}. Galaxies at redshift 2 show predominantly increasing star formation rate while for most galaxies at redshift 0.5 the star formation rate decreases with time. Finally, star formation rates are shown as a function of stellar mass in Fig. \ref{input}. The number of low mass objects decreases towards z=0.5. Most objects have very low star formation rates.\\
It is important to notice that - from the stellar population properties point of view - galaxies at z=2 and 3 and galaxies at z=0.5 and 1 are similar, and with regard to various assumptions in the fitting. The wavelength coverage is obviously redshift sensitive and conclusions may be different.\\
Note that we do not show the z$\sim$0 objects further because they are more easily accessed by other diagnostic methods such as spectroscopy, and we shall study them in a separate publication.

\begin{table*}
\caption{Properties of mock star-forming and mock passive galaxies. Values given for mock star-forming galaxies refer to the sample used in this paper, not to the entire merger tree.\protect\\
$^1$ This reflects the range in ages of the oldest population.\protect\\
$^2$ This reflects the range in mass-weighted metallicities.\protect\\
$^3$ $E(B-V)= 0.33 \cdot (Log(SFR)-2) + 1/3$  and $E(B-V) = 0$ for SFRs less than $10\,M_{\odot}/yr$ for mock star-forming galaxies. No internal dust reddening was applied to the mock passive galaxies.\protect\\
$^4$ Observed K-band magnitudes from unreddened spectra after application of randomisation with magnitude errors. For mock passive galaxies the brightest magnitude refers to the youngest and most massive object and the faintest to the oldest and least massive object at the given redshift.}
\begin{center}
\begin{tabular}{@{}ccccccccc}\hline
galaxy type & redshift & number & ages $^1$ 		& metallicity $^2$	& M* & SFR & E(B-V) $^3$ & K mag $^4$\\
& & & [Gyr] & [Z/H] &  [log $M_{\odot}$] &  [$M_{\odot}$/yr] & [Vega] \\\hline
star-forming & 0.5	  & 100	& 0.8 - 8.0				& -1.35 - (-0.36)	& 8.18 - 11.23					& 0.00 - 9.62			& 0			& 17.07 - 24.66\\\hline
star-forming & 1.0	  & 100	& 0.6 - 5.1				& -1.35 - (-0.47)	& 7.79 - 11.14					& 0.00 - 31.13			& 0 - 0.16	& 17.94 - 26.75\\\hline
star-forming & 2.0	  & 100	& 0.2 - 2.5				& -1.35 - (-0.57)	& 7.80 - 10.84					& 0.03 - 40.42			& 0 - 0.20	& 19.74 - 25.76\\\hline
star-forming & 3.0	  & 100	& 0.2 - 1.5				& -1.35 - (-0.69)	& 7.88 - 10.83					& 0.15 - 64.09			& 0 - 0.27	& 20.18 - 26.00\\\hline
passive 	    & 0.5	  & 56	& 1, 1.5, 2, 3, 4, 5, 6, 7	& 0.00		  	& 10.50 - 12.00					& 0					& 0 			& 13.76 - 19.09\\\hline
passive 	    & 1.0	  & 42	& 1, 1.5, 2, 3, 4, 5		& 0.00			& 10.50 - 12.00					& 0					& 0 			& 15.52 - 20.50\\\hline
passive 	    & 2.0	  & 21	& 1, 1.5, 2			& 0.00			& 10.50 - 12.00					& 0					& 0 				& 17.48 - 21.46\\\hline
passive 	    & 3.0	  & 14	& 1, 1.5				& 0.00			& 10.50 - 12.00					& 0					& 0 			& 18.55 - 22.60\\\hline
\end{tabular}
\end{center}
\label{mockprops}
\end{table*}%


\subsection{Passive galaxies}\label{pass}
Star formation in the semi-analytic model is never completely shut off. Therefore, each galaxy will always contain young and bright stars although their contribution to the total mass may be very small. In order to study the case of truly passive and old galaxies without any young stars, we consider a set of solar metallicity, Salpeter IMF, simple stellar populations of M05 and place them at the various redshifts. We treat the photometry of mock passive galaxies in a similar way to that of mock star-forming galaxies, i.e. the observer's frame spectra are convolved with broad-band filter bands and the resulting broad-band magnitudes are scattered with Gaussian errors with 3 $\sigma$ errorbars where $\sigma=0.1$ mag. Photometric errors in the fitting are set to 0.1 mag. The ages of mock passive galaxies range from 1 to 7 Gyr at $z=0.5$ and from 1 to 1.5 Gyr at $z=3$ (see Table \ref{mockprops} for details). Stellar masses vary between $10^{10.5}$ and $10^{12}\,M_{\odot}$ in steps of 0.25 dex. Internal reddening was not added to these passive galaxies.\\
The choice of a SSP as star formation history for passive galaxies at high redshift might not be very realistic and indeed studies by e.g. \citet{cim08} indicate that short-lived starbursts with duration times of $\sim0.1$ Gyr are required in order to reproduce some of the observed early-type galaxies at $1<z<2$. However, a SSP reflects the most extreme case which in conjunction with the mock star-forming galaxies enclose cases in which star formation might have lasted a short time before passive evolution set in. Thus the case of real passive galaxies at high redshift will lie in between these two boundaries. We refer to \citet{Longhetti08} for the study of mock passive galaxies with exponentially declining star formation histories.\\
%In light of this the results for the passive galaxies in this paper should be directly applied only to passive galaxies with no signs of recent star formation activity. For early-type galaxies that show signs of recent activity one has to expect larger uncertainties in the derived parameters due to the overshining effect of the young stellar populations.} 

%%%%%%%%%%%%%%%%%%%%%%%%%%%%%%%%%%%%%%%%%%%%%%%%%%
%%% SED-FITTING AND HYPERZSPEC
%%%%%%%%%%%%%%%%%%%%%%%%%%%%%%%%%%%%%%%%%%%%%%%%%%

\section{Spectral Energy Distribution Fitting}\label{sed}
%Hyperz, Hyperzspec, Input, Output fitting method...
Spectral Energy Distribution (SED) fitting is based on the comparison between theoretical template spectra and observed broad-band magnitudes.
The SED-fitting is carried out in the same way as described in M06 and M10. We use a modified version of the \emph{HyperZ}-code \citep{Bol2000} in which redshift can be fixed at its spectroscopic value for each object (\hs , M. Bolzonella, private communication). The fitting method is based on a $\chi^2$-minimisation\footnote{$\chi^2$ is defined as $\chi^2=\sum_{i=1}^{N_{filters}}{[\frac{F_{obs,i}-b\times F_{temp,i}}{\sigma_i}]}^2$ in which $F_{obs,i}$ and $F_{temp,i}$ are the observed and template fluxes in filter i, $\sigma_i$ is the photometric uncertainty and b is a normalisation factor. The reduced $\chi^2_{\nu}$ is $\frac{\chi^2}{\nu}$ with $\nu$ as the number of degrees of freedom.}. The $\chi^2_{\nu}$ is computed for templates covering a broad range of star formation modes, ages and metallicities. The combination of parameters that provides the minimum $\chi^2$ is the best-fit solution. \\
\hs \space provides various empirically-derived laws to treat galaxy internal reddening: Milky Way by \citet{Allen76}, Milky Way by \citet{Seaton79}, Large Magellanic Cloud by \citet{Fitz86}, Small Magellanic Cloud by \citet{Prevot84}, Calzetti's law for local starburst galaxies \citep{Calzetti} and no reddening.\\ 
Galaxy ages are constrained to be younger than the age of the Universe at each redshift.\\ 
The best fit solution yields the galaxy age, star formation law, metallicity and dust reddening. The age is defined as the time elapsed since the onset of star formation, which is therefore the age of the oldest stellar population present in the template. The normalisation between best fit template and observed SED allows us to calculate stellar mass and star formation rate using codes developed in \citet{Daddi05} and M06.\\
SED templates are based on the simple and composite stellar population models of M05. In the following subsections we describe the variety of template and filter setups in use.

%%%%%%%%%%%%%%%%%%%%%%%%%%%%%%%%%%%%%%%%%%%%%%%%%%
%%% TEMPLATE SETUPS
%%%%%%%%%%%%%%%%%%%%%%%%%%%%%%%%%%%%%%%%%%%%%%%%%%

\subsection{Model Template Setups}\label{difftemp}

\subsubsection{Wide template setup}%M06 template setup
We adopt a mixed template setup as in \citet[][hereafter wide setup]{M06} as default setup. It consists of  32 types of theoretical  spectra which cover a wide range of star formation histories - specifically: i) SSP (single starburst), ii) exponentially declining SFR (so-called $\tau$-model with $\tau=0.1, 0.3$ and $1$ Gyr), iii) truncated SFR (constant star formation for a time t with $t=0.1, 0.3$ and $1$ Gyr, no star formation afterwards) and iv) constant SFR - and metallicities of  $\frac{1}{5}\,Z_{\odot}$, $\frac{1}{2}\,Z_{\odot}$, $Z_{\odot}$ and $2\,Z_{\odot}$. A Salpeter IMF is assumed, consistent with the mock galaxies, but we also explore different IMFs.\\
Despite the increase in CPU time that is required to fit data with the wide setup, it was found by M06 that this approach gives the most accurate results for the sample of passive galaxies in their study. For example, as an independent constraint, they find that predictions for the Mg$_{UV}$ line strength from the best SED-fit agree very well with the observed line strengths in the GOODS-S sample \citep{Daddi05}.\\
However, as well known, passive galaxies are easier to fit than star-forming ones and also metallicity effects are more evident in passive spectra. Hence, the result of M06 may not be confirmed for other classes of galaxies.\\
In order to study the effect of metallicity, we also obtain best fit solutions using the wide setup in mono-metallicity form.  Furthermore, we retrieve separate solutions for the full setup in combination with different reddening laws to reveal their effect on the fitting.\\
We also consider an identical setup based on BC03 models (hereafter BC03 setup). It is well known that these models mostly differ from the M05 models in the energetics of the TP-AGB and at the young ages \citep{M11BF}. From this viewpoint, these two stellar population models represent two extremes which enclose other ones, e.g. Charlot \& Bruzual (CB07), \citep[e.g.][]{cim08,Conroy2009}. As BC03 models are widely used in the literature using them for the same set of mock galaxies allows us to provide scaling relations which help translating the results obtained with BC03 models to M05 models and vice-versa. Results are summarised in Appendix \ref{bc03results}. 

\subsubsection{Only-$\tau$ setup}
Exponentially declining SFH or $\tau$-type models are a common choice in the literature \citep[e.g. ][]{Papovich2001,S05,Pozzetti2007,Ilbert2010}. In M10 we showed that for real star-forming galaxies as well as mock star-forming galaxies at z=2 $\tau$-models do not provide a good approximation of the star formation history.\\
Here, we try to quantify the effect of this mismatch by adopting a template setup which contains $\tau$-models (only-$\tau$ setup, with $\tau=0.01, 0.05, 0.1, 0.2, 0.5, 1, 2$ and 5 Gyr) and constant star formation, solar metallicity and a Salpeter IMF as in \citet{S05} (but based on M05). Furthermore, \citet{S05} used a Calzetti type reddening law and models from BC03. We explore these templates by exploiting various reddening laws.

\subsubsection{Inverted-$\tau$ models}
Exponentially increasing SFHs (so-called {\it inverted}-$\tau$ models) were put forward in M10. It was found there that - with priors on the galaxy formation redshift and the typical $\tau$ - these models were the only ones able to recover the SFR and the reddening that was determined from independent indicators such as far-IR luminosities or the UV-corrected slope. In M10 we also used a set of mocks from this paper and show that the same setup also gave excellent results on the SFR-mass-relation of semi-analytic galaxies.\\
In this paper we extend the test to mocks at other redshifts, using the same assumptions, namely a formation redshift of $\sim$ 5, which translates into $\sim 1$ Gyr age models for fitting $z\sim3$ galaxies and $\sim 4$ Gyr age models for $z<2$.

\subsubsection{Age Grids} \label{agegrid}
%only with standard M06 setup, finer age grid more accurate
We also investigate intrinsic properties of the templates themselves, in particular the age grid. Templates normally get rebinned from 221 to 51 values in age within the \hs-code. The original age grid is equally spaced in logarithmic space. Hence, the rebinning has a larger effect on younger ages. The latest version of the code considers 221 ages without rebinning. In both cases minimum and maximum age are 100,000 yrs and 20 Gyr, respectively. We adopt the finer age grid of 221 ages as default and test the effects of using the wide setup with a rebinned age grid.\\
Furthermore, we study the effect of excluding very young ages in the fitting by limiting the minimum age to 20 Myr and 100 Myr. We showed in M10 that solutions obtained with the exclusion of the youngest ages are less biased by very young and bright stellar populations that dominate the emitted light, thus disabling their outshining effect. Similar minimum age constraints are applied by \citet{Bol2009} and \citet{Wuyts2009} in order to exclude very young solutions and to improve the mass estimate. \citet{Bol2009} verified this through tests with simulations.\\
Age grids are only varied for mock star-forming galaxies as these are affected by overshining; our mock passive galaxies are by definition devoid of this\footnote{We have verified that a minimum age constraint of 0.1 Gyr has no effect on the final results for the mock passive galaxies.}.

\subsubsection{Varying the IMF} %nur mit standard M05 setup
%only with standard M06 setup to see the effect of the IMF (e.g. with Kroupa less heavy)
A further interesting question to address is the sensitivity of the SED-fit to IMF choices as the IMF is a free parameter in population synthesis models. Commonly a conversion factor between Salpeter and other IMFs is assumed to convert stellar masses and other quantities, for example masses obtained with a Kroupa IMF are smaller by a factor of $\sim1.6$ when compared to Salpeter (M06). Here, we have the opportunity to assess whether a simple scaling is sufficient. We use the wide setup with templates based on a variety of different IMFs, namely Salpeter, Kroupa \citep{kro01}, Chabrier \citep{chab} and a top- heavy IMF (with slope $x\sim0$). For the Chabrier and top-heavy IMFs we adopt the slopes as given in \citet{vD08}.

\subsection{Wavelength range included in the fitting} \label{filters}
%only with standard M06 setup, to see at which point robustness of results breaks down
Which wavelength coverage is required in order to gain robust results from SED-fitting? How many broad-band filters should be used? These are key questions in galaxy evolution studies especially for the planning of galaxy surveys and observational proposals and for understanding the reliability of the results. In order to answer these questions we carry out a comprehensive test of various filter setups and their performance in recovering the galaxy physical properties. M06 discuss the importance of including the rest-frame near-IR in the fitting (illustrated in their Figure 9). \citet{Kannappan} also point out the dependency of stellar mass estimates on filter type and wavelength coverage. \citet{Ilbert2010} also find differences in the derived stellar masses depending on the inclusion or exclusion of the rest-frame near-IR for galaxies at $z>1.5$ and conclude that the rest-frame near-IR should be included in the fitting. \citet{S05} on the other hand conclude that the near-IR does not carry additional information with respect to a fitting carried out in the optical, using BC03 population models. \citet{vdW06} even conclude - in combination with the M05 stellar population models - that including the near-IR could even be damaging the results.\\
The light of a galaxy is a superposition of the light emitted by its stellar populations and the wavelength range in which a single stellar population emits most of its light changes with age. Hence, by sampling the entire spectrum there is generally a better hope to trace all different stellar populations in a galaxy whereas restricting the filter set to certain wavelengths may result in over-emphasising the contributions of specific populations. In order to better understand the extent of this effect, we explored the following filters sets:\\
\begin{enumerate}
\item UBVRI JHK IRAC (3.6, 4.5, 5.8 $\mu$m), UBVRI JHK, UBVRI JH, UBVRI J, UBVRI
\item UBVRI IRAC
\item BVRI JHK IRAC, VRI JHK IRAC, RI JHK IRAC
\item u' g' r' i' z' (SDSS) [z=0.5]
\item UBVRI \emph{Y} JHK IRAC
\item BRIK
\end{enumerate}
Obviously, the assumption that all mock galaxies are detected in all filter bands is an idealised case. In particular, the majority the mock passive galaxies at $z\geq2$ would not be detected in filter bands bluer than V (I for log M*$\sim10.5$) in even the deepest surveys, such as e.g. GOODS \citep{Gia2004}, because of their old age and SFH. However, this case is addressed when filter setups that omit the bluest filter bands are used in the fitting.  

\subsection{Photometric uncertainties}\label{photunc}
In order to quantify the effect of photometric uncertainties we use two different catalogues of the same 100 galaxies\footnote{without dust reddening} at $z=0.5$. One catalogue contains the magnitudes that were scattered with Gaussian errors as described in section \ref{mocks}, the other the true magnitudes without this scatter. For this exercise we focus on the mock catalogues created from unreddened spectra. We derive the stellar population properties with the exact same method using our default setup and no reddening in the SED-fitting. The exclusion of reddening in the catalogues and fitting allows us to single out the effect of the photometric uncertainties without being affected by the age-dust degeneracy.\\ 
The influence of photometric uncertainties for the sample of mock passive galaxies is determined in the same way but using only a solar metallicity SSP as template to provide the exact match in star formation history to the mock passive galaxies.

%%%%%%%%%%%%%%%%%%%%%%%%%%%%%%%%%%%%%%%%%%%%%%%%%%
%%% RESULTS
%%%%%%%%%%%%%%%%%%%%%%%%%%%%%%%%%%%%%%%%%%%%%%%%%%

\section{Results}\label{results}
\begin{figure}\includegraphics[width=84mm]{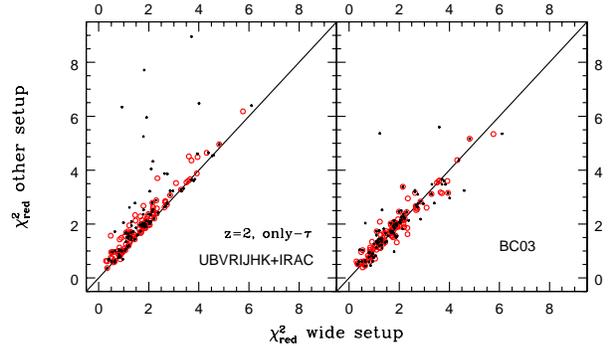}
\caption{\label{chidt} Minimum $\chi^2_{\nu}$ of best fits with alternative setups in comparison to the wide setup, using UBVRIJHK+IRAC bands for mock star-forming galaxies at redshift 2. Black symbols show the un-reddened case (internal dust reddening is not included in the mock galaxies and the fitting is carried out without dust reddening). Red symbols show the case with dust reddening. The same results are found at all other redshifts.}
\end{figure}

%comparison plots for different setups as discussed above
In this section we compare the galaxy properties derived from SED-fitting of mock galaxies to the true values. The comparison is performed as a function of template setup (star formation history, metallicity, age grid, IMF, stellar population model\footnote{see Appendix \ref{bc03results}}), filter set and wavelength coverage, and reddening law. For the semi-analytic galaxies, we first produced mock catalogues without dust reddening to better single out the contributions of the multiple stellar populations building up the galaxies. For these unreddened galaxies we also do not consider reddening in the fitting. In the following, we refer to this as the 'case without reddening' or the 'unreddened case'. Although ignoring the existence of dust when simulating star-forming galaxies is an unrealistic case it provides an insight into the impact of the age-dust degeneracy. The procedure is then repeated including dust reddening, both in the mock catalogue and the fitting. For the mock passive galaxies no internal dust reddening has been included, but both the cases of including and excluding reddening in the fitting are studied. Table \ref{overfits} gives an overview of all fits that were carried out.\\
In order to evaluate the optimal template and filter sets - intended as the one for which fitted and input properties are closest - we define a quality estimator \emph{Q} similar to a $\chi^2$:
\[
Q=\sqrt{\sum_{i}{(\Delta)^2}}\,\,\rmn{\space with}\,\, \Delta=x_{i,fitting}-x_{i,input},
\]
 i being the number of objects and \emph{x} representing age, metallicity, reddening, stellar mass or SFR, respectively. The optimal value of \q \space and $\Delta$ is zero.\\ %Additionally, we calculate \q \space in different i-bins (see Tables \ref{overfits05red} - \ref{overfits3red} for a summary).\\
A first indicator of the individual performances of the various template setups is the $\chi^2_{\nu}$ of the best fit. We show results for the mock star-forming galaxies at z=2 in Fig. \ref{chidt} as an example. For nearly all objects the wide setup provides a better $\chi^2_{\nu}$ in comparison to the only-$\tau$ setup due to the wider template library with regard to metallicity and star formation histories. However, we showed in M10 that models with the smallest $\chi^2_{\nu}$ are not necessarily a better physical solution. This is further confirmed here by the fact that a BC03-based setup shows a better $\chi^2_{\nu}$ in some cases although the mock galaxies are based on the M05 models and therefore their SED is intrinsically different.\footnote{We summarise the results for fits with the BC03-setup to M05-based mock star-forming galaxies in Appendix \ref{bc03results} and vice versa in Appendix \ref{pegase}}. This example highlights again how misleading the mere consideration of the minimum $\chi^2_{\nu}$ can be.\\
%It is well known that fits including reddening have better $\chi^2_{\nu}$ because of the additional degree of freedom. \\
In the following we focus on the ability of different template setups and wavelength coverage to recover the age, metallicity, SFH, reddening and stellar mass. In order to ease the comparison between different setups and fitting parameters we always provide the results for the wide setup as reference in each figure. Furthermore, we list the ranges containing 68\% of the solutions along with the median in appendix \ref{overres}. 

\subsection{Star-forming galaxies}\label{sfgresults}
% all as function of template setup (wide, only-tau, inverted), wavelength coverage, redshift

\subsubsection{Age}\label{ageresults}
\begin{figure*}\includegraphics[width=144mm]{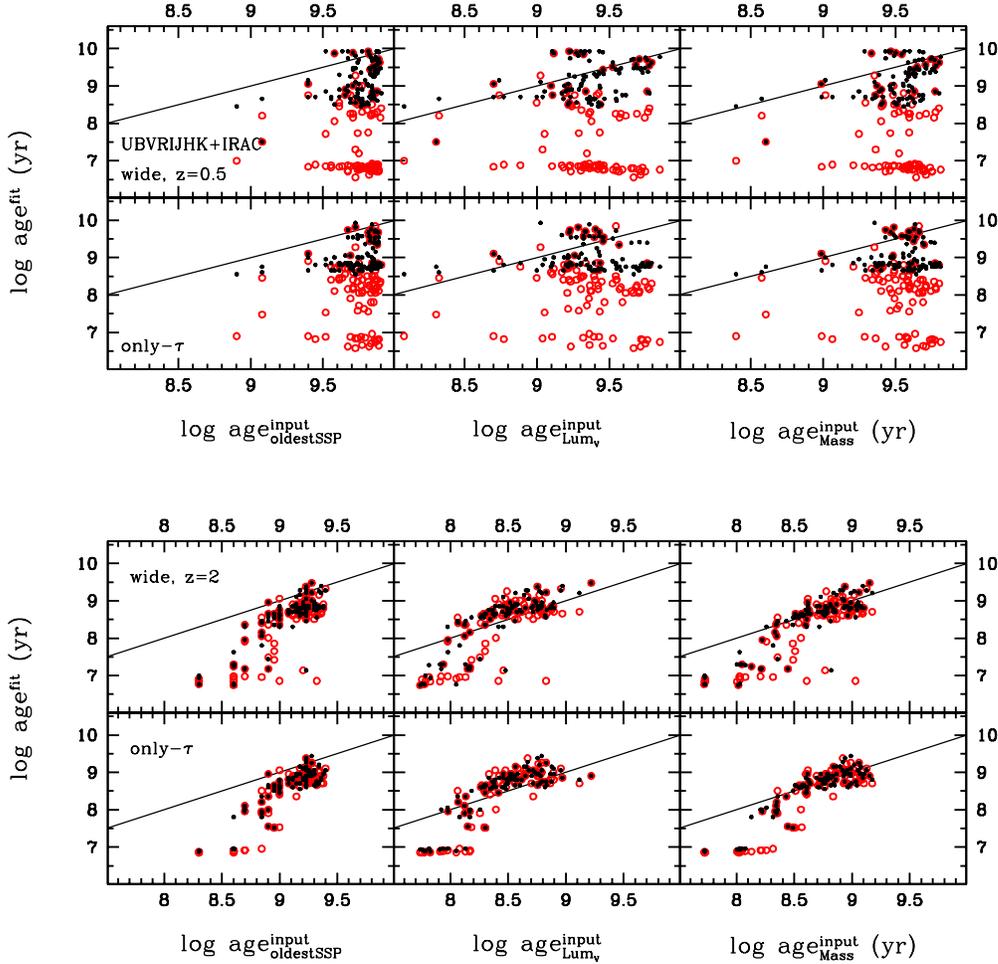}\\
\caption{\label{agedt05} Best fit ages (y-axis) compared to input ages (x-axis) of mock star-forming galaxies for the wide and only-$\tau$ setups at redshift 0.5 and 2 (top and bottom panels, respectively). Symbols as in Fig. \ref{chidt}. Left: Comparison with the age of the oldest population. Middle: comparison with $V$-band luminosity-weighted ages. Right: Comparison with mass-weighted ages.}
\end{figure*}

\begin{figure*}\includegraphics[width=124mm]{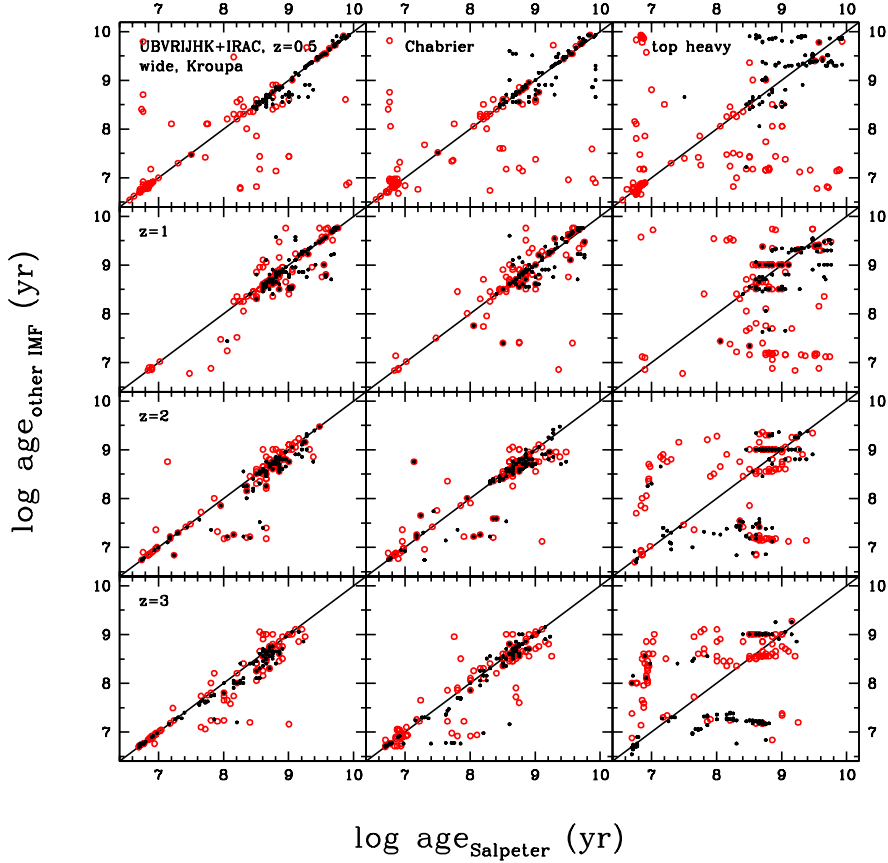}
\caption{\label{ageIMF} Comparison of ages derived from SED-fitting with templates of different IMFs. Symbols as in Fig. \ref{chidt}.}
\end{figure*}

\begin{figure*}\includegraphics[width=144mm]{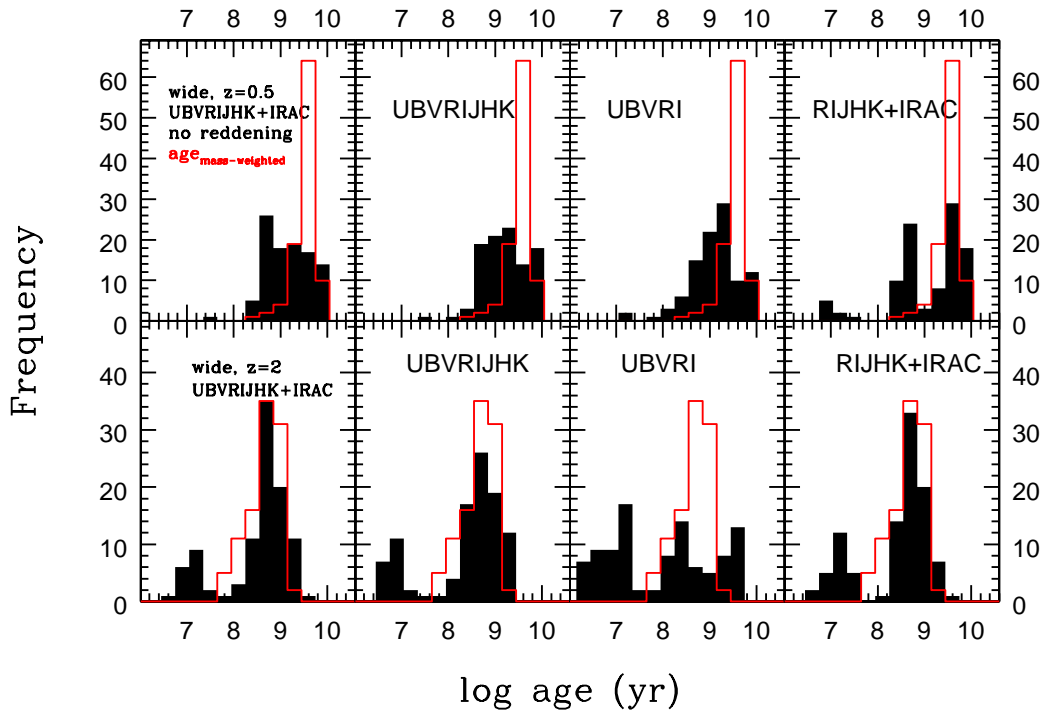}
\caption{\label{agedf} Ages derived from SED-fitting as a function of wavelength for redshift 0.5 and 2 in the case without dust reddening. The red histogram shows the distribution of the mass-weighted age.}
\end{figure*}

The comparison of fitted ages with intrinsic ages allows us to identify which \textit{age} (mass-weighted age, luminosity-weighted age\footnote{Ages are weighted by the $V$-band luminosity of the corresponding SSP models. We also consider the bolometric luminosity. Overall, the latter provides younger ages in comparison to the V-band luminosity weighted ages, but the overall trends are similar.}, age of the oldest population) is actually recovered by the fit. As discussed in M10, by definition, the age parameter $t$ of the templates and the fitting denotes the age of the oldest population that is present, as it marks the starting of star formation (see M10 for details). However, as one fits the integrated luminosity the true age is difficult to extract, particularly for galaxies with extended star formation due to the overshining effect. For example, it is known that fitting single-burst models to data give 'luminosity-weighted' quantities, which mostly reflect the properties of the most luminous generation.\\
In Fig. \ref{agedt05} we show the comparison for the wide and only-$\tau$ setups at redshift 0.5 and 2. We find that ages derived for dust-free spectra agree best with mass-weighted ages, particularly at high redshift. This is somewhat a new conclusion, and suggests that the use of composite templates allows to obtain ages that are closer to the mass-weighted ones than to the luminosity-weighted ones. The youngest fitted ages ($<\sim10^8$ yr) are underestimated most. With a wide setup mass-weighted ages are recovered in 68\% of the cases between -0.91 dex and 0.07 dex in the unreddened case (-2.83 dex and -0.37 dex in the reddened case) for old galaxies with little on-going star formation at low redshift. The median offsets between recovered and mass-weighted age are -0.35 dex and -1.78 dex in the unreddened and reddened case, respectively. At higher redshift mass-weighted ages are recovered better (see Table \ref{sfoverres2}).
At high redshift fitted ages are similarly comparable to the oldest ages while most are underestimated for the oldest galaxies (at z=0.5). 
$V$-band luminosity-weighted ages are predominantly overestimated for objects with higher SFRs (at z=2) and predominantly underestimated for objects with little star formation (at z=0.5), the latter is due to the overshining. \\
%from plot that compares ages between different templates: Plot taken out as differences can be seen in comparison plot with true ages!
Since the range of available ages in the fitting narrows towards higher redshift, differences in best fit age between template setups and between recovered and true age decrease towards higher redshift, particularly for ages $\geq 10^8$ yr.\\
For the reddened case, the age-dust degeneracy is strong and causes ages to be underestimated (by up to nearly 3 dex in the worst cases). The effect is less severe at higher redshift because of the smaller available age range. Here, only the youngest ages (below $10^8$ yr) are underestimated. 
For young and dusty star-forming galaxies (those at high redshift) the reason is twofold. Firstly, a small contribution stems from the age-metallicity degeneracy. Template metallicities are much higher than mock galaxy metallicities (compare Fig. \ref{inputmetal}). Thus, bluer colours are reproduced by younger ages.\\
Secondly and dominating, template SFHs do not match those of the semi-analytic galaxies resulting in the overestimation of some mass-weighted ages (at $z=2$).\\
For older galaxies with little on-going star formation (those at low redshift) and in the reddened case very young ages ($< 10^8$ yr) and a large amount of dust are obtained with the SSP component included in the wide template setup. The fit is clearly fooled by the overshining in combination with the age-dust degeneracy \citep[see][for a detailed description of this degeneracy and its effects]{ren06}. Semi-analytic galaxies at $z\leq1$ are hardly reddened due to their low SFRs (see section \ref{mocks}). Allowing for dust reddening in the fit drives the ages to younger values since the addition of dust mimics redder colours (see also Fig. \ref{ebvdt}, this effect is less strong for $z=2$ and 3 as the simulated galaxies are dust reddened). With the only-$\tau$ setup most of these objects are fitted with significantly less dust, longer $\tau$, higher metallicity and older ages. \\
Very young ages from the only-$\tau$ setup are due to the SFH and metallicity mismatch. In general the age determination is better at high redshift (see Tables \ref{sfoverres} and \ref{sfoverres2} for median values and 68\% ranges). Deviations at low redshift are large because of on-going star formation and the wide age range covered by the stellar populations in a single galaxy (Fig. \ref{inputage}).\\ 
In order to single out the effect of metallicity on the derived ages, we used mono-metallicity template setups, keeping the variety of star formation histories and IMF of the wide setup. Overall, the effect is small and therefore, it is not shown here. When reddening is switched off the effect is very clear: when the template metallicity is high, sub-solar metallicity galaxies are fitted with a younger age to compensate this mismatch. When reddening is introduced into the procedure, the age-dust degeneracy partly compensates metallicity effects. Only when the metallicity is overestimated the most, the number of objects with a younger best fit age increases significantly, particularly for the metal poorest objects (at high redshift).\\ %\textbf{The median offsets between recovered and mass-weighted age vary from $-0.34$ to $-0.91$ ($-1.55$ to $-2.49$) in the unreddened (reddened) case at the lowest redshift from the lowest metallicity template setup to the highest metallicity template setup (see Table). }\\
%\citet{M09} and Maraston \& Str\"omb\"ack 2010 have shown that even very small differences in the shape of the SED are powerful enough to rectify long standing problems. 
We now investigate how the differences in the model-SED due to different IMFs affect the derived best fit ages. These differ substantially between the various IMFs (Fig. \ref{ageIMF}), especially for a top-heavy IMF. Again, the case without reddening helps understanding the effect. At $z=3$ and $2$ top-heavy IMF models give systematically older ages than Salpeter IMF models. The excess of massive stars and thus $UV$ light for very young ages of the top-heavy IMF models is compensated by on older age. The same is true at low-redshift where ages for a top-heavy IMF are systematically older than ages for a Salpeter IMF. Ages derived with a Chabrier and Kroupa IMF are much closer to those derived with Salpeter. At intermediate ages (i.e. in the intermediate stellar mass range) the SED of a Kroupa and Chabrier IMF is slightly redder because of the effect on the evolutionary flux. This is compensated by younger ages. When reddening is included scatter increases, particularly at low redshift, due to the age-dust degeneracy. At high redshift ages obtained with top-heavy IMF templates are now much older than those of Salpeter IMF templates (and the fitted reddening is higher).\\
Finally, we study the dependence of the derived ages on the adopted wavelength range in the fitting which is shown for the unreddened case in Fig. \ref{agedf} for a few common filter sets. Since the emission of stellar populations of different ages peaks at different wavelengths one expects that a narrow wavelength range in the fit biases the result towards particular ages. In the unreddened case the effects are small for the oldest galaxies (z=0.5). Excluding IRAC filters and using only optical wavelengths in the fitting makes derived ages somewhat older. For higher redshift and intrinsically younger objects the opposite is the case. The more the wavelength coverage is focussed towards the rest-frame UV, the younger the derived ages. Neglecting blue filter bands has a slightly rejuvenating effect for the oldest galaxies, for younger ones ($z\sim2$) the effect is smaller because the rest-frame UV is excluded and underlying older stellar populations are emphasised in the fit. The inclusion of reddening has little impact on the general dependence of derived ages on wavelength coverage and is therefore not shown. At low redshift, however, the age-dust degeneracy dominates.\\
In summary, fitted ages compare best to mass-weighted ages and not to luminosity-weighted ages, as widely believed. The difficulty in recovering the oldest age remains which is mostly related to mismatches in the star formation history. In conclusion and perhaps not surprisingly, SSPs or $\tau$-models with low $\tau$'s should not be used when dealing with star-forming objects.\\
Since using mono-metallicity template setups has a very small effect on the derived ages compared to using a wide set of metallicities because age effects dominate over metallicity effects in star-forming galaxies, it is sufficient to use mono-metallicity template setups for star-forming galaxies.\\
The choice of IMF affects ages such that the wrong IMF can be compensated by a different age. The largest effects are seen for the IMF that is most discrepant (top-heavy IMF here) to the input IMF.\\
The wavelength coverage plays a crucial role in the determination of ages such that a restriction in wavelength coverage results in younger ages at high redshift. At low redshift ages are underestimated due to the age-dust degeneracy and the wrong SFH. Because of the mismatched SFH the derived age resembles the real one, e.g. the oldest population present, in hardly any case.\\
Median offsets between recovered and mass-weighted ages for various fitting setups are listed in Tables \ref{sfoverres} and \ref{sfoverres2}.
%Overall, the most important reason why ages are poorly determined is the mismatch between template and real SFH.
  
\subsubsection{Metallicity}\label{metalresults}
\begin{figure}\includegraphics[width=84mm]{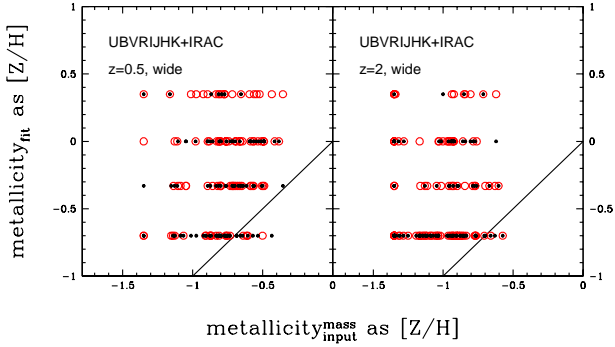}
\caption{\label{metalM06} Comparison between recovered and mass-weighted metallicity of mock star-forming galaxies at redshift 0.5 and 2 for the wide setup. Black and red symbols refer to the cases without and with dust reddening, respectively. Template metallicities are -0.7, -0.33, 0 and 0.35 (from a fifth solar metallicity to twice solar metallicity). The line of equality is shown.}
\end{figure}

We know that most template metallicities are too high compared to input metallicities. Consequently, metallicity is overestimated for most objects, particularly at high redshift (Fig. \ref{metalM06}). This is the case even when reddening is switched off and demonstrates the age-metallicity degeneracy. The highest metallicity templates of the wide setup are mainly chosen when reddening is allowed in the fit because of the added degeneracy with dust. These are best fit with a young and short starburst (SSPs or small values of $\tau$) and larger amounts of reddening.  In a statistic sense the fit is sensitive to the metallicity, such that the number of best fit templates with low metallicities is higher at high redshift. The opposite is the case for high metallicities.\\
%wavelength coverage
The derived metallicities depend only little on the wavelength coverage, therefore we do not show it. The lack in wavelength coverage affects the most when only optical filter bands are used such that most galaxies are best fitted with the highest metallicity templates.\\
Overall, for star-forming galaxies, metallicities are poorly recovered because ages are poorly recovered, i.e. ages are underestimated and metallicities are overestimated, which in turn is driven by the poor match of the SFH and the overshining effect.

\subsubsection{E(B-V)}\label{ebvresults}
\begin{figure*}
\includegraphics[width=124mm]{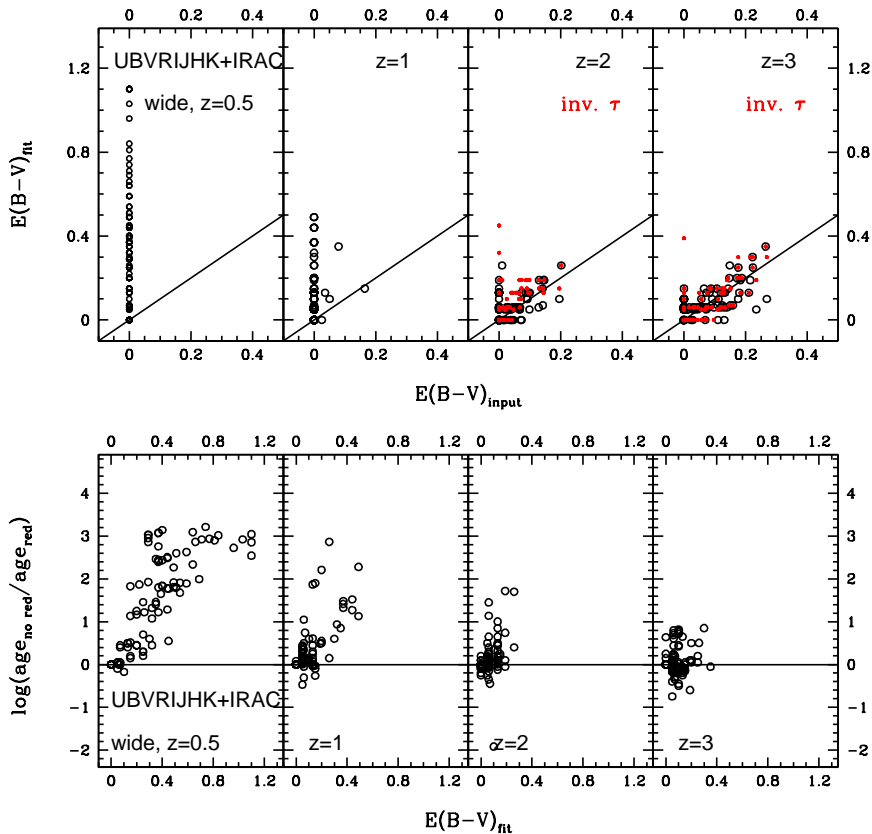}
\caption{\label{ebvdt} Upper panel: Difference between derived and input E(B-V) as a function of redshift (0.5, 1, 2 and 3, respectively, from left to right) for the wide setup (black open circles) and for the inverted-$\tau$ model (red dots, redshift 2 and 3). Lower panel: Difference in derived ages with and without internal reddening as a function of fitted reddening.}
\end{figure*}

We present the comparison between derived and input reddening for the wide setup and illustrate the age-dust degeneracy in Fig. \ref{ebvdt}. The reddening is very well recovered at $z\geq2$ for the wide and inverted-$\tau$ setups. Median offsets between recovered and true E(B-V) are 0 and 0.01, respectively. 68\% of the solutions have offsets between $\sim0$ and $\sim0.1$ (Table \ref{sfoverres2}). The only-$\tau$ setup provides slightly larger reddening values. This is the same result that M10 found for real $z\sim2$ star-forming galaxies. They showed that dust reddening obtained from a fit with inverted-$\tau$ models agrees best with those derived from the rest-frame UV (compare their Figs. 16 and 20). \\
At $z<2$ on the other hand reddening is heavily overestimated for all setups, e.g. the median offset for a wide setup is 0.29 and 68\% of the solutions lie within $\Delta E(B-V)=0$ and 0.59 (see Table \ref{sfoverres}).
Accordingly, we find that fitted ages are younger at low redshift in the case with reddening compared to the case without reddening. When  reddening is overestimated by $E(B-V)\sim0.4-1.2$ ages are lower by up to $\sim3$ dex at redshift 0.5. This difference decreases towards redshift 3 where ages are also older than in the unreddened case (lower panel in Fig.\ref{ebvdt}).\\%We explore the effect of specific reddening laws in more detail in section \ref{rl}.\\
Because the ability to recover reddening depends only little on the wavelength coverage, we only summarise the main points: 1) the exclusion of near-IR filter bands has almost no effect because dust reddening absorbs light predominantly at rest-frame UV wavelengths. 2) For that reason the exclusion of the bluest filter bands has the largest effect on the derived reddening, such that reddening is even more overestimated at low redshift and underestimated at high redshift.\\
Compared to the full age grid shown in Fig. \ref{ebvdt} a cap on minimum age reduces the overestimation in E(B-V) from maximal $\Delta E(B-V) =1.1$ to $0.45$ at $z=0.5$ for all template setups. The median offset is improved by 0.15. The maximum age difference is now $\sim 2$ dex at the lowest redshifts. At higher redshift the minimum age restriction improves the derived reddening only little and ages differ less between cases with and without reddening. Rebinning of the age grid does not improve the recovery of reddening at any redshift.\\
Metallicity has only a small effect on the derived reddening, hence, we do not show it. E(B-V) is slightly more overestimated at high redshift for the highest metallicity template setup. Overall, the scatter is increased.\\
In summary, reddening for both inverted-$\tau$ and wide setups is well recovered for galaxies that are dust reddened; E(B-V) is recovered in 68\% of the cases within $-0.01$ and $\sim0.10$. This depends little on wavelength coverage, metallicity and age. For intrinsically dust-free galaxies (those at z=0.5), reddening is heavily overestimated (68\% within $\Delta E(B-V)= 0$ and 0.59) and ages are underestimated because of the age-dust degeneracy.

\subsubsection{Stellar Mass}\label{massresults}

\begin{figure*}\includegraphics[width=144mm]{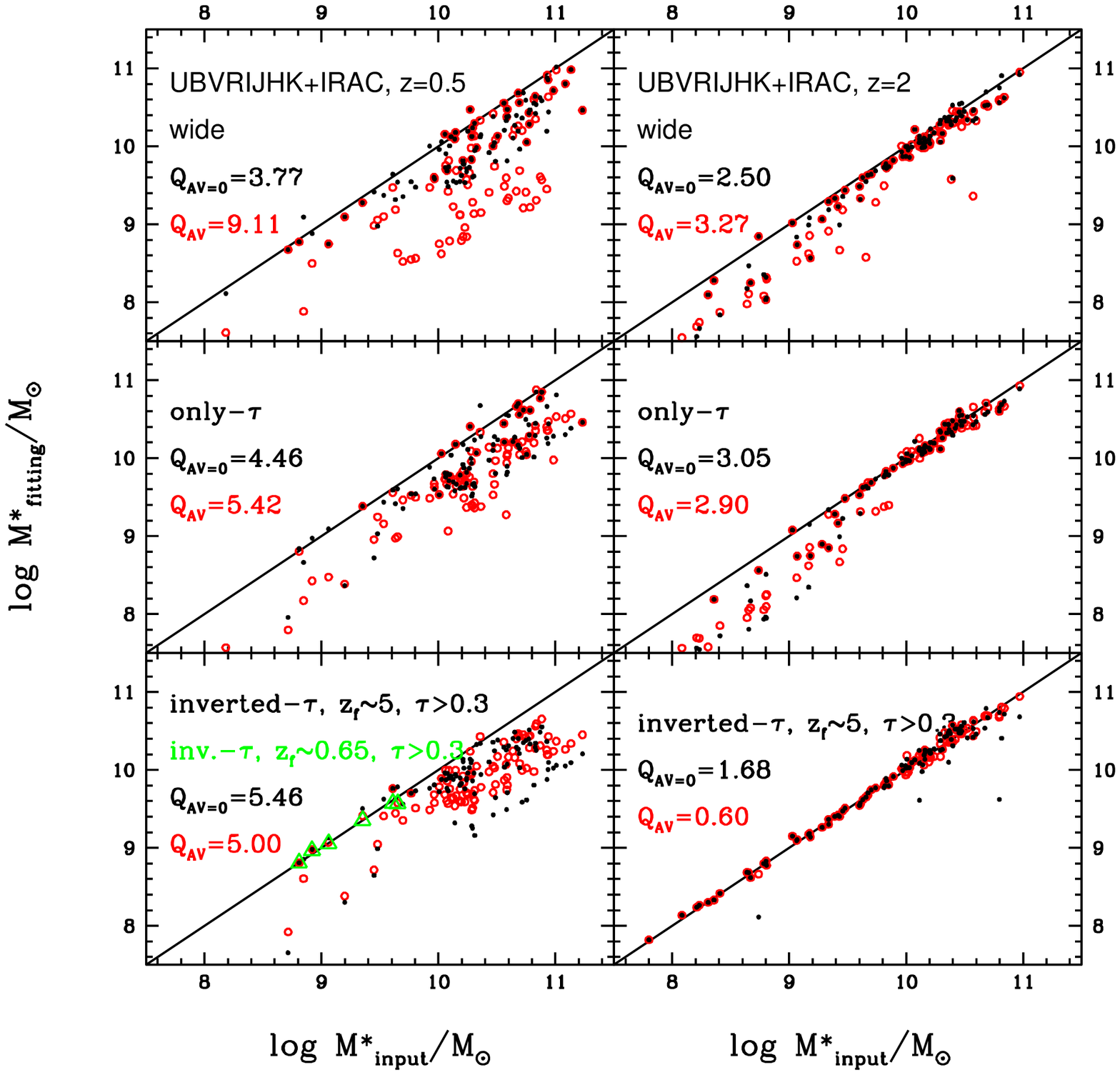}
\caption{\label{massdt05} Stellar mass recovery at redshift 0.5 and 2 as a function of template setup, namely from top to bottom:  wide, only-$\tau$, inverted-$\tau$. Red circles refer to cases with reddening, black dots to no reddening. Quality factors are given for the entire mass range for reddened and unreddened cases, respectively. Green triangles show the mass estimate for the youngest objects at z=0.5 with increasing SFH (see Fig. \ref{inputsfh}) when an inverted-$\tau$ model of age 1.1 Gyr is used in the fitting.}
\end{figure*}
The choice of template setup also affects the mass recovery (Fig. \ref{massdt05}). %We can summarize the main effects for the template setups described as: \\
%1)
With the wide template setup the mass recovery at redshift 2 and 3 is very good, particularly for masses between $10^{10}$ and $10^{11} M_{\odot}$. Masses are underestimated for lower-mass galaxies. These are small disks with increasing SFHs like the ones shown in panel a) of Fig. \ref{inputsfh} left-hand side. Indeed, their mass can be perfectly recovered using the inverted-$\tau$ models (bottom panel in Fig. \ref{massdt05}, see below). Including dust affects the result only little. Median offsets between true and recovered mass are $\sim-0.1$ dex at z=2 and 68\% of the solutions show offsets between $\sim-0.3$ dex to 0 dex in the unreddened case and $\sim-0.5$ dex  to $\sim0$ dex when reddening is included. For older galaxies with little on-going star formation (z=0.5) on the other hand, masses are underestimated. For these the median offsets with and without reddening in the fit are $\sim-0.6$ dex and $\sim-0.3$ dex, respectively. 68\% of objects show offsets in stellar mass between $\sim-0.5$ dex to $\sim0$ dex in the unreddened case and $\sim-1.4$ dex to $\sim-0.2$ dex in the reddened case. When dust is included, this effect is mostly due to the age-dust degeneracy, whereas without reddening a different set of degeneracies dominates. Firstly, no available SFH provides a real match to the mock SFHs. As seen in Fig. \ref{inputsfh} the typical SFH at low redshift is bimodal in the sense that star formation decreases after an initial period of increasing star formation. On-going star formation causes overshining, thus hiding most of the old underlying populations. This has the largest effect at redshift 0.5 because the age distribution of the stellar populations building up a particular galaxy is much wider (see Fig. \ref{agedt05}). Secondly, template metallicities are discrete and for the bulk of stellar populations of about half the objects still too metal rich (compare Fig. \ref{inputmetal} and Fig. \ref{metalM06}). The blue colour of a galaxy is matched by a template of low metallicity in combination with a longer star formation time scale and an older age, or a higher metallicity, shorter star formation time scale and a younger age. The latter leads to underestimate the mass. SSPs are the main cause for very young ages and consequently very low stellar masses in the wide template setup. Masses are underestimated by up to 0.8 dex.\\ 
It is important to note that - if instead of picking up the best fit for those objects for which the mass is underestimated, one would average over neighbouring solutions (in the range of $\chi^2_{\nu}+1$) - their mass estimate could be improved by up to 0.4 dex. Although masses are significantly improved for some of these objects, carrying out the procedure for all objects only results in an average improvement of 0.06 dex in the unreddened case and 0.18 dex when dust reddening is included.\\%For objects with SFRs around 1 $M_{\odot}$/yr this is compensated by a younger age, effectively causing an underestimation in mass. For objects with very low or sufficiently high star formation rate (SFR $< 0.5$ or SFR $> 2 M_{\odot}$/yr, respectively) 
%2)
% stellar mass diff for metallicity template setups
\begin{figure*}\includegraphics[width=144mm]{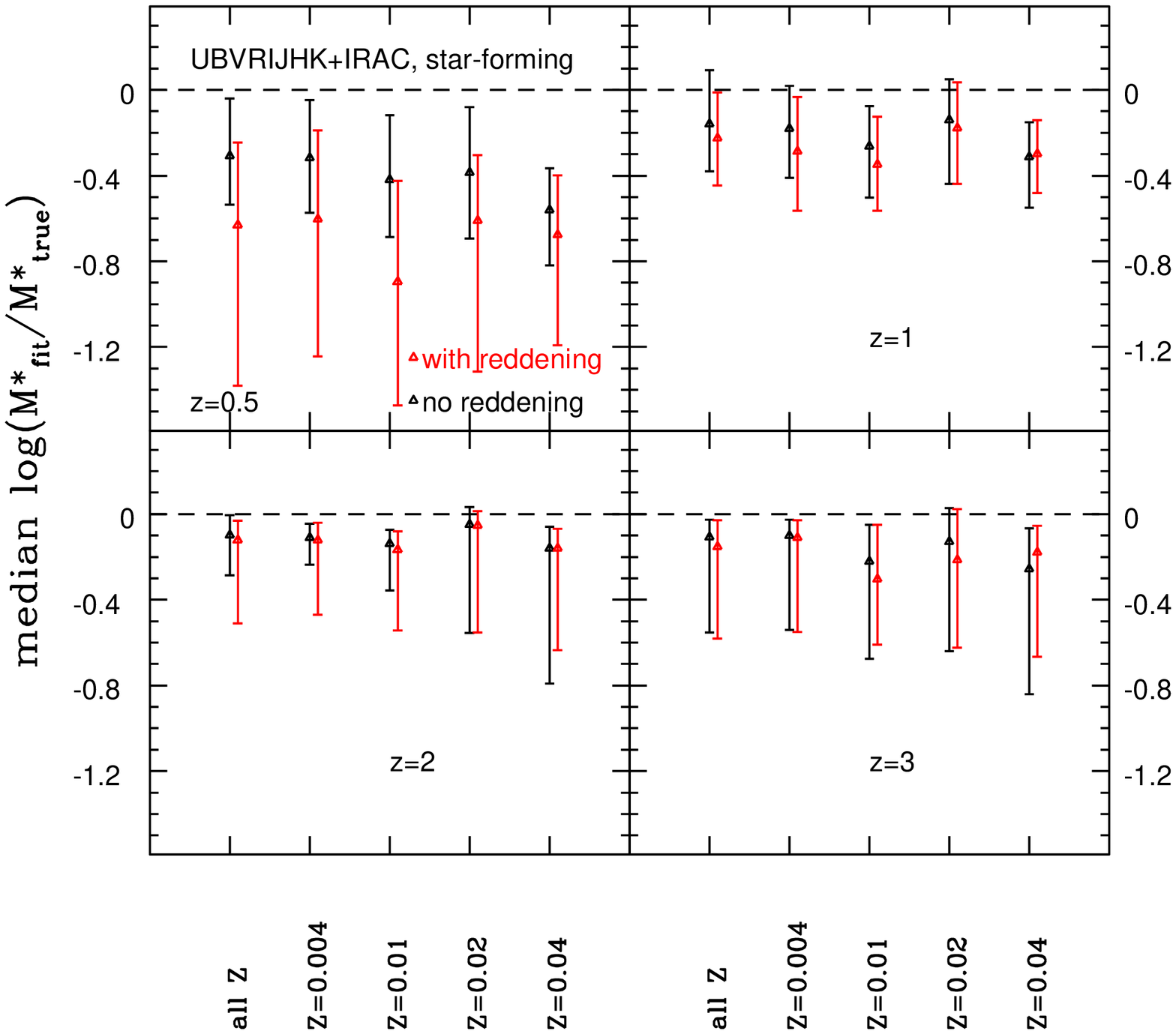}
\caption{\label{massmetal} Median recovered stellar mass of mock star-forming galaxies for mono-metallicity wide setups as a function of redshift, hence galaxy age (upper left to lower right). Metallicity in each panel increases from left to right. Red and black triangles refer to cases with and without reddening, respectively. Error bars reflect 68\% confidence levels.}
\end{figure*}
An only-$\tau$ setup can recover stellar masses at high redshift similarly well in the high-mass range, and due to the different star formation history, similarly badly in the low-mass range. For older galaxies with little on-going star formation at low redshift masses are underestimated. However, this setup including reddening performs slightly better than the wide setup.\\
%3)
At all redshifts masses are better recovered when the mock star-forming galaxies are unreddened, because there is no age-dust degeneracy (Fig. \ref{massdt05}). At $z=0.5$ masses are recovered within -0.8 dex and +0.3 dex for all template setups.  When reddening is included the SED-recovered stellar masses are underestimated by up to $\sim$1.6 dex for the least dust reddened and oldest galaxies in our sample. This stems from underestimating the age due to the age-dust degeneracy (Fig. \ref{agedt05}).\\
%4)
The mass recovery improves with overall decreasing galaxy ages and age spread, for all setups which perform in a very similar way for $z\geq1$. Masses larger than $10^{9.5}\,M_{\odot}$ are very well recovered for all template setups. Masses lower than $10^{9.5}\,M_{\odot}$ are underestimated by $\sim -0.7$ dex at $z=3$.\\
Here, it is important to note that the mock galaxies have SFR of at most $30 M_{\odot}$/yr (see Fig. \ref{input}), whereas real galaxies can have higher star formation rates \citep[e.g.][]{Daddi2007a}. Hence, these conclusions may not be extended to real galaxies, and indeed in M10 we showed that for $z\sim 2$ star-forming galaxies with higher SFRs, an only-$\tau$ setup underestimates the age, hence underestimates the stellar mass and overestimates the star formation rate. In M10 we have also verified that an inverted-$\tau$-type star formation history with high formation redshift is best for the derivation of physical properties such as reddening and SFR and for the mock galaxies of this work for stellar mass (their Fig. 24). Here we reinforce the result by a starring mass recovery for mock galaxies at redshift 2 and 3 (Fig. \ref{massdt05} bottom right) with inverted-$\tau$ models.\\
Obviously, the same models with the same priors do not allow a good mass recovery for older galaxies at lower redshifts because they do not trace their correct SFH. However, if we change priors and use inverted-$\tau$ models with formation redshift (i.e. age) closer to the age of the oldest population, thus simulating objects like the one shown in the bottom-left of Fig. \ref{inputsfh}, the stellar mass is recovered very well (green triangles in Fig. \ref{massdt05}). However, the SFHs of real galaxies are not known and a distinction between objects of increasing and decreasing SFR cannot be made prior to the fitting. A focussed investigation on these models and on how to apply them to different redshifts and types of galaxies is postponed to a dedicated paper.\\
Nowadays, in order to simulate more realistic star formation histories, several authors adopt a two-component template, such that a second burst at an arbitrary time is superimposed on an underlying exponentially declining star formation history 
\citep{Papovich2001,Gallazzi2005,Salim2007,Pozzetti2007,Walcher2008,Muzzin09,Noll2009}. In this way, the stellar mass can be maximised. \citet{Pozzetti2007} report an up to 40\% increase in stellar mass using two-component models over models with simpler SFHs. This composite model maybe a much better choice at low redshifts.\\
%mono-metallicity setups
\begin{figure}\includegraphics[width=84mm]{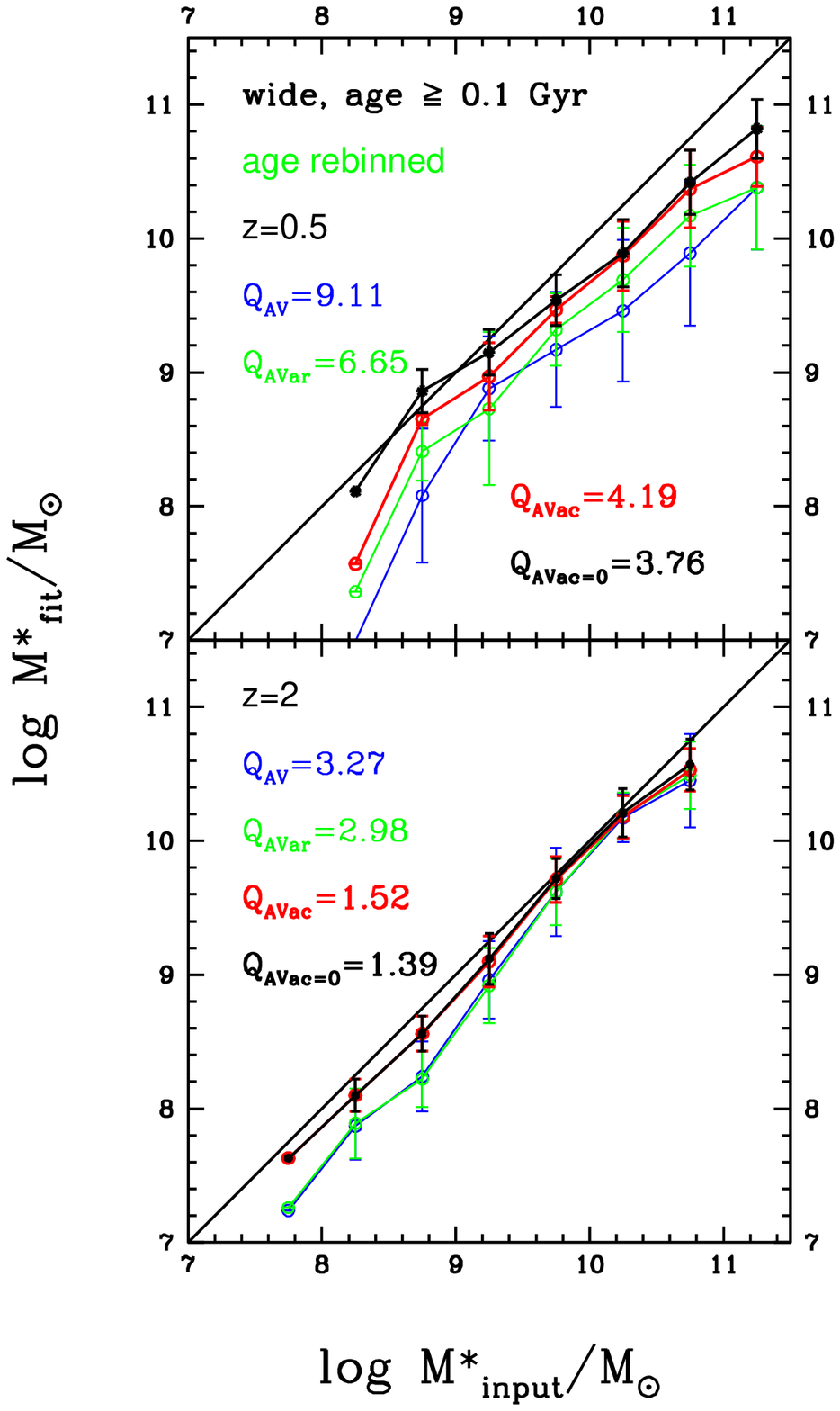}
\caption{\label{massAC05} Average recovered stellar mass in given input mass bin (binsize 0.5 dex) for the wide setup with a minimum age of 0.1 Gyr and a rebinned age grid. Red points and lines represent the reddened case, black stands for the unreddened case, blue thin lines show recovered masses obtained with the full age grid including reddening, green thin lines refer to a setup with rebinned age grid and reddening. Errorbars are one standard deviation. Quality factors are given for the entire mass range.}
\end{figure}
In Fig. \ref{massmetal} we show the mass recovery as a function of the metallicity adopted in the templates. For star-forming galaxies in the dust free case the lowest metallicity setup performs best at each redshift. The true mass is still underestimated by a median of 0.3 - 0.1 dex. The mass recovery becomes worse with increasing metallicity because metallicities are too high compared to the mock (see Fig. \ref{inputmetal}). This is distorted when reddening is included and masses are now underestimated by a median 0.6 - 0.1 dex. Overall, as already concluded by  \citet{Bol2009}, the differences between different metallicity setups are small. Consequently, fitting with a mono-metallicity template setup is sufficient for star-forming objects even when the metallicity is wrong. \\
%age grids
Since the age-dust degeneracy is the main reason for underestimated ages which in turn cause underestimated masses we quantify the effect of putting a prior on the minimum age of templates in the wide setup or choosing a coarser age grid. An age restriction is often applied in the literature \citep{Bol2009,Wuyts2009} and we have also explored its effect in M10, though there we could conclude that this trick is not sufficient to ensure a correct derivation of galaxy properties and - especially - does not allow to make progress in understanding galaxy evolution.
We found that among the explored age grids, a minimum age of 0.1 Gyr improves the mass estimate the most (Fig. \ref{massAC05})\footnote{We have also explored larger minimum ages - 0.5, 1 and 5 Gyr - for z=0.5 star-forming galaxies and found that these have a negative effect on the mass estimate.}. The resulting regulation of both the overshining and the age-dust degeneracy is particularly effective for the oldest galaxies (at low redshift) and in the reddened case. Here, the improvement is 0.4 dex on average, but can be much larger for single objects. Especially, at $z>1$ stellar masses are remarkably well recovered with this artificial age constraint. Hence, if one is merely interested in estimating the stellar mass setting an artificial age constraint appears to be a good idea. We found that switching off reddening and applying a minimum age cut has similar effects.\\
%\textbf{starts making a difference for high-z objects, because they have higher percentage of young stars that dominate light and request young template, hence chi2 worse for some objects, but masses get improved upon, at least for low mass end, when age is not contrained too much, meaning 0.02 is fine, 0.1 tends to have the negative opposite effect.}
%IMF effects
\begin{figure*}\includegraphics[width=154mm]{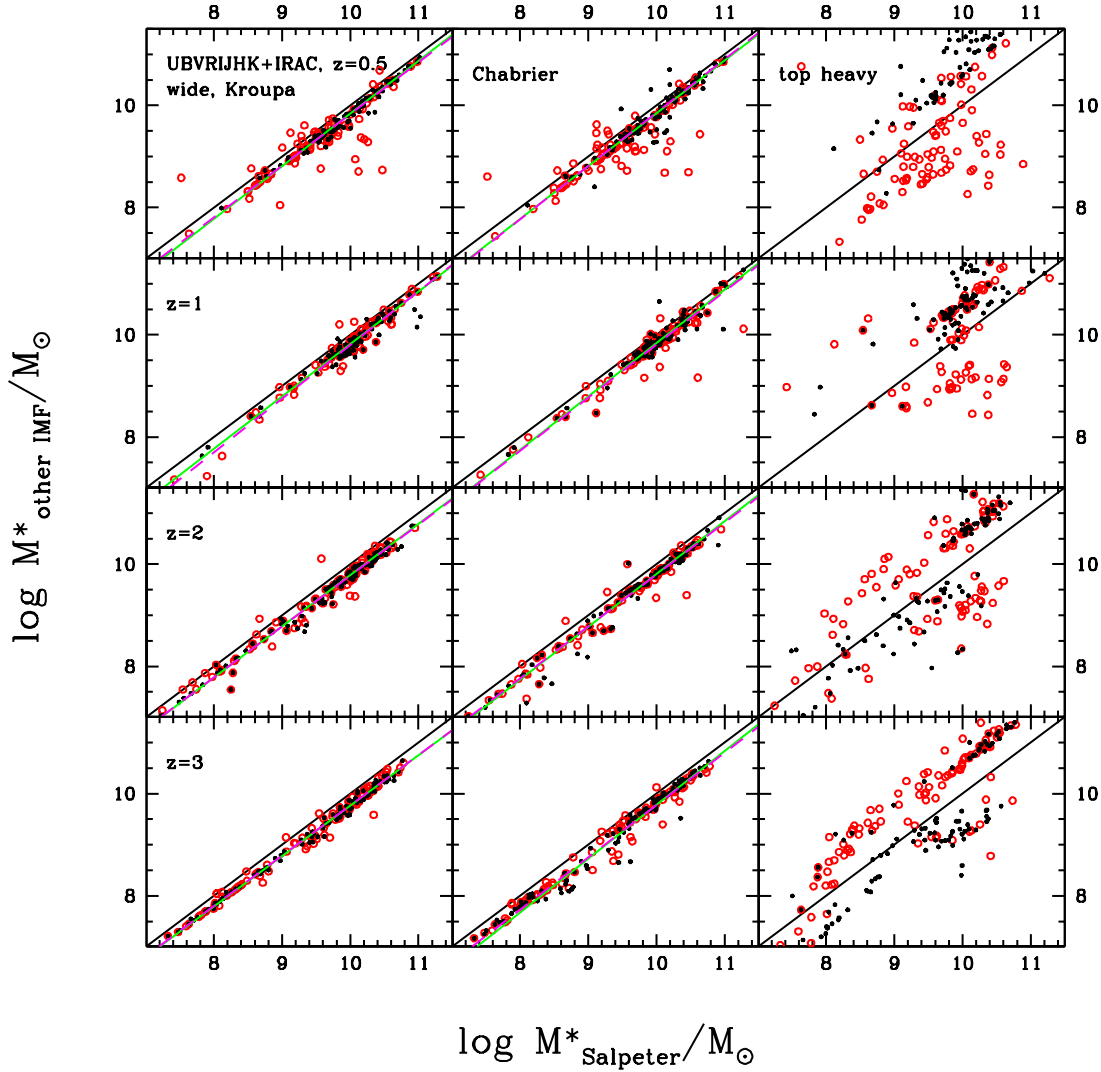}
\caption{\label{massIMFd} Stellar masses of mock star-forming galaxies derived using templates with IMFs different from the input Salpeter IMF. Red marks solutions obtained with reddening, black the unreddened case. Green solid lines represent fits to the unreddened case, magenta dashed lines fits to the reddened case. There is little difference between them (see also tables \ref{scale} and \ref{scalepass}). For obtaining scaling relations we used the entire merger tree for the linear fits. The coefficients of the fits are provided in section \ref{scalerel}.}
\end{figure*}
\begin{figure*}\includegraphics[width=144mm]{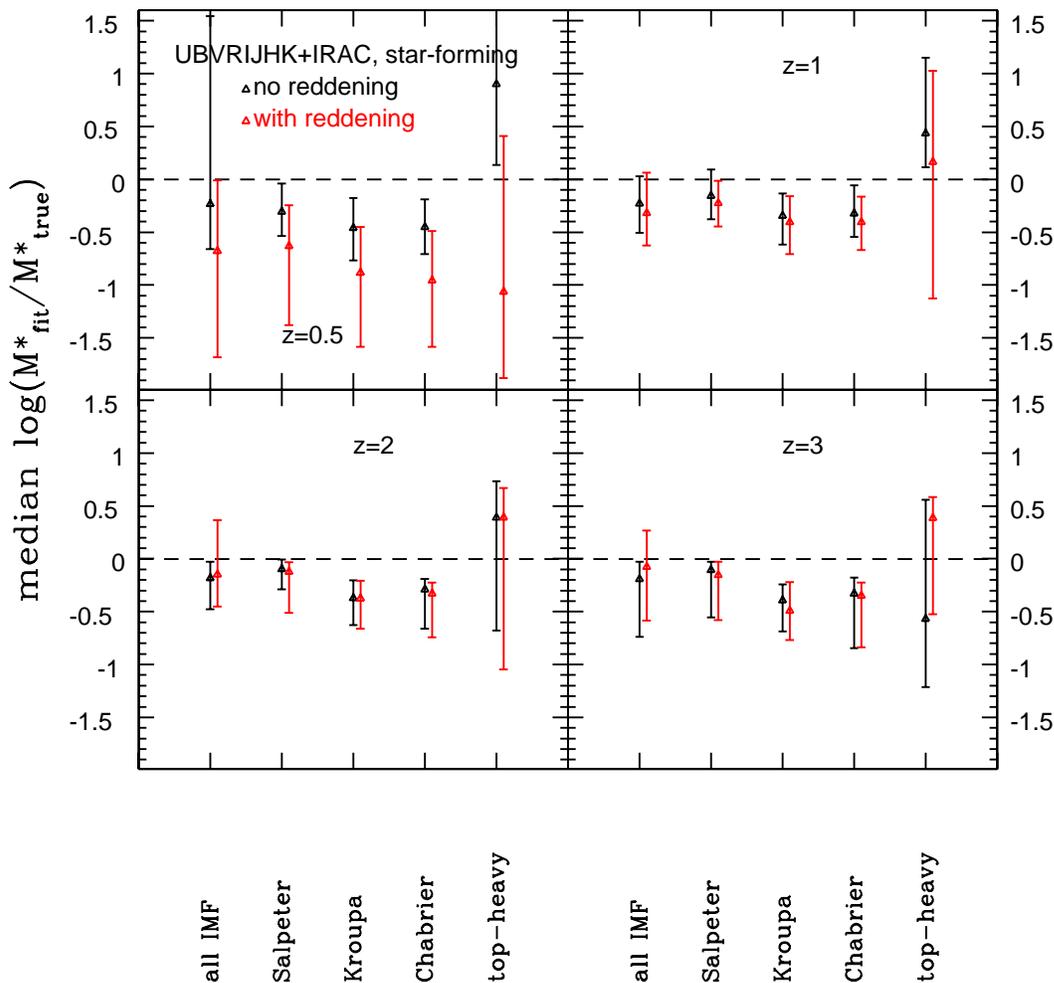}
\caption{\label{massIMF} Median stellar mass recovery as a function of redshift and IMF (wide template setup and UBVRIJHK+IRAC wavelength coverage for each). 68\% confidence levels are shown as error bars. Redshift increases from the top left to the bottom right. Red and black mark solutions obtained with and without reddening, respectively.}
\end{figure*}
The difference in the stellar mass recovery when one uses an incorrect IMF is small (Fig. \ref{massIMFd}). Cases for different IMFs differ the most at $z=0.5$ for the reddened case - a clear effect of the age-dust degeneracy. A top-heavy IMF - the most discrepant to a Salpeter IMF - fails to reproduce the correct stellar mass (Fig. \ref{massIMF}). At each redshift we find a large scatter around the true masses.\\
However, there is no clear offset between these IMFs\footnote{For verification we used the entire merger tree, but in Fig. \ref{massIMF} only show the sample of 100 galaxies.} (even without dust) for galaxies with little star formation, unlike commonly assumed in the literature. This demonstrates that other free parameters in the fitting, such as age and SFH are able to compensate for the wrong IMF. For young galaxies with high star formation an offset exists. \citet{Bol2009} state statistical differences of stellar mass between the various IMFs as $log\,M^*_{Salpeter} \simeq log\,M^*_{x} + y$ with $y=0.23$ and $0.19$ for Chabrier and Kroupa IMF ($x$), respectively, but acknowledge that the best fit value for other parameters can significantly differ. We list our scaling relations in section \ref{scalerel}.\\
Although Figs. \ref{massIMFd} and \ref{massIMF} show that the choice of the IMF is felt in the fit and the correct IMF recovers masses best, the right IMF cannot be identified by choosing the fit with the minimum $\chi^2_{\nu}$ among all solutions. In fact, very often even a top-heavy IMF is picked. This confirms again that the minimum $\chi^2_{\nu}$ does not necessarily provide the best physical solution and confirms the notion that picking up the IMF from SED-fits is virtually impossible.\\
%wavelength coverage
\begin{figure*}\includegraphics[width=144mm]{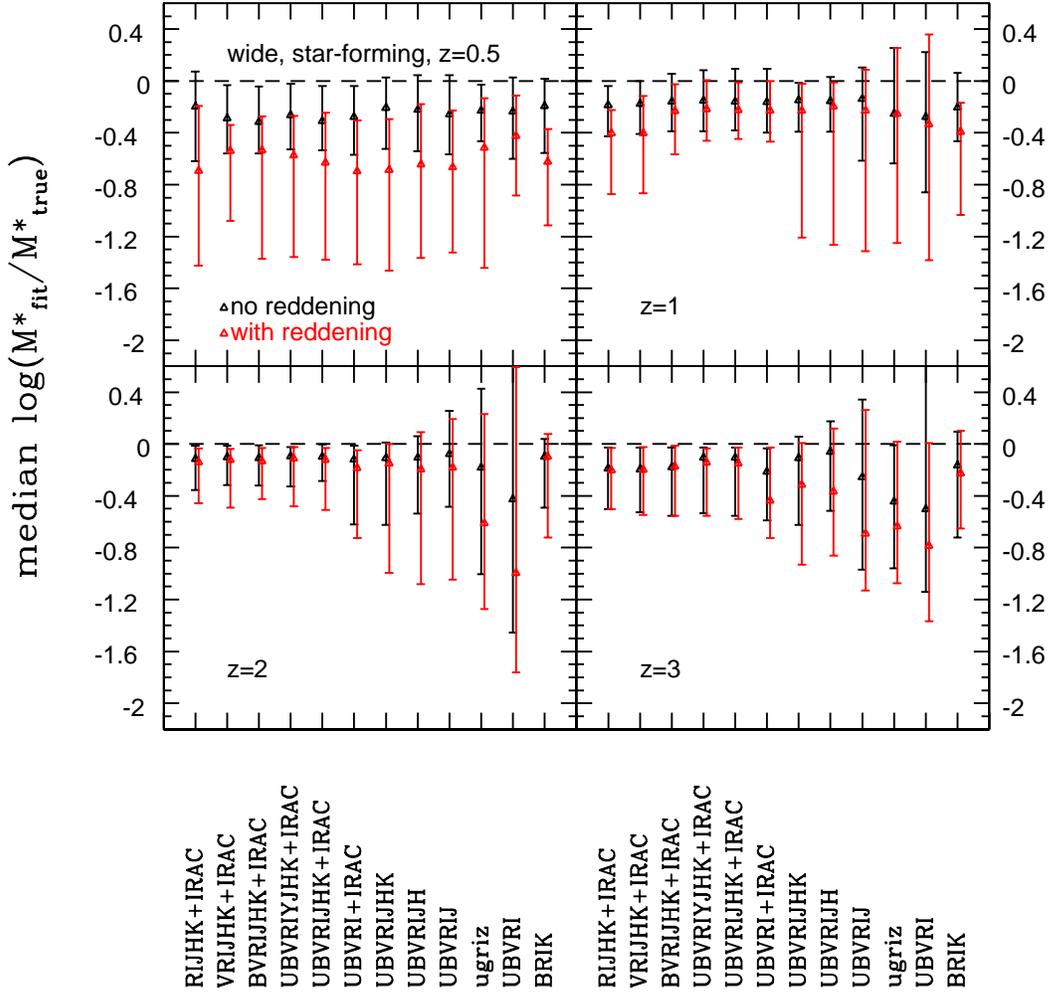}
\caption{\label{massdf05} Stellar mass recovery (median) of star-forming galaxies as a function of wavelength coverage and filter set at redshifts 0.5, 1, 2 and 3, for the wide template setup. The filter setup is varied from left to right in each panel, redshift increases from top left to bottom right. Red represents reddened solutions, black unreddened ones. 68\% confidence levels for each setup are shown.}
\end{figure*}
\begin{figure*}\includegraphics[width=124mm]{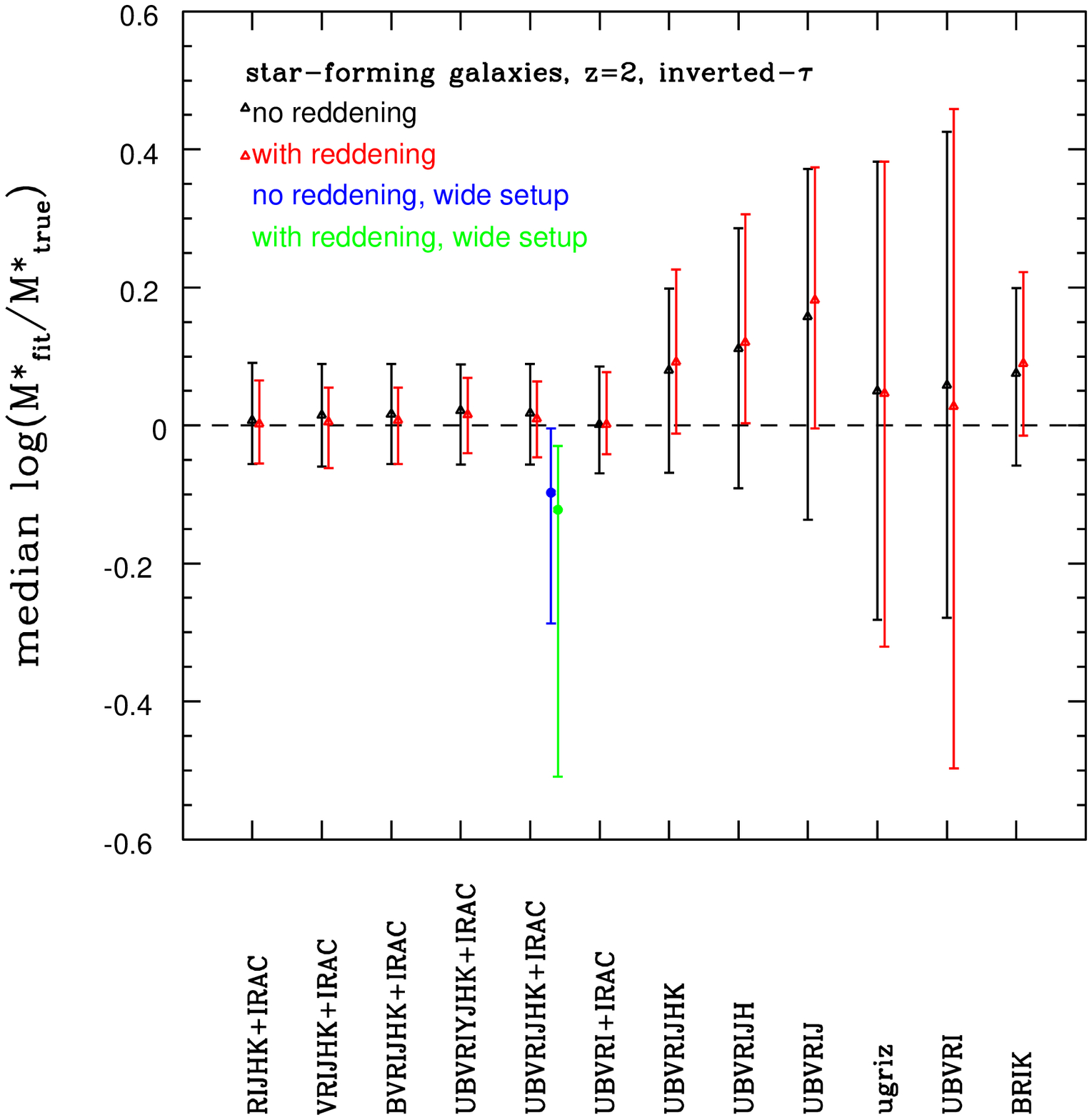}
\caption{\label{massdf2NEW} Median differences between true and recovered stellar mass as function of wavelength coverage and filter set for the inverted-$\tau$ templates at redshift 2. Blue and green symbols refer to the mass recovery obtained with the wide setup without and with reddening, respectively.}
\end{figure*}
We focus now on the effect of the assumed wavelength coverage in the fitting, on stellar masses for the wide setup at each redshift and the inverted-$\tau$ models at redshift 2 (Figs. \ref{massdf05} and \ref{massdf2NEW}). \\
The dependence on wavelength coverage is weak for old galaxies with little on-going star formation at z=0.5 -  masses are similarly underestimated for each filter setup\footnote{Note that excluding SSPs in the fitting setup improves the mass estimate in the reddened case by $\sim0.2$ dex for the various filter setups. In the unreddened case the improvement is not significant.}. This is obviously driven by the overshining and the wrong SFH which the filter setup cannot help rectifying. For $z>1$, where galaxies are younger and the SFR is higher, we find that the broader the wavelength coverage, the better the mass recovery, particularly when reddening is involved. When red filter bands are excluded in the fit, masses are underestimated by up to 2 dex. The coverage of the rest-frame near-IR appears to be crucial - as concluded in M06. The same conclusion was reached by \citet{Kannappan,Bol2009,Lee2009} and \citet{Ilbert2010}. \citet{S05},\citet{vdW06} conclude the opposite. \cite{Muzzin2009b} merely find an improvement in the confidence levels for the derived parameters when including the rest-frame near-IR.\\
\citet{Pozzetti2007} (for $z>1$) and \citet{Walcher2008} (for $z<1.2$) find that the lack of near-IR filter bands (observed-frame) in the fit leads to larger stellar masses compared to estimates which included them. We find a decrease instead. We assign this different behaviour to the differences in stellar population model and SFHs used in the fitting setup.\\
For $z\geq2$ we find that excluding filter bands covering the rest-frame UV allows one to recover stellar masses on average equally as well as the full filter set. Whereas the near-IR contains information from older, less luminous stellar populations and allows us to get a robust estimate on stellar mass, the UV contains mostly information about younger, more recently formed stellar populations. Hence, for a good stellar mass recovery, the bluest filter bands are less important and may even be damaging. Clearly, a filter setup that only covers the rest-frame UV is useless for the mass recovery of high redshift galaxies. Remarkably, the combination of optical (BRI) and only one near-IR filter (K) recovers masses reasonably well. This could be useful for economising the request of telescope time.\\
The inclusion of a further filter band (\emph{Y} with $\lambda_{eff} \sim 10000\AA$) does not improve the stellar mass estimate independently of redshift and reddening. In general, the mass recovery at redshifts $z\geq1$ is extremely good, provided one uses a filter set that covers the rest frame near-IR because this helps to pick the right star formation history (see M05). \\
In M10 we showed that inverted-$\tau$ models recover the stellar masses of high redshift star-forming galaxies best. We show here (Fig. \ref{massdf2NEW}) that this is robust against variations in filter setup because this model is less dependent on age. Only for the most restricted wavelength ranges and those without rest-frame near-IR a significant scatter around the true masses is found.\\
As a summary of this whole section, the stellar masses of old galaxies with little residual star formation are usually underestimated because ages are underestimated which in turn is due to a mismatch in SFH between template and galaxy and the overshining effect. This effect is worse when dust reddening is included because of the age-dust degeneracy. This depends only very little on metallicity and wavelength coverage. An artificial age constraint improves the mass estimation.\\
The masses of young galaxies with high star formation are slightly underestimated when age is a free parameter in the fit. Masses are perfectly recovered with inverted-$\tau$ templates with proper priors on formation epochs and $\tau$, as shown in M10.
\begin{figure*}\includegraphics[width=164mm]{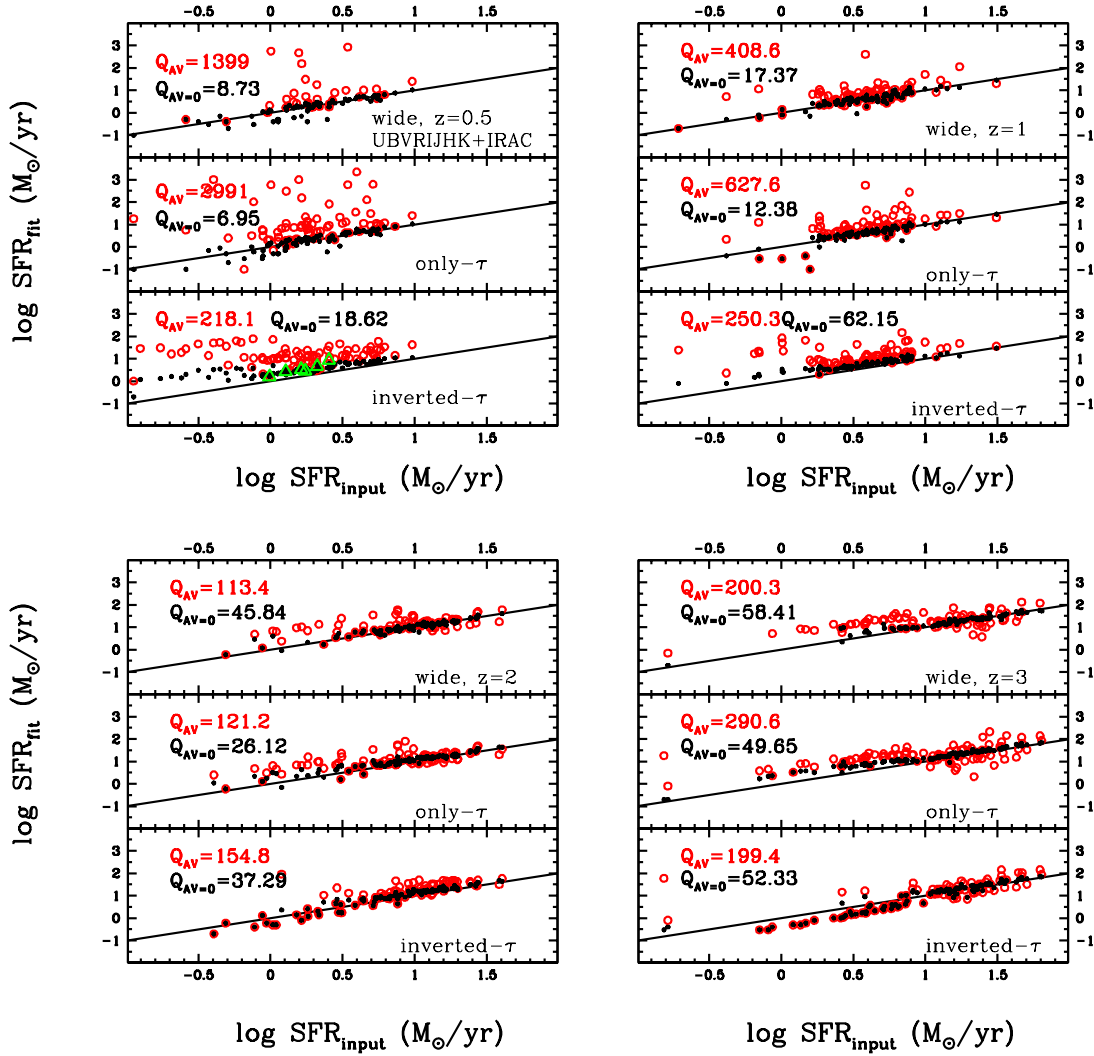}
\caption{\label{sfrdt05} The recovery of SFR as a function of template setup, namely wide, only-$\tau$ and inverted-$\tau$ from top to bottom. Redshift increases from top left to bottom right. Red circles refer to cases with reddening, black dots to no reddening. Quality factors are given for the entire SFR range for reddened and unreddened case, respectively. Green triangles show the SFR estimate for the youngest objects at z=0.5 with increasing SFH when an inverted-$\tau$ model of age 1.1 Gyr is used in the fitting. These are the same objects as in Fig. \ref{massdt05}. Objects with SFR=0 are not shown. SFR=0 indicates that the best fit solution is an SSP or a truncated SF model with an age larger than the truncation time.}
\end{figure*}
\begin{figure}\includegraphics[width=84mm]{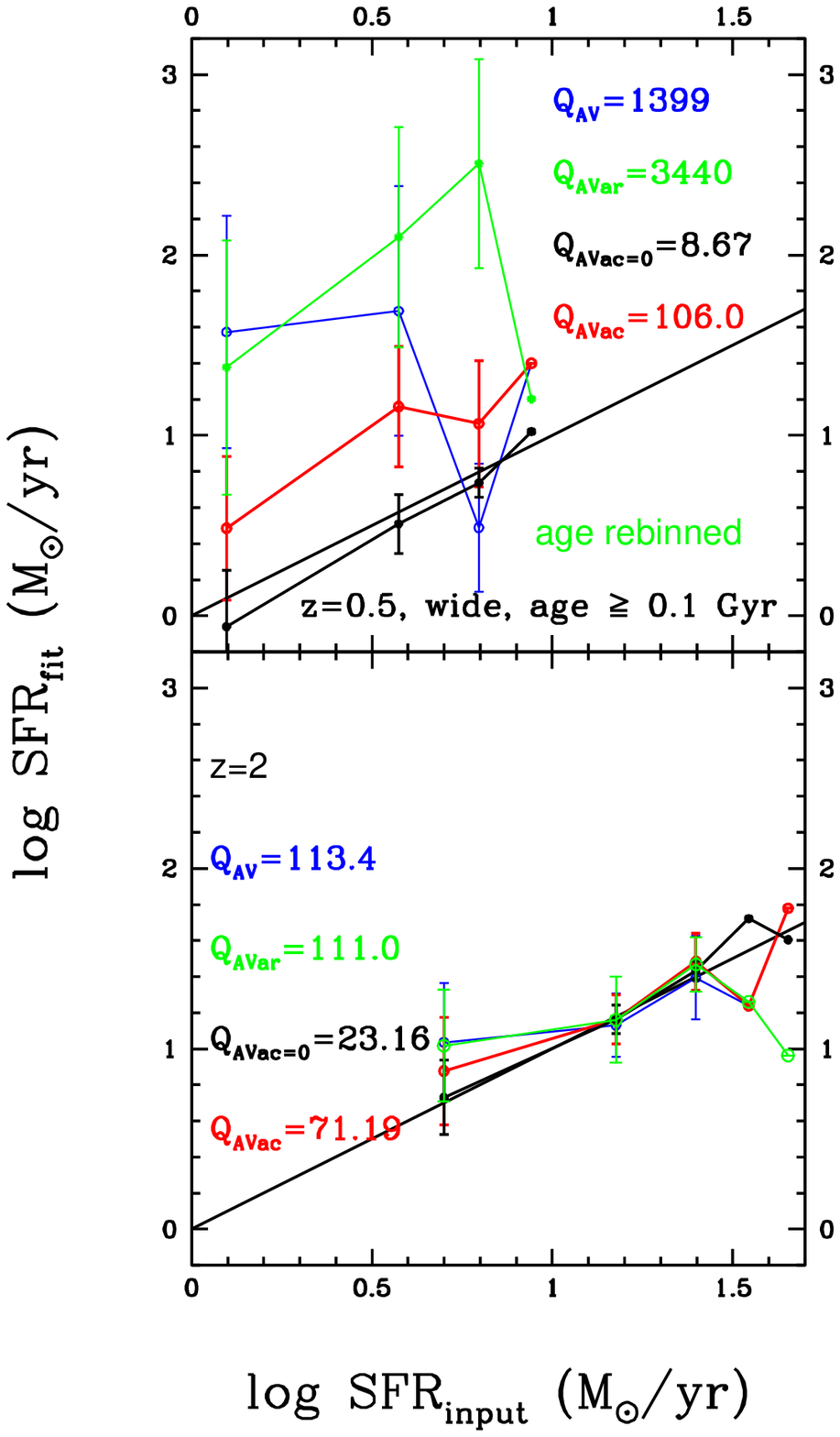}
\caption{\label{sfrAC05} Averaged recovered SFRs (y-axis) compared to known binned SFRs (x-axis) as a function of age grid. Representatively, we show results for redshift 0.5 and 2 only. Red dots and lines represent solutions with reddening, black ones those without reddening, both when the minimum age is 0.1 Gyr. Blue thin lines show the SFR recovery with a wide setup and reddening. Green lines refer to a wide setup with rebinned age grid and reddening in the fit. Binsizes vary from $2.5\,M_{\odot}$/yr to $20\,M_{\odot}$/yr from $z=0.5$ to $z=3$. Errorbars are one standard deviation. Quality factors are given for the entire SFR range for reddened and unreddened cases, respectively.}
\end{figure}
Dust reddening and metallicity affect the whole result very little. Masses are increasingly underestimated for narrower wavelength coverages when the wide setup is used. Here, the rest-frame near-IR is crucial for a robust mass estimate. Masses derived using inverted-$\tau$ models are less sensitive to wavelength coverage. Overall, a wide wavelength coverage also implies a smaller scatter.\\
The SED-fit and the mass estimate is sensitive to the choice of IMF of the fitting template, but the correct IMF cannot be identified by means of the overall minimum $\chi^2_{\nu}$.\\
Ranges including 68\% of the solutions for the stellar mass recovery are listed in Tables \ref{sfoverres} and \ref{sfoverres2}.

\subsubsection{Star formation history and star formation rate}\label{sfhresults}

\begin{figure*}\includegraphics[width=144mm]{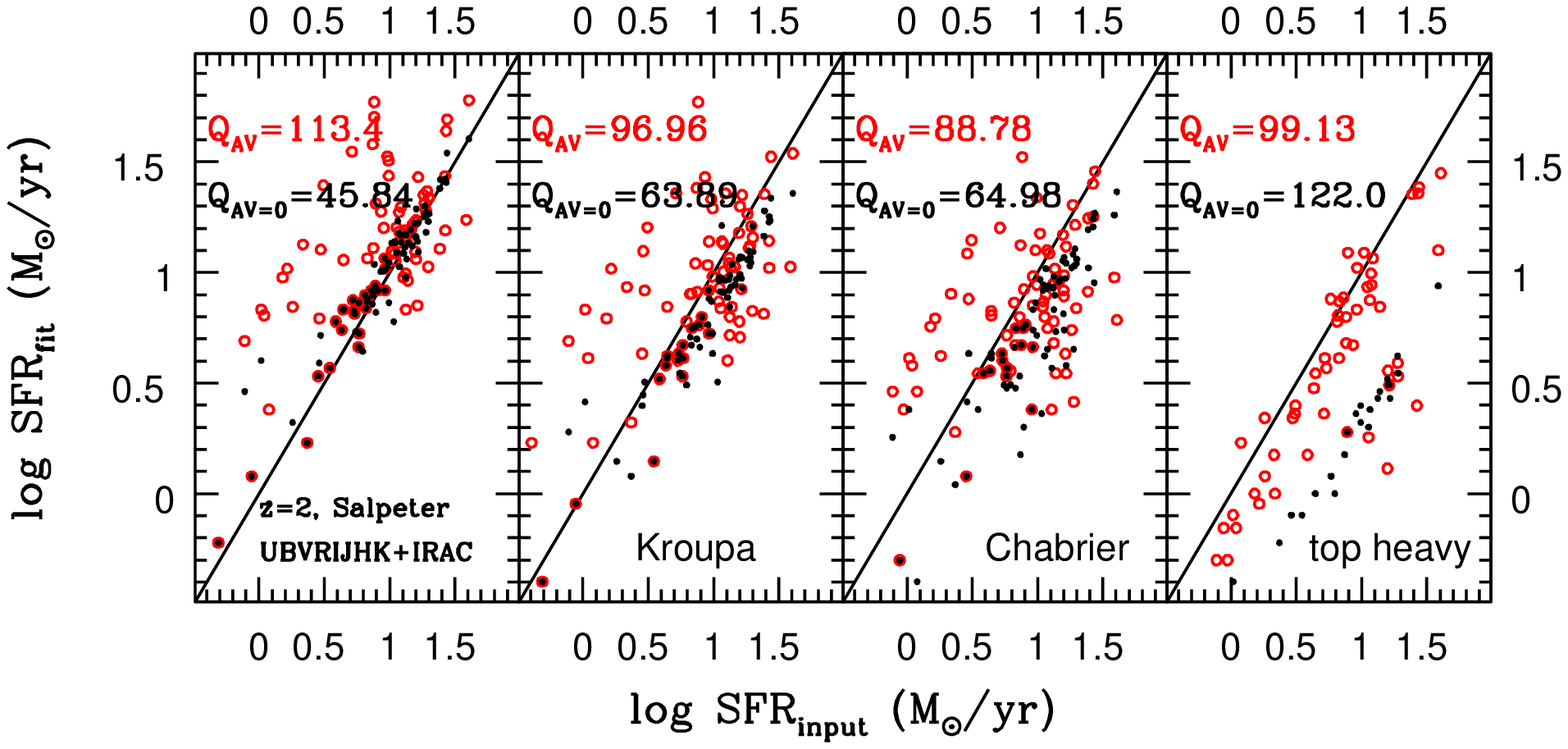}
\caption{\label{sfrIMF} SFR recovery as a function of template IMF at redshift 2. Colours are the same as in Fig. \ref{massdtrl05}. SFR recovery for a Salpeter IMF is shown in Fig. \ref{sfrdt05}. SFR=0 cases are not shown in this plot.}
\end{figure*}
\begin{figure*}\includegraphics[width=184mm]{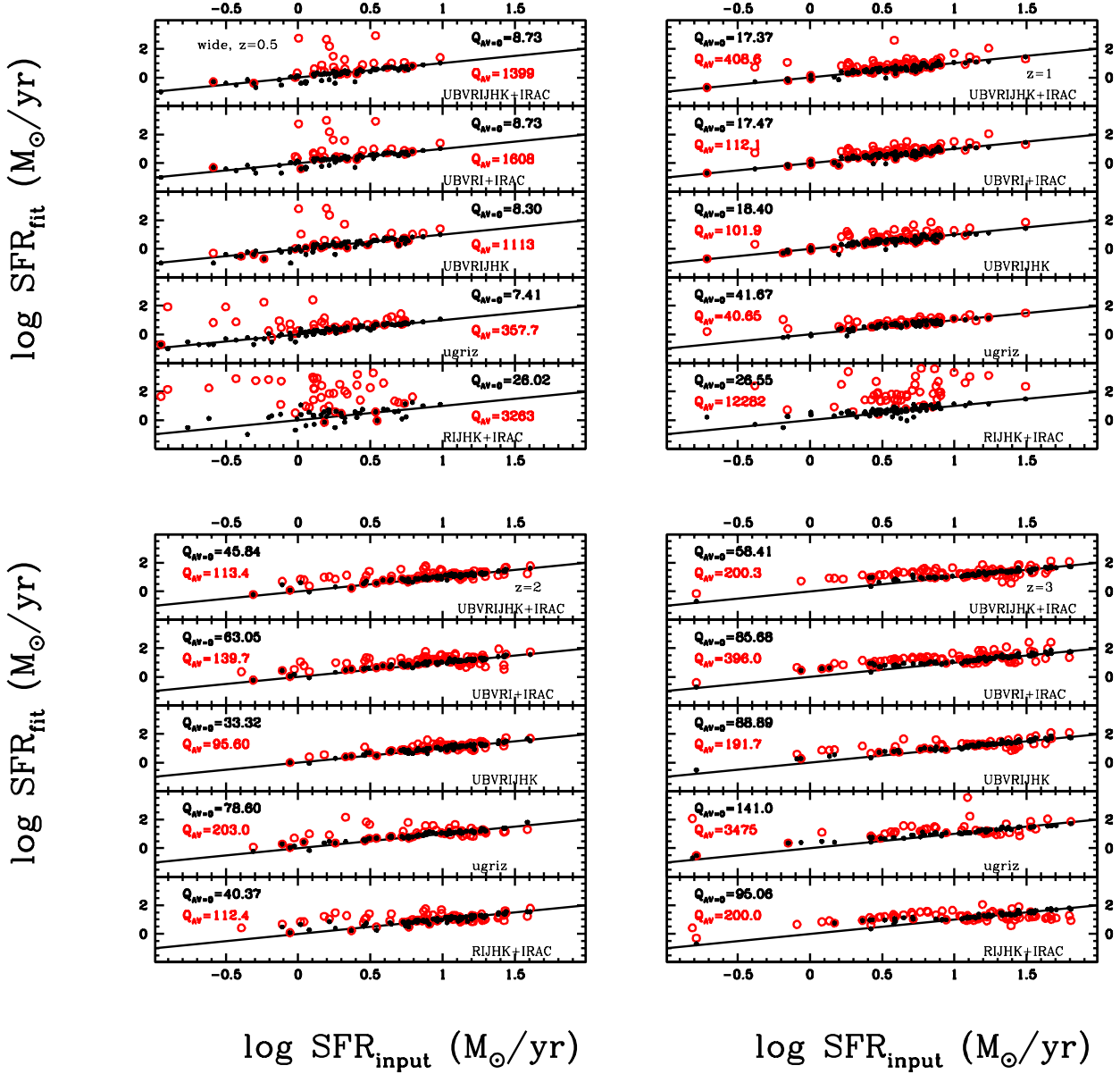}
\caption{\label{sfrdf05} SFR recovery as a function of wavelength coverage and redshift (increasing from top left to bottom right). The filter setup is varied from top to bottom in each panel. Objects with SFR=0 are best fit with a SSP or a truncated model with age larger than the truncation time and are not shown in this plot. Red circles refer to the reddened case, black dots to the unreddened case.}
\end{figure*}
\begin{figure}\includegraphics[width=84mm]{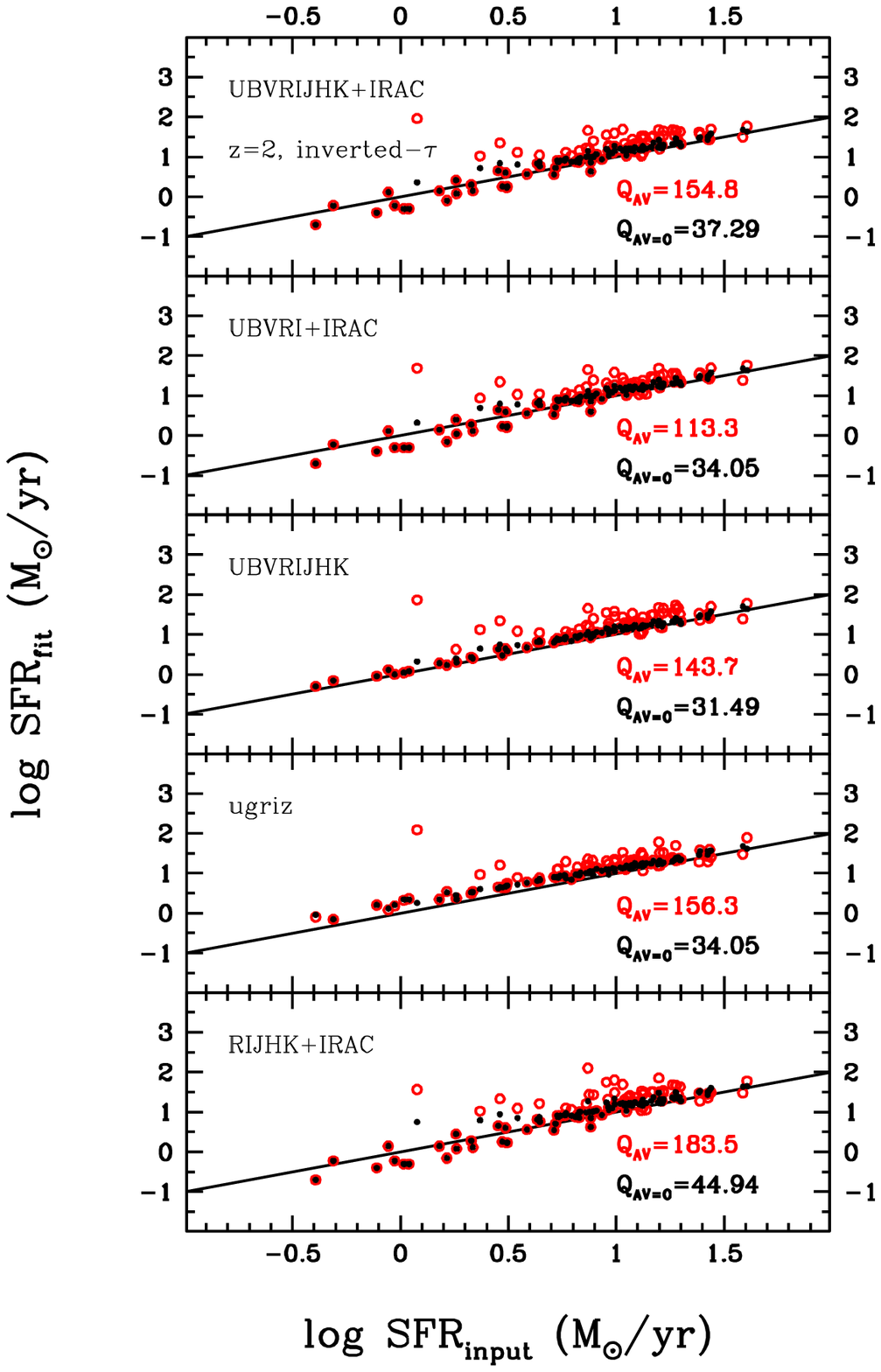}
\caption{\label{compsfrNit} SFR recovery using inverted-$\tau$ models for mock star-forming galaxies at redshift 2 as a function of wavelength coverage. The filter setup is varied from top to bottom. Red circles are with reddening, black dots are without reddening.}
\end{figure}
Fig. \ref{sfrdt05} shows that -  similarly to stellar masses - SFRs agree very well with input SFRs only when reddening is switched off (median $\Delta$ log SFR $\approx0$, see Tables \ref{sfoverres} and \ref{sfoverres2}). This is nearly independent of template setup. When reddening is involved, however, some SFRs are drastically overestimated at $z\leq1$. As was shown in M10, such very high star formation rates in combination with low stellar masses are a pure artifact of SED-fitting. The inverted-$\tau$ setup allows for the best estimates. The wide setup reproduces the SFR best at $z=1$ but this may just be due to the particular composition of the mock galaxies at these redshifts. At redshift 3 for the wide and only-$\tau$ setup SFRs are also underestimated at the high SFR end because of SFH mismatches. Low SFRs are overestimated for the same reason. The inverted-$\tau$ setup can recover the SFRs well for objects with increasing SFH. In M10 we showed that SFRs derived from inverted$-\tau$ models are in excellent agreement with those derived with the UV-corrected-slope method for high-redshift star-forming galaxies in the GOODS-S sample (see Fig. 19 in M10). Note that in M10 we used only the Calzetti reddening law in the fit while in this paper we show SFRs that are derived with all reddening laws which results in small differences in SFR for some objects. \\
%age constraints
The accuracy with which one estimates SFRs also profits from the constraint in age and the limited age-dust degeneracy. Particularly, low star formation rates are overestimated less at $z=0.5$ (Fig. \ref{sfrAC05}). Only allowing ages larger than 0.1 Gyr gives the best result, although some degree of overestimation is still present. Simply rebinning the age grid, however, does not work, SFRs are even more overestimated. At higher redshift the improvement is less than at $z=0.5$. Again a rebinned age grid performs worse and leads to underestimation of the highest SFRs.\\
The effect on stellar mass and SFR caused by the age constraint is the same for each template setup (excluding the inverted-$\tau$ setup because its age is fixed).\\
%sfr IMF effects
SFR estimates using templates with a Kroupa or a Chabrier IMF instead of a Salpeter IMF show qualitatively the same behaviour (in the case with reddening). In the unreddened case SFRs are predicted correctly from fits with these three IMFs. In Fig. \ref{sfrIMF} we show the star formation rates derived with a Salpeter, Kroupa, Chabrier and top-heavy IMF for redshift 2 galaxies as an example. SFRs for a top-heavy IMF are generally lower than SFRs for a Salpeter IMF and also lower than the true value. This leads to a clear offset in the no reddening case, independently of redshift. When reddening is included, SFRs increase as the age decreases.\\
%SFR plots for IMF
Fig. \ref{sfrdf05} demonstrates the effect of wavelength coverage (or lack thereof) on the determination of the star formation rate. SFRs are generally well recovered in the unreddened case partly because of compensating effects. Deviations from this are due to a lack in wavelength coverage (RIJHK+IRAC at $z=3$). In the reddened case SFRs are overestimated at almost all redshifts. While the mass recovery is nearly insensitive to the neglection of blue filter bands, the SFR estimate is hampered. This is not surprising since the most recently formed stars dominate the light emission in the UV which therefore plays an important role in the estimation of SFRs. At $z=3$ SFRs are also underestimated at the high end.\\
In the reddened case, SFRs are best recovered with filter setups that cover the broadest wavelength range, while in the unreddened case a dependence with wavelength coverage is negligible. \\
We show the derived SFRs from inverted-$\tau$ models in Fig. \ref{compsfrNit} for a selection of filter setups. Similarly to the mass derivation, the SFR estimates do not differ much with filter setup and wavelength coverage. SFRs are very well reproduced in the unreddened case. In the reddened case, SFRs are overestimated for all filter setups when SFHs between mock galaxy and template do not match.\\
As already concluded in M10, the SFR determination is driven by the correct SFH. The good SFR recovery in the un-reddened case at $z=0.5$ despite the mismatch in SFH is due to compensating effects with underestimated ages and stellar masses.
This again reinforces our conclusion, that the {\it simultaneous} recovery of mass, age, SFR, etc. for galaxies from SED-fitting is not possible without knowing the exact star formation history.
%\afterpage{\clearpage}

\subsection{Passive galaxies}\label{passgalresults}

\subsubsection{Age}\label{ageresultspass}
\begin{figure*}\includegraphics[width=144mm]{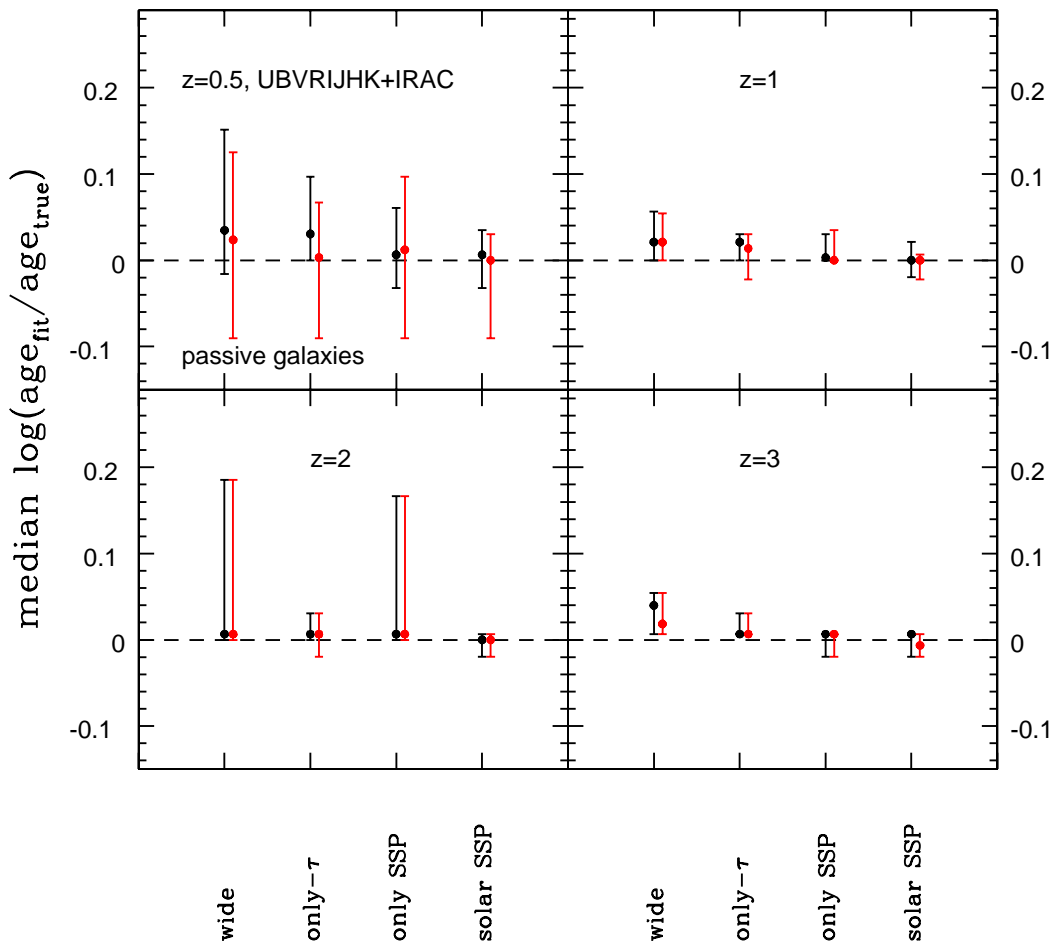}
\caption{\label{oldagedt} Median recovery of age for mock passive galaxies as a function of redshift and template setup. Black and red symbols refer to the cases without and with reddening, respectively. 68\% confidence levels are shown.}
\end{figure*}

\begin{figure*}\includegraphics[width=144mm]{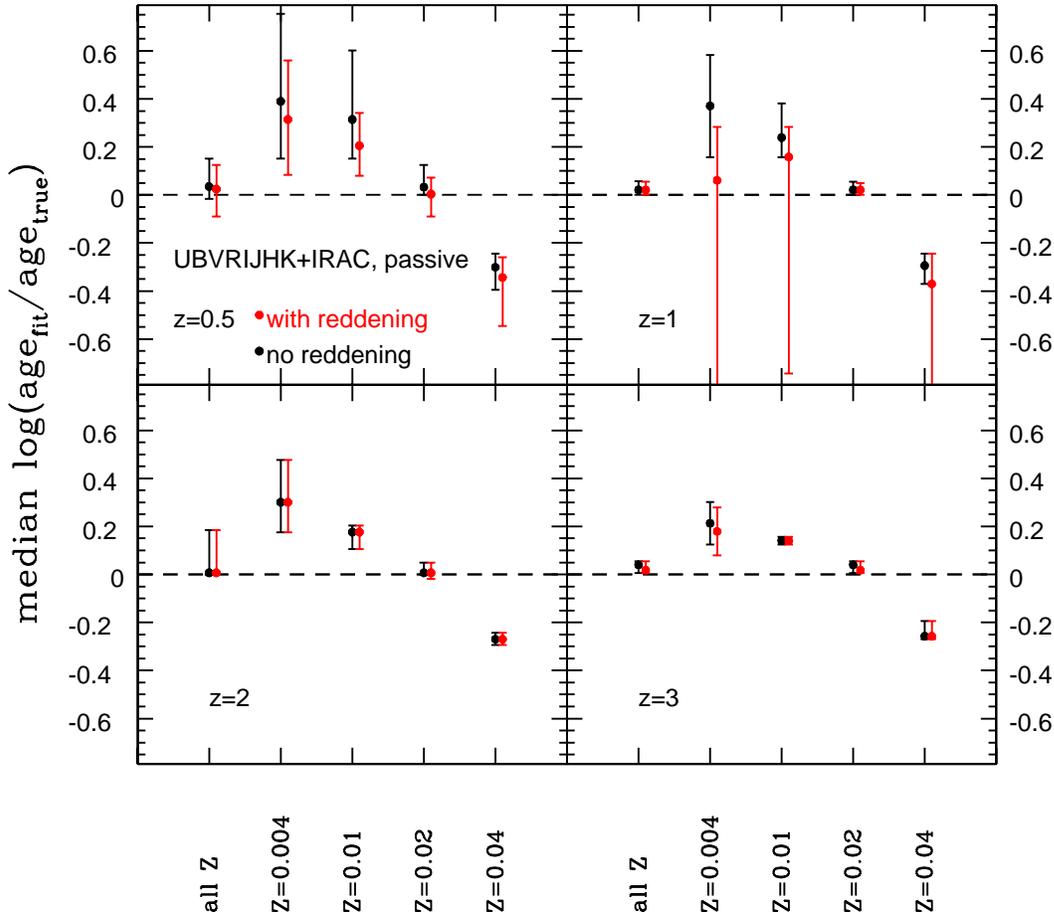}
\caption{\label{oldagedmetal} Median deviation of fitted age from true age of mock passive galaxies as a function of redshift and metallicity. Black dots and 68\% confidence levels refer to the unreddened case, red to the case that includes reddening in the fit. Note the change in scale between redshift 0.5 and 1 and redshift 2 and 3.}
\end{figure*}

\begin{figure*}\includegraphics[width=144mm]{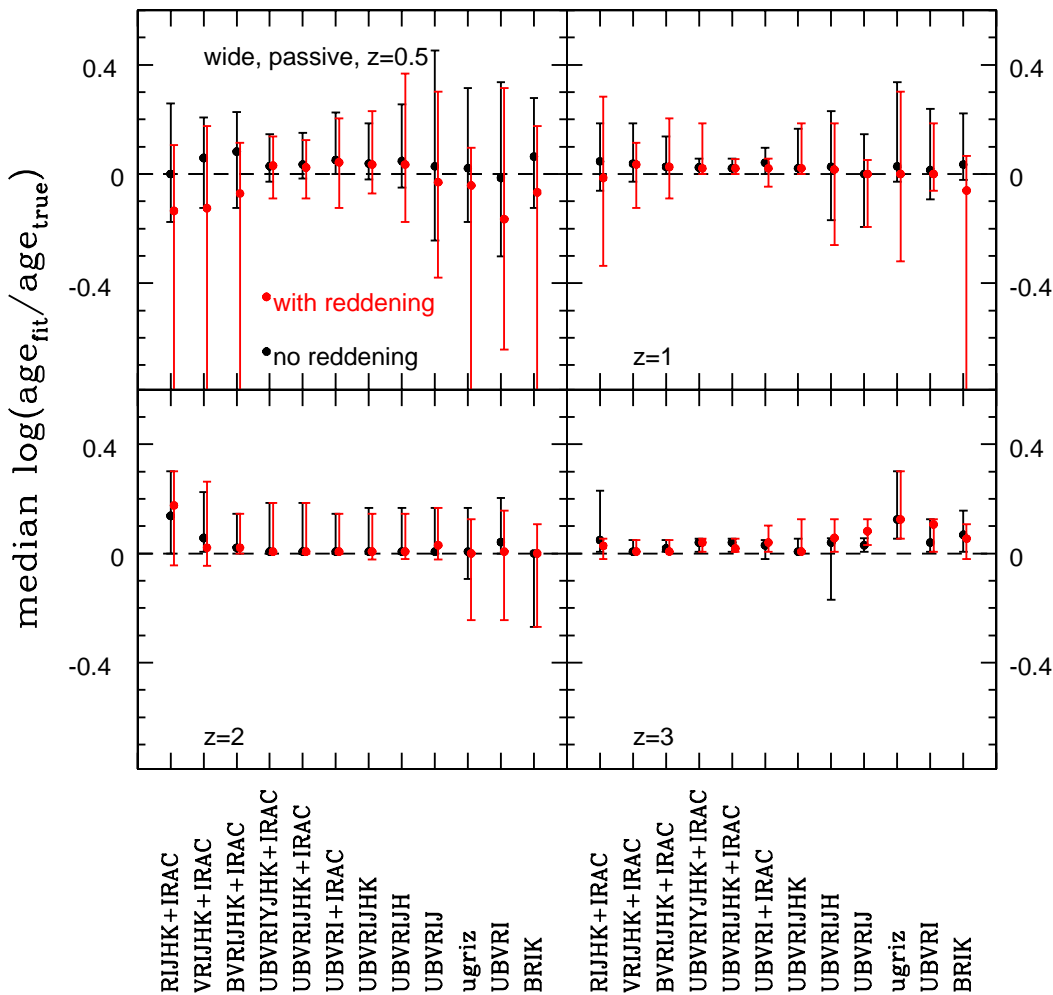}
\caption{\label{oldagedf} Median deviation of fitted age from true ages of mock passive galaxies as a function of wavelength coverage at redshifts from 0.5 to 3 using the wide setup. Black dots and 68\% confidence levels refer to the unreddened case, red to the case that includes reddening in the fit.}
\end{figure*}

Fig. \ref{oldagedt} shows the age recovery of mock passive galaxies as a function of template setup for wide, only-$\tau$, only SSPs (but all four metallicities) and only solar SSP template setups. Overall, the age determination is much better than was the case for star-forming galaxies (Fig. \ref{agedt05}). A wider range of star formation histories and metallicities in the fitting leads to generally overestimated ages in the unreddened case at redshifts 0.5 and 1 with the mismatched SFH as the biggest driving factor. For the wide setup at z=0.5 ages are recovered within $\sim0$ dex to $\sim0.15$ dex in 68\% of the cases when reddening is switched off (see Tables \ref{poverres} and \ref{poverres2}). Overall, the age recovery improves with templates with the correct metallicity (i.e. only-$\tau$ and solar SSP). Additionally, matching the SFH correctly recovers the age best. Small deviations from the true age still occur when fitting with the correct template for two reasons: 1) a slight mismatch in age between template age grid and galaxy age\footnote{For ages of 1 and 1.5 Gyr the closest matching age of the template age grids are 1.015 Gyr  and 1.434 Gyr.} and 2) photometric uncertainties.\\ 
Since the choice of SFH is limited in the only-$\tau$ setup and the metallicity is the correct one, short $\tau$'s which resemble a SSP closest are chosen. The longer period of star formation is then compensated by an older age. Within the wide setup a variety of SFHs exists besides the correct solution of a solar metallicity SSP. Both $\tau$ and truncated models closely resemble a SSP as long as the timescale $\tau$ for star formation or truncation time $t$ are short enough compared to the age. The differences between them are smaller than the imposed photometric uncertainties. Furthermore, a wider variety of metallicities is available. Thus, it is not surprising that the age-metallicity degeneracy and a degeneracy between age and SFH cause the average recovered age to be slightly older than for the only-$\tau$ setup and that the dispersion is larger. The median offset in age for the wide setup is maximally $\sim0.04$ dex.\\
Comparing results for the only-SSP and solar-SSP setups - thus using the correct SFH - enables us to isolate the effect of the age-metallicity degeneracy. This is small on the median, but the scatter induced by using wrong metallicities is much larger (Tables \ref{poverres} and \ref{poverres2}).\\
Similarly to the mock star-forming galaxies, including reddening in the fit has the largest effect for the oldest galaxies (z=0.5) due to the age-dust degeneracy. But overall, the effect of reddening is small.\\
%metallicity effects
Another way to decouple metallicity from SFH effects is to use mono-metallicity wide setups for which we show the results in Fig. \ref{oldagedmetal}. Obviously, ages derived with a solar metallicity setup are best recovered. The remaining scatter of $0.05-0.1$ dex stems from the SFH mismatch. Template setups with sub-solar metallicity overestimate the age to compensate their bluer colour. For the lowest metallicity, every galaxy is fit with the maximum age possible at the given redshift (or close to that age) in order to compensate the underestimated metallicity. In addition, all $\chi_{\nu}^2$ are larger than 2, most are larger than 10, and fits are getting worse towards higher redshift. Clearly, neither age nor SFH can compensate such a large discrepancy in metallicity. The situation is reversed for super-solar metallicity templates. At low redshift the scatter is largest because the covered age range is wider. Ages around 1 Gyr are very well defined by the TP-AGB emission, older ages are more difficult to distinguish because of little evolution in the colours during the RGB-dominated epoch. The introduction of reddening has the largest impact at low redshift. Particularly at z=1 the age-dust degeneracy partly compensates and reverses the effect of the age-metallicity degeneracy and causes an underestimation in age. At high redshift, reddening has no impact.\\
Most importantly Fig. \ref{oldagedmetal} shows that even when using all metallicities ages and metallicities are very well recovered. This means that when we do not know the metallicity it is better to use a wide range for it, as concluded in M06.\\
Mismatches in the IMF have rather small effects for the mock passive galaxies which is why we do not show it. Ages derived with different IMFs are similarly well recovered and do not show an offset between each other because differences in SED-shape due to the IMF are small for intermediate-mass stars, hence intermediate and old ages. The scatter increases for the oldest galaxies. The effect of reddening is negligible.\\
%wavelength dependence
We investigate the wavelength dependence of the derived ages using a wide setup in Fig. \ref{oldagedf}. Median ages are recovered within $\pm0.1$ dex in the unreddened case with little dependence on the wavelength coverage at low redshift. At higher redshift filter setups that do not cover the rest-frame near-IR and the red part of the optical generate a larger scatter. As already mentioned, the mock passive galaxies at these redshifts are 1 - 2 Gyr old, the age range in which the near-IR is dominated by the TP-AGB emission. Without this crucial information, the fit is more prone to degeneracies between age, metallicity and SFH. Interestingly, a filter setup containing only four filter bands but including the near-IR and thus sampling the SED shape over a wider wavelength range, performs in most cases equally as good as other filter sets. The lack of rest-frame UV data at z=2 and 3, causes ages to be overestimated because of star formation history mismatches.\\ 
The inclusion of reddening in the fit has its biggest impact at low redshift. Here the age-dust-degeneracy causes a large scatter and an on average underestimated age for shorter wavelength coverages. Particularly, the large scatter shows that the age recovery fails when the information of the blue and reddest filter bands is missing. Overall, ages are best derived with the full wavelength coverage. Compared to star-forming galaxies, ages of passive galaxies can be better determined in the fit due to a well-defined 4000 \AA \space break and the absence of multiple generations.\\
In summary, mismatches in SFHs are contributing the most to an offset in the derived ages. Thus the best SFHs for passive galaxies are SSPs and exponentially declining SFRs. As metallicity can be recovered (see also section \ref{metalresultspass}), SED templates should cover a range in metallicities. Ages are best determined using the full wavelength coverage. Note that a {\it ugriz} filter setup at $z\sim0.5$ performs well, albeit with larger scatter, which is mostly due to the low number of objects. This is an important finding as we are obtaining galaxy properties (stellar masses in particular) of BOSS (Baryon Oscillation Spectroscopic Survey) galaxies using this setup (Maraston et al. \textit{in prep.}). We list median offsets between recovered and true age as well as 68\% ranges in Tables \ref{poverres} and \ref{poverres2}.
%age diff for different metallicity template setups.

\subsubsection{Metallicity}\label{metalresultspass}
\begin{figure}\includegraphics[width=84mm]{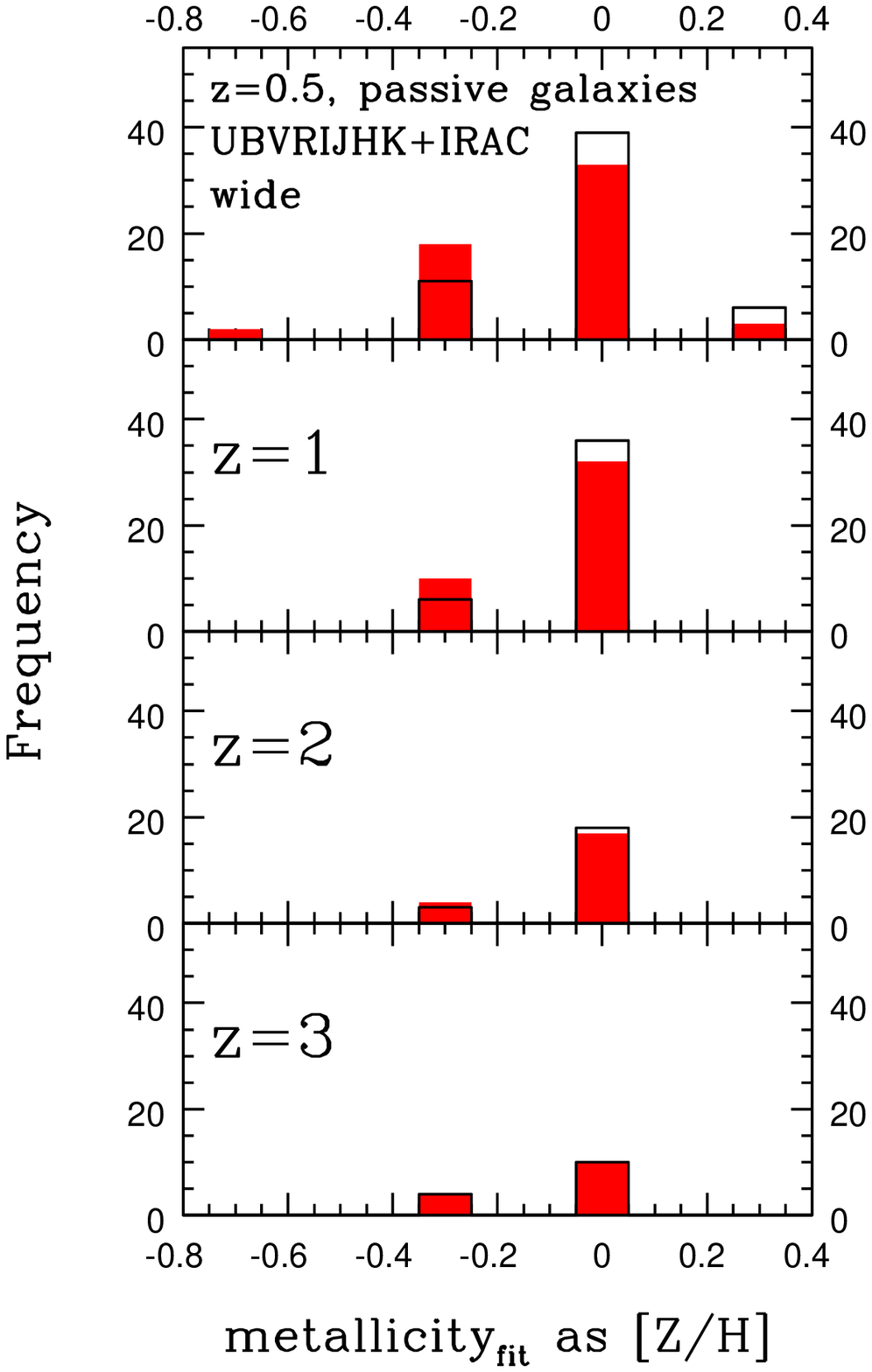}
\caption{\label{oldmetal} Recovered metallicities of mock passive galaxies at redshifts from 0.5 to 3. Black and red are without and with reddening, respectively.}
\end{figure}

For nearly passive objects at $z\sim2$, M06 found that metallicity plays an important role in allowing a robust determination of the stellar population parameters. We confirm this result in section \ref{ageresultspass}. Although a wide choice of metallicities is provided, the fit closely recovers the true metallicity for most objects (Fig. \ref{oldmetal}). When reddening is introduced, the age-dust-metallicity degeneracy affects mostly the oldest galaxies (at low redshift).\\
%wavelength coverage
Derived metallicities depend on the wavelength coverage in the fitting such that at low redshift a lack of rest-frame near-IR and red optical coverage causes metallicities to be overestimated in more cases, clearly, by compensating underestimated ages. The rest-frame near-IR is crucial in breaking the age-metallicity degeneracy as already concluded in M06. When blue filter bands are excluded metallicities are underestimated for a few more objects. These effects are smaller at higher redshift. Including reddening results in underestimation of metallicity for more objects when the wavelength coverage is restricted. Again the effect is largest at low redshift.\\
In conclusion, for nearly passive and aged galaxies it is important to fit with a wide choice of metallicities because metallicity can be recovered. The best template setup for metallicity recovery is the wide setup. The broadest wavelength coverage results in the best metallicity estimates.

\subsubsection{E(B-V)}\label{ebvresultspass}
\begin{figure}\includegraphics[width=84mm]{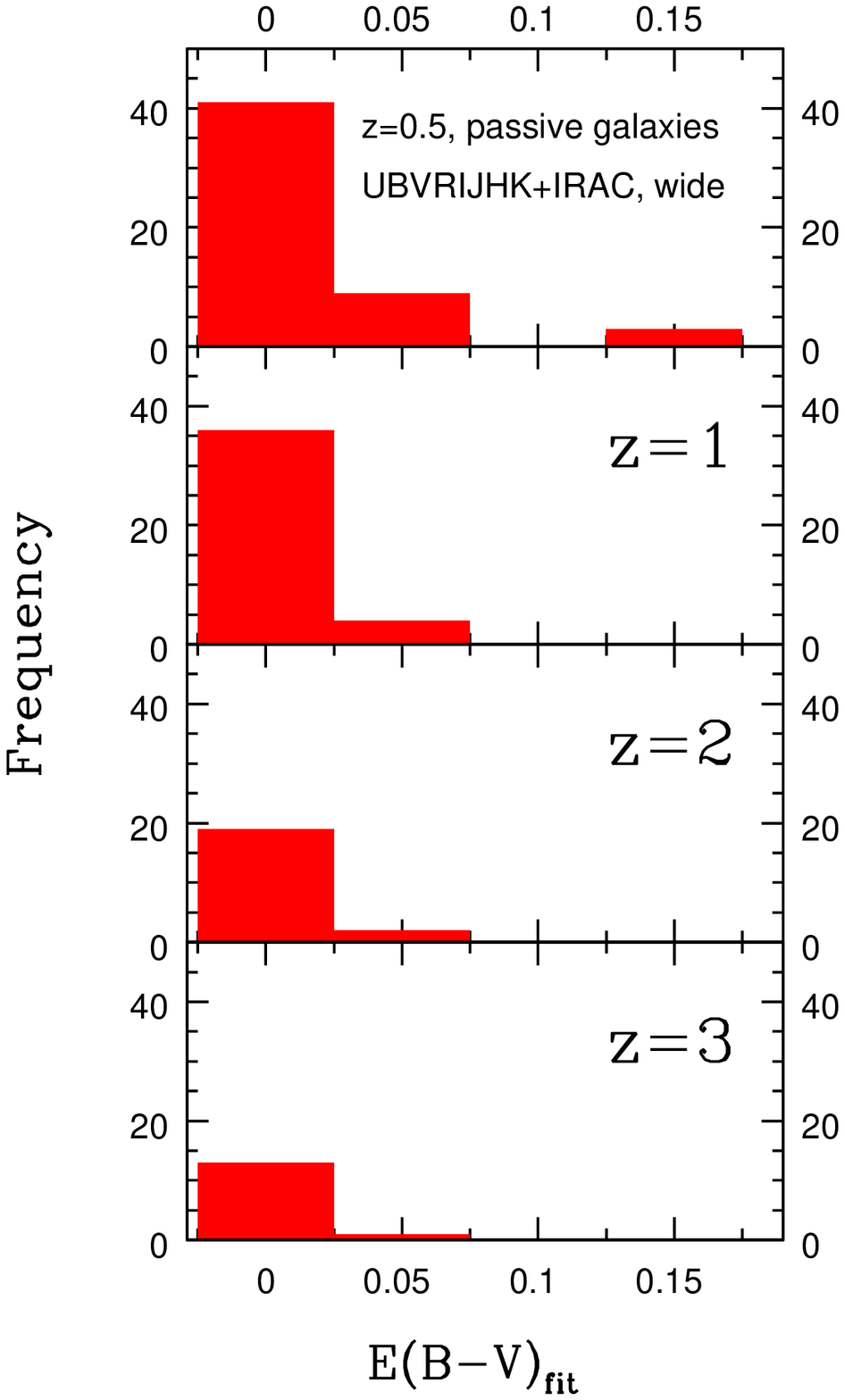}
\caption{\label{oldebv} Recovered E(B-V) of mock passive galaxies at redshifts from 0.5 to 3.}
\end{figure}

Since we did not add reddening to the passive galaxies, any reddening resulting from the fitting procedure is overestimated. But generally, the reddening is correctly identified for most objects (Fig. \ref{oldebv}, Tables \ref{poverres} and \ref{poverres2}). At z=0.5 68\% of recovered reddening show offsets between 0 and 0.07 from the true reddening. Independently of template setup, the largest effect is found at low redshift where the age-dust degeneracy is more efficient as the allowed age range is larger.\\
%wavelength coverage
More objects are wrongly identified to be reddened as wavelength coverage decreases (in both red and blue wavelength) but the effect is small.

\subsubsection{Stellar Mass}\label{massresultspass}
\begin{figure*}\includegraphics[width=144mm]{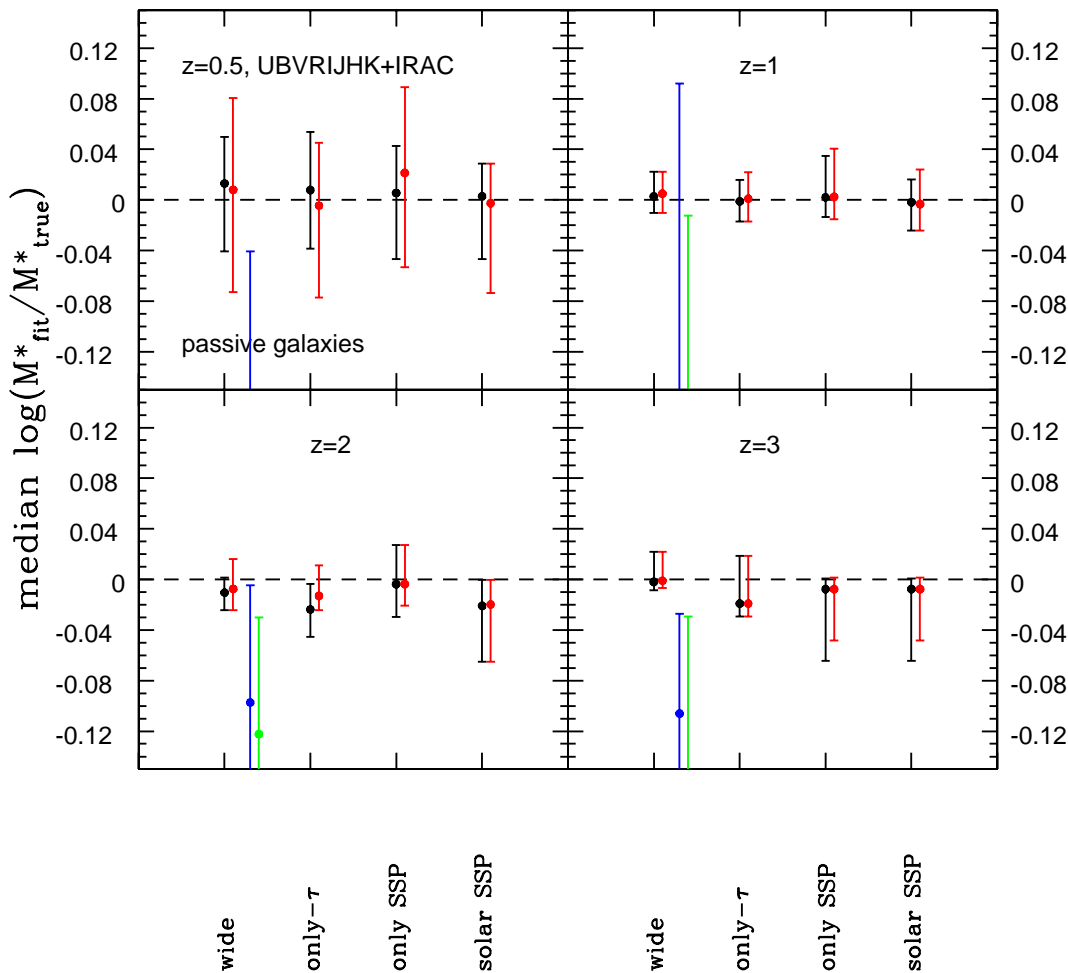}
\caption{\label{oldmassdt} Median mass recovery of mock passive galaxies for different template setups as a function of redshift. Red dots and 68\% confidence levels refer to the inclusion of reddening in the fitting, black symbols to the unreddened case. Blue and green symbols refer to the mass recovery obtained with the wide setup for mock star-forming galaxies without and with reddening, respectively.}
\end{figure*}

%For the mock passive galaxies the effects on stellar mass estimates are:\\
The mass recovery for the mock passive galaxies is very good (Fig. \ref{oldmassdt}) and the median offset in mass is maximally $\pm\sim0.02$ dex and 68\% ranges are within $\pm\sim0.10$ for all template setups (see Tables \ref{poverres} and \ref{poverres2}). As usual, differences between the reddened and unreddened case are small. Scatter is largest at low redshift. Deviations from the true stellar mass for the wide setup are due to degeneracies between star formation histories, metallicity, age and dust (when included in the fit). Offsets in stellar mass even when using the correct template are caused by a slight mismatch in age between template and galaxy for the youngest ages and photometric uncertainties (see also sections \ref{ageresultspass} and \ref{photuncres}, respectively). When fitting with the solar metallicity SSP these small age mismatches cannot be compensated by a different choice of metallicity and SFH (or dust at high redshift). The contribution of dust increases the errors on the derived ages at low redshift because of the age-dust degeneracy. Clearly, in the other template setups the mass is corrected by the choice of an incorrect metallicity and/or SFH and, in the reddened case, dust.\\
% stellar mass diff for metallicity template setups
\begin{figure*}\includegraphics[width=144mm]{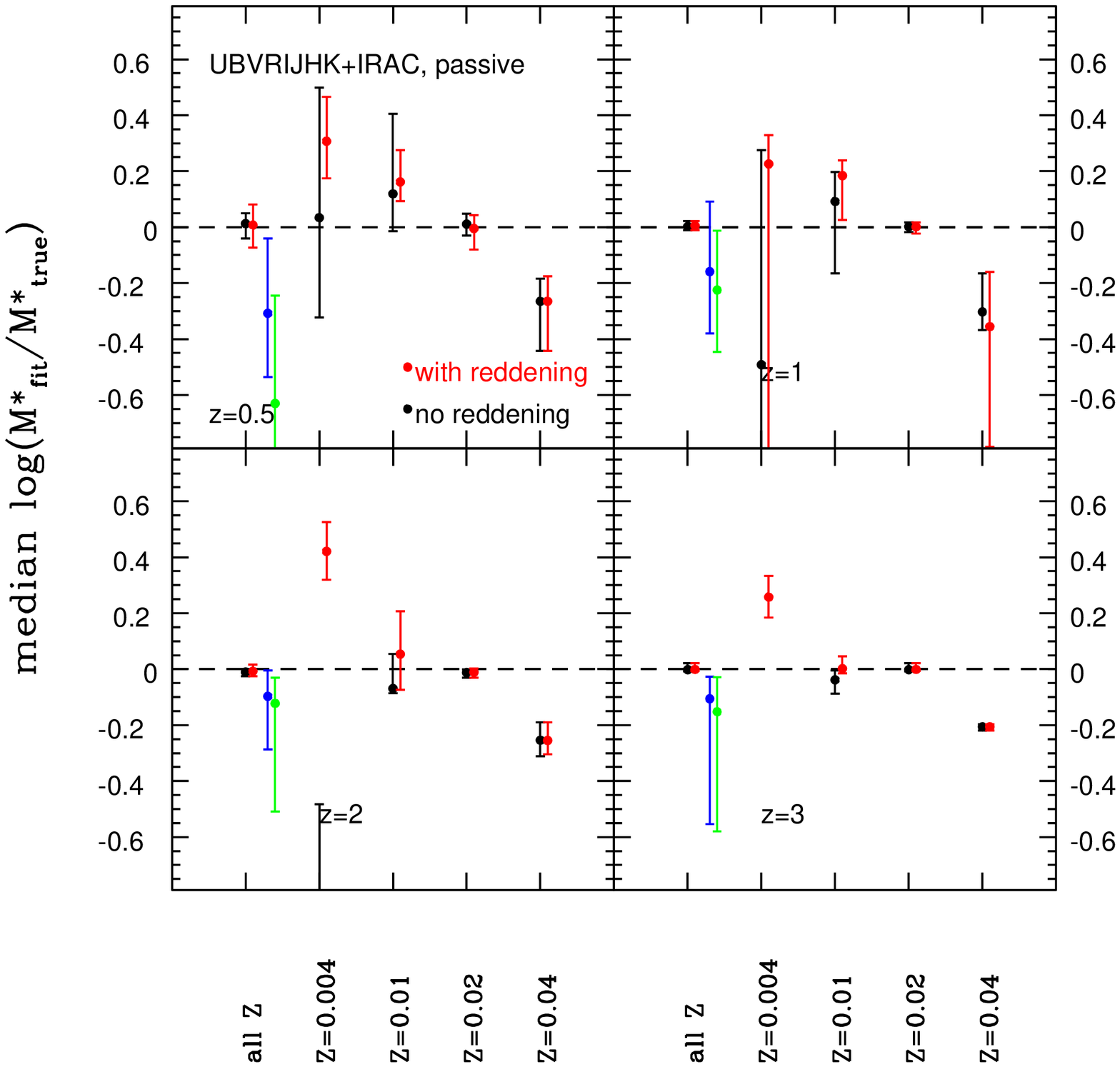}
\caption{\label{massmetalpass} Median recovered stellar mass with 68\% confidence levels for mono-metallicity wide setups as a function of redshift and therefore galaxy age (upper left to lower right). Metallicity in each panel increases from left to right. The input metallicity is solar. Missing points at Z=0.004 without reddening lie at $-1.51^{+1.03}_{-0.68}$ at z=2 and at $-3.31^{+0.47}_{-0.54}$ at z=3. Symbols are the same as in Fig. \ref{oldmassdt}.}
\end{figure*}
Metallicity effects are further explored in Fig. \ref{massmetalpass}. Stellar masses follow the same trend with metallicity as ages (compare to Fig. \ref{oldagedmetal}). For lower metallicities for which ages are overestimated, stellar masses are larger. For higher metallicities masses are smaller. In the unreddened case, masses derived from the lowest metallicity template are significantly underestimated as a direct consequence of the metallicity mismatch and its effect as described in section \ref{ageresultspass}. This shortcoming in metallicity can only be partly compensated by age and not at all by SFH (which would just make the SED even bluer). At $z>0.5$ the fit with the lowest metallicity setup fails completely ($\chi^2{\nu}>50$) and masses are not recovered. Adding dust improves the fits since some of the required red colour of the observed SED can be compensated by dust but ages are still overestimated. This results in overestimated masses.\\
%mass IMF effects
\begin{figure*}\includegraphics[width=144mm]{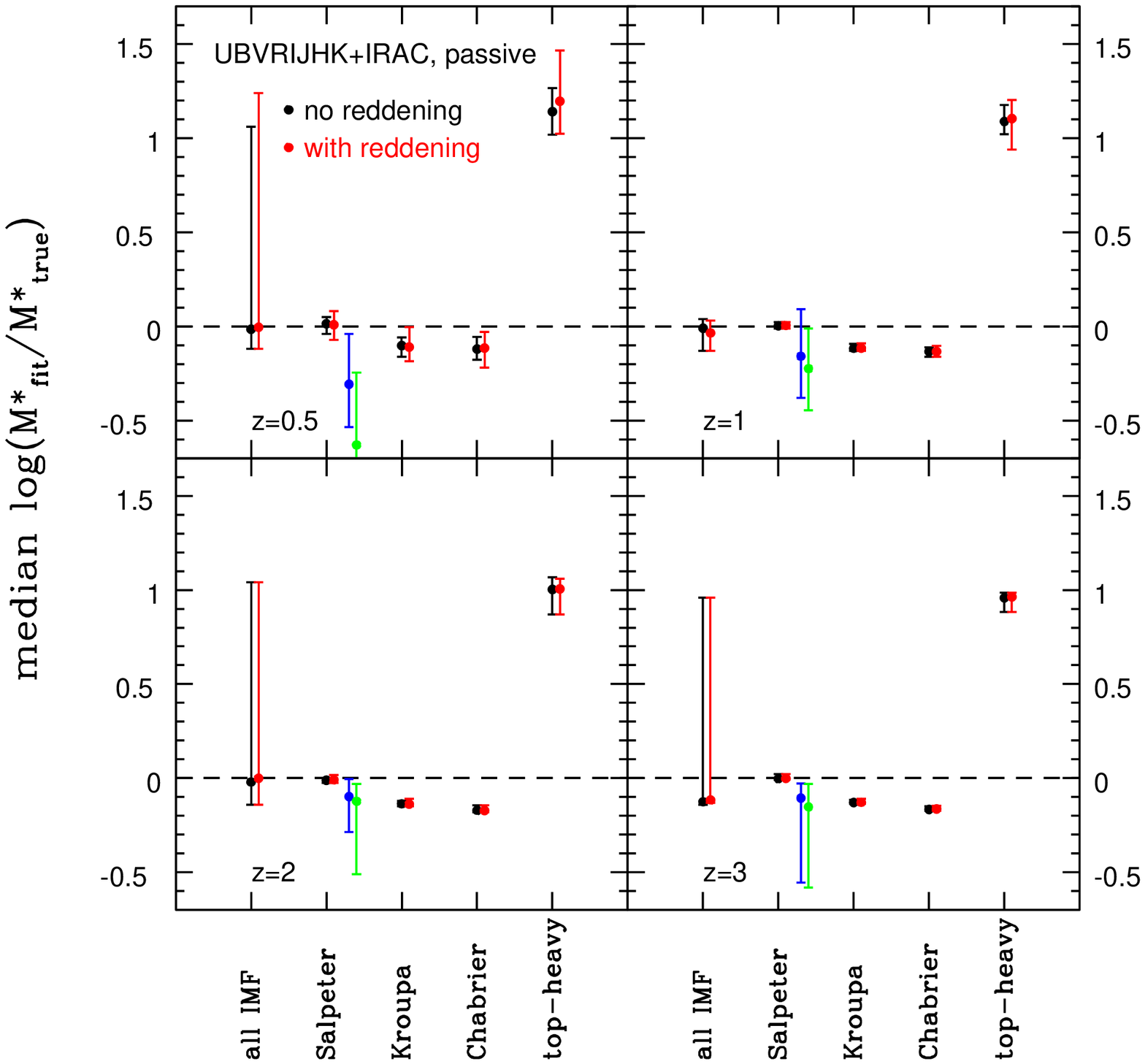}
\caption{\label{massIMFpass} Median stellar mass recovery as a function of the IMF assumed in the fitting template for a wide template setup and UBVRIJHK+IRAC wavelength coverage. The input IMF is Salpeter. Errorbars are 68\% confidence levels. Redshift increases from the top left to the bottom right. Symbols are the same as in Fig. \ref{oldmassdt}.}
\end{figure*}
Stellar masses of mock passive galaxies are best determined with a template setup based on the same IMF used as input, namely a Salpeter IMF (Fig. \ref{massIMFpass})\footnote{The inclusion of dust reddening in the fitting has very little impact on these results and merely increases the scatter at lower redshift.}. Although ages derived with a Kroupa and Chabrier IMF template setup are very similar to those for the Salpeter setup, stellar masses are systematically underestimated by $\sim 0.1-0.2$ dex. The stellar masses obtained with a top-heavy IMF template setup on the other hand are largely overestimated (by $\sim1$ dex on average). These differences can be explained with the different mass-to-light ratios ($M/L$) for the various IMFs. As show in \citet{M98} bottom-light IMFs have lighter $M/L$ because of a lower amount of dwarf M stars whereas a top-heavy IMF is heavier because of the large amount of massive remnants. %If the same light is fitted with the same age for each template IMF than the difference in the IMFs will show in the derived stellar masses. For example, the M/L of a Kroupa and Chabrier IMF SSP of age 5 Gyr and solar metallicity in the K-band are smaller than those of a Salpeter IMF (M/L=0.55 for Kroupa, M/L=0.86 for Salpeter), hence the stellar masses are smaller. For a top-heavy IMF the opposite is the case resulting in larger stellar masses.\\ %(\textbf{\emph{@ Claudia: Do we have the M/L for the top-heavy IMF somewhere? Couldn't find it on the webpage}}). \\
It is encouraging that the choice of IMF is felt in the fit, but an identification of the correct IMF based on the minimum $\chi^2_{\nu}$ is not possible as shown in Fig. \ref{massIMFpass}. The large scatter stems mainly from the top-heavy IMF. Unfortunately, no comparison can be made with the true stellar mass for real galaxies.\\
%mass wavelength coverage
\begin{figure*}\includegraphics[width=144mm]{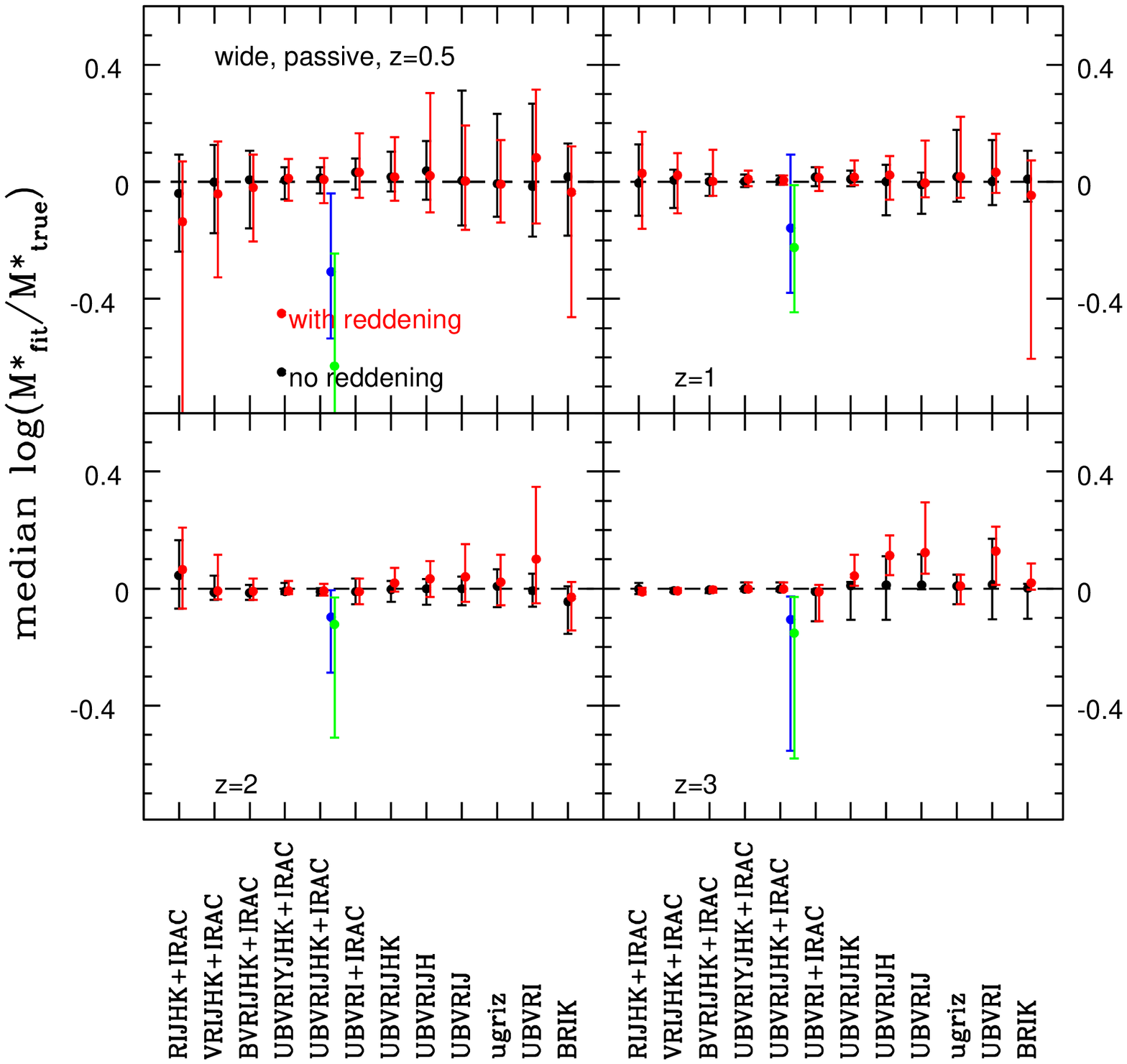}
\caption{\label{massdf05pass} Median differences between true and recovered stellar mass as function of photometric filters used in the model fitting, at redshifts 0.5, 1, 2 and 3. The filter setup is varied from left to right in each panel, redshift increases from the top left to the bottom right. Given are the 68\% confidence levels for each setup. Symbols are the same as in Fig. \ref{oldmassdt}.}
\end{figure*}
The median stellar mass recovery of passive galaxies depends little on the wavelength coverage used in the fit, but the scatter increases when fewer filter bands are used in the fitting (Fig. \ref{massdf05pass}, Tables \ref{poverres} and \ref{poverres2}). Masses are best recovered with the full wavelength coverage when the rest-frame near-IR is included. Stellar masses are recovered within $0.05$ dex when reddening is not included in the fit. Similar to the ages, the inclusion of reddening affects the most at low redshift where masses are underestimated for filter setups lacking blue filter bands. Towards higher redshift the exclusion of the rest-frame near-IR wavelengths in the fit introduces larger errors in the mass estimation. The information stored in the rest-frame near-IR is crucial in order to recover the right stellar mass.\\
In summary, stellar masses of passive galaxies can be recovered within $\sim0.02$ dex for a wide coverage in metallicities, SFHs and wavelength and the exclusion of reddening in the fitting. Restricting the template setup in metallicity has the strongest effect on the mass estimate. The dependence on wavelength coverage is small though not negligible and the scatter decreases when the rest-frame near-IR is included.
%\subsubsection{Star formation history}\label{sfhresultspass}
%In the previous sections we have seen that the stellar population parameters of galaxies are best reproduced if the template matches the SFH, age and metallicity correctly. However, even when a wider choice of SFHs is allowed in the fit, ages, masses and metallicities are very well recovered because of compensating effects.

\subsection{Measuring the strength of the latest starburst}\label{burst}
As widely discussed in M10 and section \ref{sfgresults}, a young stellar population overshines underlying older populations. Naturally, the question arises whether the derived stellar population properties reflect those of the latest burst. For this purpose we simulated passive galaxies located at redshift 1 that consist of a large percentage of an old stellar population and a low percentage of a younger population, both with solar metallicity. The old component is 5 Gyr old, for the young component we take ages of 1 Myr, 10 Myr, 100 Myr and 1 Gyr. We combine old and young components as 99 and 1 \% and 90 and 10 \%. Each galaxy has a total mass of $10^{11}\,M_{\odot}$. Thus, recovered stellar masses of $10^{10}\,M_{\odot}$ and $10^9\,M_{\odot}$ would reflect the mass of the burst. Ages of $\sim4.5$ (for 10\% young) and $\sim4.95$ Gyr (for 1\% young) would reflect the mass-weighted age of the galaxies. The fitting is carried out using the broadest wavelength coverage and a wide setup. Results for stellar mass and age are listed in Table \ref{bursttab}.\\
Clearly, even just one percent of a relatively younger stellar population with respect to the dominant one has the power to hide $30-40$\% of the stellar mass locked up in old stars. When the population is just 1 Myr old it even hides nearly 99\%. For these ages, the fit renders precisely the mass of the burst. When reddening is allowed in the fit, an age of 10 Myr is already young enough to hide nearly 99\% of the remaining old stellar population. In both cases is the mass of the burst recovered within $\sim0.3$ dex rather than the mass of the entire galaxy.
The situation is very similar when the latest stellar population contributes 10\%. The recovered mass is closer to the burst mass for the youngest bursts. Ages are very close to the age of the young component. When reddening is involved, the age-dust degeneracy even causes the recovered mass and age to be lower than those of the burst component which clearly is a mere artefact stemming from the many degeneracies. Derived ages and masses are closest to that of the old population when the young population is 1 Gyr old.\\
Thus, a wide age gap between old and young stars is the biggest challenge in the fitting. This explains the worse recovery of the physical properties of the mock star-forming galaxies at z=0.5.\\
\citet{Serra2007} found in a similar exercise that the stellar ages inferred from Lick-indices measurements of two-component composite stellar populations are driven by the younger less massive component.
\begin{table}
\caption{Derived ages and stellar masses for a $10^{11}M_{\odot}$ galaxy composed of a 5 Gyr old population and a low percentage - 1 or 10\% - of a younger population. Burst characteristics are listed in the first column, while the 2nd and 3rd columns list the derived ages and masses.}
\label{bursttab}
\begin{center}
\begin{tabular}{@{}lcc}\hline
Burst & derived age & derived M* \\
no reddening &  & [$M_{\odot}$]\\\hline
1\% 1Gyr         &3.75 Gyr &  10.85\\\hline
1\% 100Myr    &4.5 Gyr&  10.86\\\hline
1\% 10Myr      &4.5 Gyr&  10.74\\\hline
1\% 1Myr        &15 Myr&  9.23\\\hline
10\% 1Gyr      &2.75 Gyr&  10.83\\\hline
10\% 100Myr &404 Myr&  10.42\\\hline
10\% 10Myr    &8.7 Myr&  9.91\\\hline
10\% 1Myr      &6.6 Myr& 9.93 \\\hline\hline
Burst & derived age & derived M* \\
+ reddening & & [$M_{\odot}$]\\\hline
1\% 1Gyr         &3.5 Gyr&  10.90\\\hline
1\% 100Myr    &4.5 Gyr&  10.86\\\hline
1\% 10Myr      &7.9 Myr&  9.28\\\hline
1\% 1Myr        &13.8 Myr&  9.30\\\hline
10\% 1Gyr      &2.75 Gyr&  10.83\\\hline
10\% 100Myr &6.3 Myr&  9.69\\\hline
10\% 10Myr    &8.7 Myr&  9.91\\\hline
10\% 1Myr      &6.6 Myr& 9.93 \\\hline
\end{tabular}
\end{center}
\end{table}%

\subsection{Reddening laws}\label{rl}
%mass plots for dt rl
\begin{figure}\includegraphics[width=84mm]{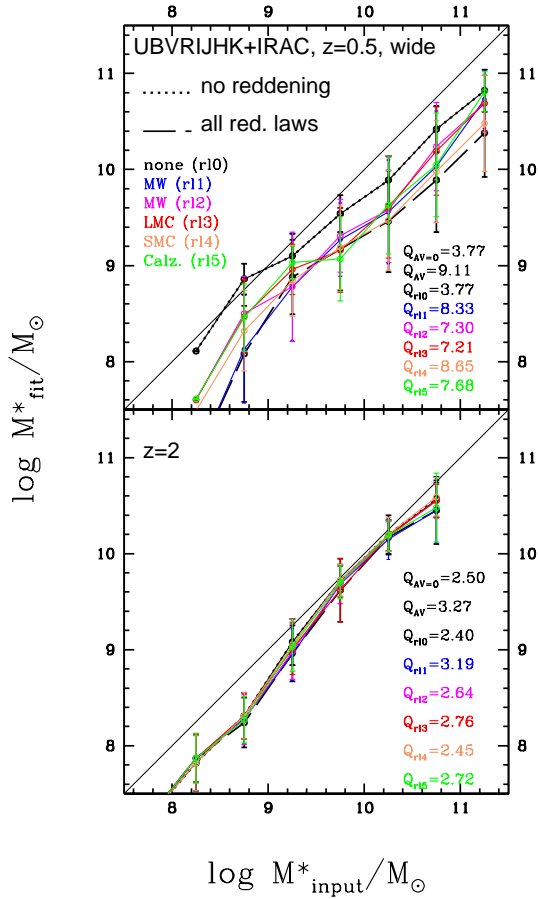}
\caption{\label{massdtrl05} Average mass recovery as a function of reddening law at redshift 0.5 and 2. The dotted black line shows the mass recovery when no reddening is used in either spectra or fitting. The dashed thick black line shows the result when the best fit out of all reddening laws is chosen for each object. The coloured thin lines represent the results applying only one reddening law on the reddened spectra. Errorbars refer to one standard deviation.}
\end{figure}
%SFR plots for dt rl
\begin{figure}\includegraphics[width=84mm]{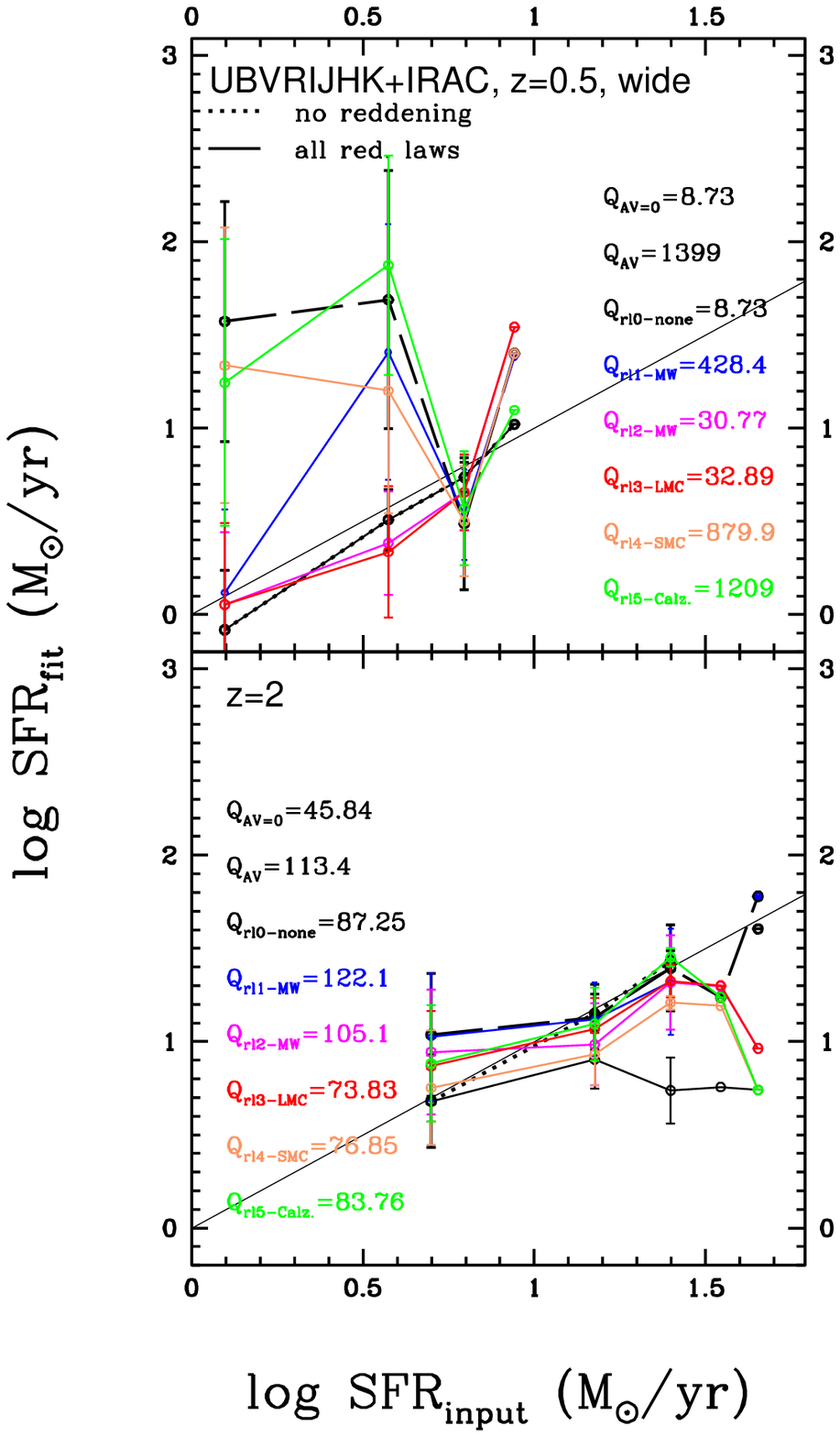}
\caption{\label{sfrdtrl05} Average SFR recovery as a function of reddening law at each redshift where redshift increases from the top left to the bottom right. We show solutions for a wide setup only. Colours and lines are the same as in Fig. \ref{massdtrl05}. Note that the data point at the highest SFR at z=0.5 and the two points at the highest SFR at z=2 consist of only one object each and are merely shown for completeness. Errorbars refer to one standard deviation.}
\end{figure}
%In section \ref{ebvresults} we have shown that reddening is overestimated for most mock star-forming galaxies. 
In this section we explore the dependence of the fitting result on the adopted reddening law. The question is: Can a better overall result be achieved by considering several reddening laws or is using just one - which makes the fit faster - sufficient? \\
Since we do not add dust reddening to passive galaxies, we investigate the case only for star-forming galaxies. Naively one would expect the Calzetti reddening law to provide the best fit and most accurate result for all reddened objects since the reddening prescription for the semi-analytic galaxies is based on it (section \ref{mocks}).
We find that the best mass estimate is achieved when no reddening is used in the fitting. This is also true at higher redshift where galaxies are dust reddened.\\
%mass
The largest difference in mass recovery between the reddening laws occurs at $z=0.5$ where there is little reddening anyway (Fig. \ref{massdtrl05}). The MW law by Allen and the SMC law give the worst mass recovery. This is also reflected in the results for the best fit among all reddening laws. In particular, the SMC reddening curve is much steeper and higher than the Calzetti law \citep[compare Fig. 7 in ][]{Bol2000}. On average, the exact type of reddening law is less important at high redshift but variations in the quality factors imply larger effects for individual objects. \citet{Papovich2001,FS2004,Pozzetti2007,Muzzin09} and \citet[][b]{Muzzin2009b} also found the effect of different extinction laws on stellar masses to be small.\\
%SFR
SFRs on the other hand are much more sensitive to the type of reddening law (Fig. \ref{sfrdtrl05}). We can summarize this in two points: \\
1) When the reddening is a fitting parameter and galaxies are dust reddened, the fitting code is able to pick out the right reddening law for most objects, the Calzetti law in our case. Fitting without reddening leads to underestimated SFRs.\\
2) When galaxies have little reddening (z=0.5 case) and reddening is a free parameter in the fit, the age-dust degeneracy causes SFRs to be overestimated. SFRs are best recovered with no reddening in the fit in this case. \\
In conclusion, stellar masses are best derived without reddening in the fitting at each redshift when the SFHs cannot be matched. For the best SFR estimate on the other hand all reddening laws should be used in the fitting for high redshift, dust reddened objects. For objects without dust reddening, SFRs are best obtained without reddening in the fitting. These conclusions stress again that the simultaneous recovery of masses and SFRs is very difficult and can be achieved only when the assumed SFH is the correct one, as we found at z=2 with inverted-$\tau$ models and a proper parameter selection (M10).
%The specific type of reddening law loses its importance when a prior on minimum age is introduced. For the SFRs the dependence becomes weaker than before but is still present. Results at higher redshift remain unchanged.\\
%We find similar results for mass and SFR recovery as a function of reddening law for all template setups.

\subsection{The effect of photometric uncertainties}\label{photuncres}
\begin{figure*}\includegraphics[width=84mm]{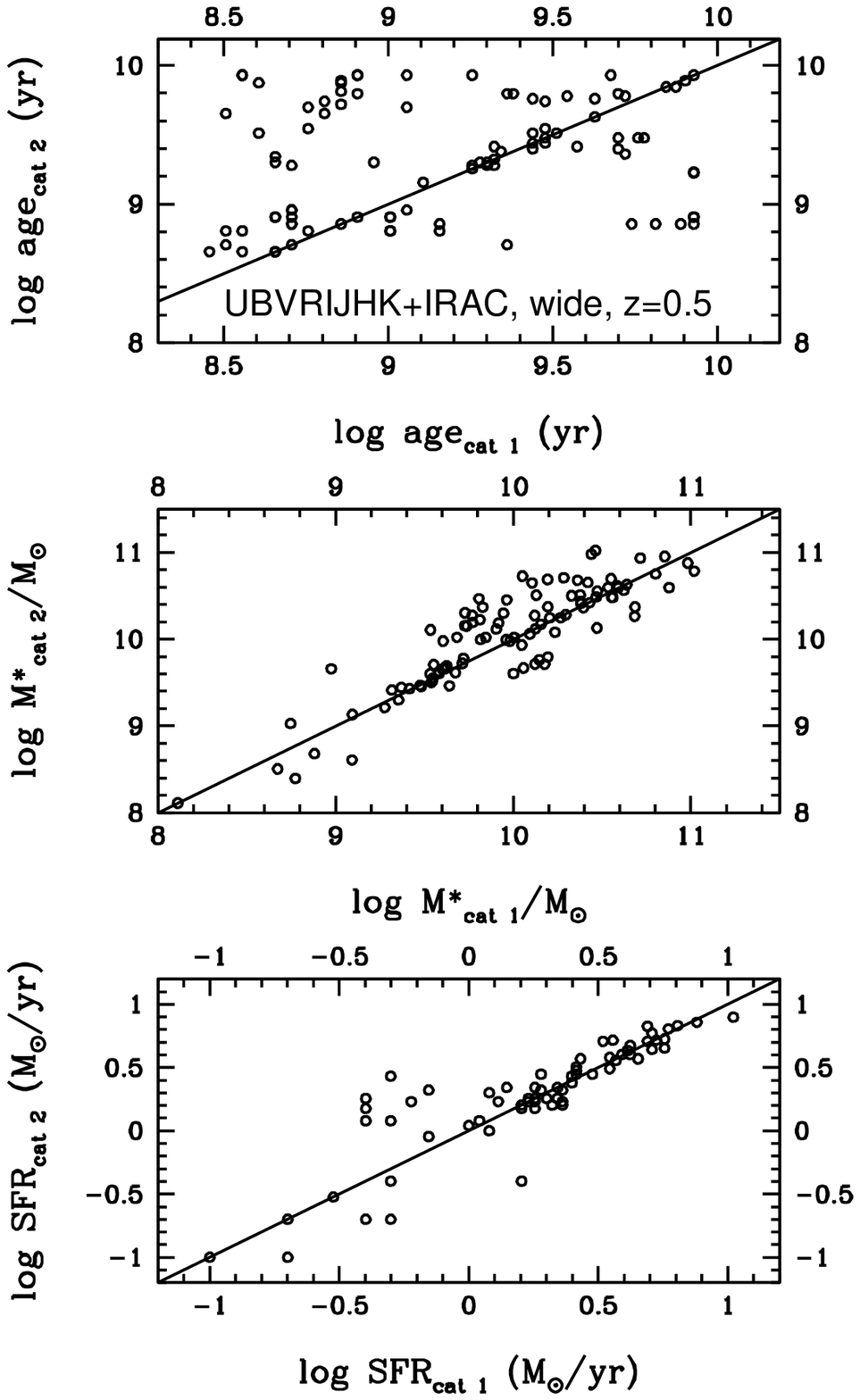}
\includegraphics[width=84mm]{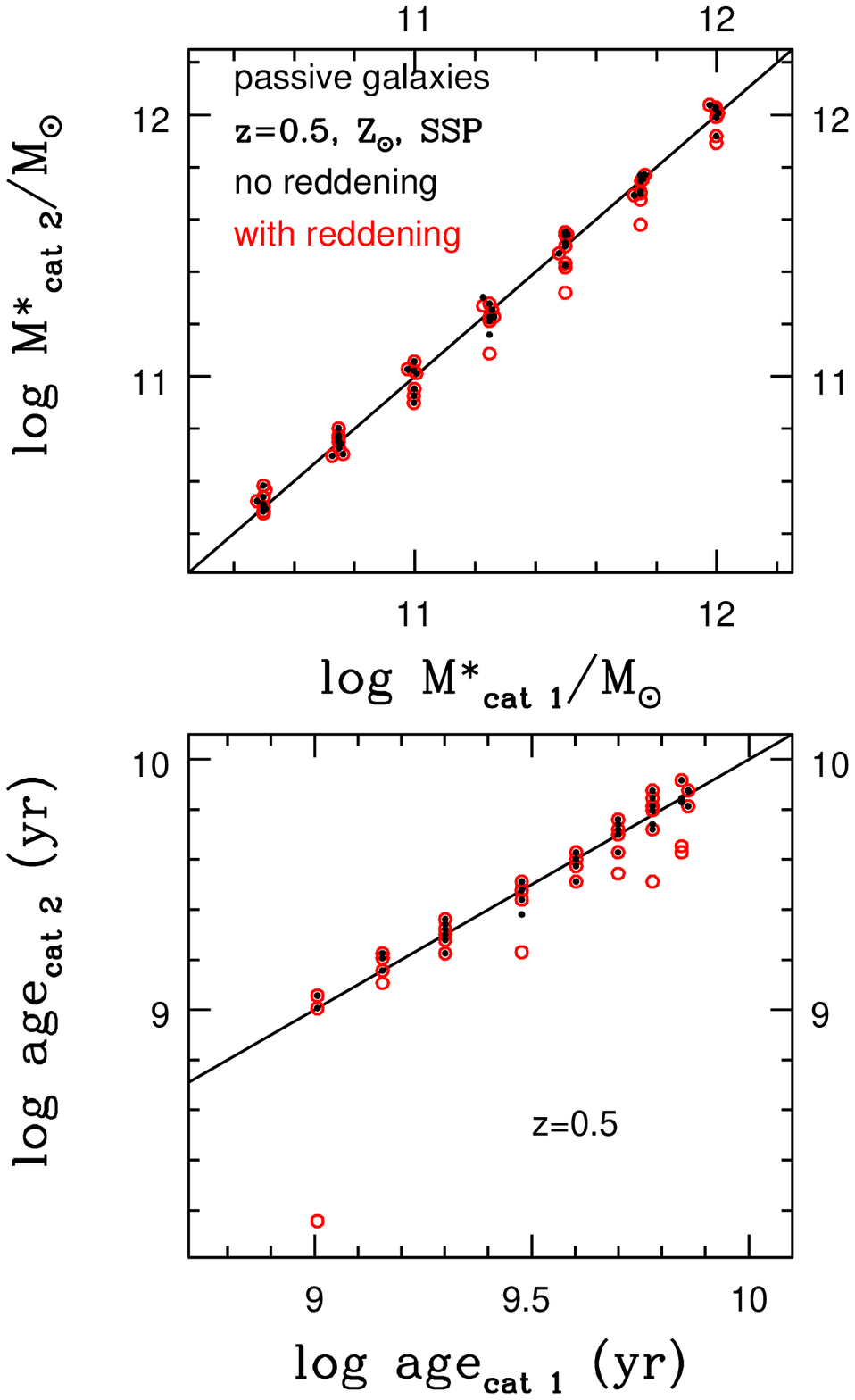}
\caption{\label{agepum} Left: Effect of photometric uncertainties for mock star-forming galaxies at z=0.5 for a wide template setup and broad wavelength coverage in the fit. From top to bottom: ages, stellar masses and star formation rates, respectively. Right: The same for mock passive galaxies. Red and black symbols refer to the cases with and without reddening in the fit. We show galaxies at redshift 0.5 because for these the differences from the true values are largest.}
\end{figure*}
Comparing the results obtained when using original magnitudes and those scattered using a Gaussian error\footnote{Original magnitudes were scattered with their three sigma photometric errors.}  in the fit enables us to study the effect of photometric uncertainties. In order to single out this effect we only consider the case without dust reddening. We focus on the older galaxies with little or no on-going star formation (mocks at z=0.5) for which we find the largest deviations between true and recovered stellar population properties. Star-forming galaxies are fit with the wide setup. Passive galaxies are fit with a solar metallicity SSP only. \\
% mock star forming: age
Fig. \ref{agepum} (top left) shows that ages can be very discrepant, reaching 1.5 dex difference. %The dependence on wavelength coverage is small.\\
% mass
Masses differ on average by $0.08\pm{0.27}$ dex and up to $\sim0.7$ dex (Fig. \ref{agepum}, middle left). %The scatter increases for less extended filter sets for which the difference in mass can be as large as 1.2 dex for single objects. 
The overall trend of underestimating the masses of older galaxies with little on-going star formation remains unaffected by the photometric uncertainties.\\
%SFR
The mean difference between the derived SFRs is small for the highest SFRs (see Fig. \ref{agepum}, bottom left). At the low SFR end deviations are larger (up to 1 dex). \\%Scatter decreases with wavelength coverage leading to a maximum deviation of 0.7 dex because the emphasis of blue filter bands in a filter setup that covers only the optical compared to one that extends to the near-IR. In this way, the amount of newly formed stars can be estimated without any underlying old population that would dominate the light towards the near-IR. Masses are underestimated for the same reason.\\
%passive galaxies
Fig. \ref{massdf05pass} and Fig. \ref{agepum} (right) shows that for passive galaxies the effect is small. The largest deviations in stellar mass are found at z=0.5 amounting to maximal $\sim 0.2$ dex. The inclusion of reddening is negligible. Ages are more affected in the reddening case, particularly at the youngest and oldest ages, because of the age-dust degeneracy. Older ages are hard to distinguish from each other in the fitting as the spectra are very similar.\\
We further explored what effect the size of photometric errorbars has on results for stellar masses for both types of galaxies in the reddened and unreddened case when a wide wavelength coverage and wide template setup is used. For this we investigated the following for the mock star-forming galaxies: 1) using the 3 $\sigma$ photometric errors where $\sigma$ is obtained from the COSMOS and GOODS-South surveys to scatter the original magnitudes (instead of 3 $\sigma$ with $\sigma=0.1$ mag) and carrying out the fit with the survey photometric errors as described in section \ref{mocks}. 2) Using the 3 $\sigma$ photometric errors where $\sigma=0.1$ to scatter the original magnitudes and carrying out the fit with $\sigma=0.1$ for each band. 3) Doubling and tripling the photometric errors in the fit when using survey photometric errors and $\sigma=0.1$ mag photometric errors in the fit when magnitudes were scattered with 3 $\sigma$ errorbars where $\sigma=0.1$ mag. Case 3) is also investigated for the mock passive galaxies.\\
We find the largest effect for mock star-forming galaxies at the lowest redshift for which the median offset between derived and true stellar mass can differ by up to $\sim$0.15 dex in the reddened case and up to $\sim$ 0.07 dex in the unreddened case between the various options leaving masses still underestimated overall. Increasing the errorbars in the fit to double and triple the values plays only a small role. The effects are negligible at high redshift and for mock passive galaxies. Hence, uncertainties on the star formation histories dominate the results.

\section{Comparison to the literature}\label{litcomp}
%comparison to other literature, Wuyts
Other works in the literature exist on this topic. \citet[][ hereafter W09]{Wuyts2009} address the efficiency of SED-fitting for a smoothed particle hydrodynamical merger simulation of two disk galaxies before (disk phase), during (merger phase with triggered star formation) and after the merging (spheroid phase) spanning a total time of 2 Gyr, but not fixed to a particular redshift. They construct the SEDs at various snapshots in time for different viewing angles, place them at redshifts between 1.5 and 2.9 and study the effects of SFH, dust attenuation, metallicity and AGN contribution on the recovery of age, reddening (E(B-V)), visual attenuation ($A_v$), stellar mass and SFR. Their SED-fitting uses solar metallicity templates for three SFHs, namely SSP, constant SFR and a $\tau$-model with $\tau=0.3$ Gyr. The templates are based on a Salpeter IMF and BC03 stellar population synthesis code. $A_v$ varies between 0 and 4 in steps of 0.2 but is only applied to the constant SFR and $\tau$-model template. Ages are constrained to be larger than 50 Myr and younger than the age of the Universe. The fitting is carried out with \emph{HyperZ} and a Calzetti reddening law.\\
Our semi-analytic mock galaxies show considerable star formation for $z\geq1$ and thus compare to the disk (and merger phase) of W09\footnote{We do not distinguish our mock galaxies in disks and mergers although they might have undergone a merging event shortly before we observe them}. For this we predominantly consider the wide setup with age constraint and the inverted-$\tau$ setup (both with reddening). We compare the results of the wide setup without reddening for our mock passive galaxies for $z\leq1$ to their spheroid phase. For the purpose of comparison, we define $\Delta$ as the difference between recovered and true value\footnote{logarithmic values for age, mass and SFR} analogue to W09. Negative values of $\Delta$ correspond to underestimation of the according quantity, positive values to overestimation. As true age for the mock star-forming galaxies we assume the mass-weighted age. We list our median values for $\Delta$ together with those of W09 in Table \ref{wuytscomp}.\\
W09 find slight overestimation of ages in the disk phase and underestimation in the merger phase. The ages we derive are underestimated at $z=1$ and $2$ and overestimated at $z=3$ for the wide setup with age constraint. Ages are more underestimated towards lower redshift due to the wide range in population ages comprised within those galaxies (compare Fig. \ref{inputage}). W09 seem to determine better ages as they consider only a small age-range, no metallicity effects and set an age constraint. We do not compare the age for the inverted-$\tau$ models as those have a constant age which is more comparable to the oldest age. While E(B-V) is underestimated in all phases in W09, our derived reddening values are  overestimated at lower redshift and very well recovered at higher redshifts. Our scatter is smaller than that of W09 at all redshifts for the two setups. Stellar masses at $z\geq2$ are equally underestimated as in W09's disk and merger phase. Our mass estimate for $z\geq2$ improves if we only consider objects in the same mass range as W09 (log M*$>9.8$). However, masses are best recovered with the inverted-$\tau$ setup where they show almost no offset with smaller scatter. The median stellar mass offset is larger for galaxies at redshift 1. In both disk and merger phase W09 underestimate the SFRs. We overestimate the SFRs at intermediate redshifts. At z=3 the median offset in SFR is small. Our scatter is somewhat smaller than that of W09. \\
Since we did not put reddening onto the mock passive galaxies, we only compare the fit values for the stellar masses, ages and SFRs with W09. Even in the case with reddening, a value of E(B-V)=0 is correctly recovered for most objects. Both ages and stellar masses are slightly overestimated in our case, while they are underestimated in W09. However, in each case, offsets and scatter are very small. This confirms that stellar masses of passive galaxies can be much better determined by SED-fitting than those of star-forming galaxies.\\
While W09's reddening and SFRs are underestimated in the spheroid phase, both quantities are very well recovered for our passive galaxies. We assign this to the fact that the passive galaxies we simulated are single bursts while galaxies in the spheroid phase in W09 are clearly precessed by considerable star formation and thus still suffer from overshining.\\
For a direct comparison we reproduce Fig. 8 of W09 for the basic trends and correlations between these physical  properties of galaxies and show our results for mock star-forming galaxies in Fig. \ref{trends}. As in W09, in summary, those correlations are: 1) Age and stellar mass show a strong correlation, i.e. an underestimation in age will lead to an underestimation in mass and vice versa. 2) This is anticorrelated with dust reddening, the so called age-dust degeneracy. 3) It is also anticorrelated with metallicity, the so called age-metallicity degeneracy. 4) SFR and dust reddening are tightly correlated, consequently age and SFR are anticorrelated. Furthermore, we find evolution in these with redshift. Ages and stellar masses are increasingly underestimated whereas reddening and SFRs are increasingly overestimated towards low redshift.\\
%Lee
\citet{Lee2009} focus on the robustness of the derivation of physical parameters of Lyman break galaxies at redshifts $\sim 3.4$, $4.0$ and $5.0$ (U-, B-, and V-dropouts, respectively). They create mock galaxies from a semi-analytical model using BC03 stellar population models with a Chabrier IMF and apply the Lyman break selection criterium to identify star-forming objects at high redshift. Their SED-fitting employs Chabrier IMF BC03 templates with three metallicities (0.2, 0.4 and 1 $Z_{\odot}$) and exponentially declining SFRs with $\tau=0.2 - 15$ Gyr. The Calzetti law is used for dust extinction. They find that the means of the intrinsic and reproduced mass distributions differ between 19 and 25\% (for U- and V- dropouts, respectively), such that recovered masses are underestimated. The corresponding difference for our $z=3$ objects is $\sim17$\% (or 0.08 dex) using the wide setup with reddening and $\sim15$\% (or 0.07 dex) when the minimum age is 0.1 Gyr. For the inverted-$\tau$ setup the difference between the means of true and recovered stellar mass distributions is $<1$\% (0.004 dex).\\ 
\citet{Lee2009} also report larger discrepancies for mean age and SFR in comparison to the true values, mean SFRs (averaged over 100 Myr) are underestimated by $\sim 62$\% and mass-weighted mean ages are overestimated by about a factor of 2. They assign this behaviour to two effects: the mismatch between real and template SFHs and the overshining effect of a young stellar population. Furthermore, they acknowledge the influence of the age-dust degeneracy on the derived SFRs. \citet{Lee2009} try to solve the SFH mismatch in using two-component templates consisting of two time-separate burst of star formation with different values of $\tau$. However, we have shown in M10 and the previous sections that using inverted-$\tau$ models recovers the mass for mock star-forming galaxies at $z=2$ and 3 perfectly. SFRs are slightly overestimated and show a larger scatter at the low SFR end where the exponentially increasing SFH of the template is not matched by the simulated galaxies. Based on the SFHs of the mock star-forming galaxies as shown in Fig. \ref{inputsfh} we believe that the best model to recover the masses and SFRs of star-forming objects at $z>3$ is the inverted-$\tau$ template (using appropriate formation redshifts and ages). Because of the fixed age, the overshining by young components is ignored in the fit. Derived ages will therefore always be older than input luminosity- or mass-weighted ages and compare better to the oldest age.\\
%Longhetti & Saracco 2008
\citet{Longhetti08} study the systematics involved in the mass estimates from SED-fitting of mock early-type galaxies between redshift 1 and 2. They simulate early-type galaxies with Salpeter IMF solar metallicity BC03 models of exponentially declining SFH ($\tau=0.6$ Gyr) and ages between 1.7 and 4.2 Gyr. Furthermore, they add secondary bursts at different times with mass fractions of 5 and 20\% and star formation time scale of $\tau=0.1$ Gyr. Dust obscuration is considered to be $A_V=0, 0.2$ and 0.5 using the Calzetti law. SED-fits are performed using \emph{hyperz} and solar metallicity $\tau$- models of BC03 with $\tau=0.1, 0.3, 0.6, 1.0$ Gyr. They also test the effects of using different stellar population models. \citet{Longhetti08} conclude that overall stellar masses of early-type galaxies cannot be estimated better than within a factor of 2-3 of the true value at fixed known IMF. We find that the stellar masses of the mock passive galaxies in our work at z=1 can be recovered within a factor of $\sim1.05$ ($\sim0.02$ dex) at known IMF. Including uncertainties on the IMF (but only considering among Kroupa, Chabrier and Salpeter) masses are recovered within 0.2 dex. Our result is better probably just because we have really zero on-going star formation in our passive galaxies while \citet{Longhetti08} by considering a $\tau$ model and second bursts keep a small amount of star formation, which confuses the derived properties for the reasons we amply discussed above.\\
%Bol2009: comparison between salpeter and chabrier with mock and zCOSMOS galaxies, age and reddening similar for 2 bestfit SEDs, negligible offsets and very small dispersion... (M05, BC03, CB07)
Lastly, \citet{Bol2009} find a typical total uncertainty caused by different choices of reddening law, metallicities, SFHs and stellar population models in stellar mass estimates of $\sigma\simeq0.2$ dex.

\section{Homogenising derived properties via scaling relations}\label{scalerel}
%scaling relations
For many purposes, combining results from different authors and data sets is required. In this case, homogeneity of the analysis method is essential. However, different models, templates, IMFs and fitting procedures are used in the literature and we have demonstrated in the previous sections their impact on the results. Here, we have the opportunity to obtain 'scaling relations' which allow the transformation of the derived physical properties from one set of fitting parameters to another. Based on the previous sections we have derived such relations for stellar mass for a variety of parameter choices using a least squares fit. They are listed in table \ref{scale} for star-forming galaxies and in table \ref{scalepass} for passive galaxies. In order to gain statistical robustness we use results for the entire merger tree. Some examples are shown in Fig. \ref{massIMFd} as solid green and dashed magenta lines.\\
Note that these relations provide information only about statistical differences, the difference in stellar mass between various fitting setups for a single object can significantly deviate from this.
\begin{table}
\caption{Scaling relations for mock star-forming galaxies in the form of $logM^*_{wide Salpeter}=a_1+a_2*x$ where $x$ stands for $logM^*$ of the various fitting setups. The coefficients $a_1$ and $a_2$ represent the unreddened case, $b_1$ and $b_2$ represent the reddened case.}
\begin{center}
\begin{tabular}{@{}clcccc}\hline
redshift & x & $a_1$ & $a_2$ & $b_1$ & $b_2$\\\hline
0.50  &  wide Chabrier   &   0.4814  &    0.9666	 &    0.5559  &    0.9581 \\\hline
1.00  &  wide Chabrier   &   0.5913  &    0.9563  &    0.5826  &    0.9601 \\\hline
2.00  &  wide Chabrier   &   0.5125  &    0.9673  &    0.3675  &    0.9831 \\\hline
3.00  &  wide Chabrier   &   0.7223  &    0.9485  &    0.4631  &    0.9755 \\\hline
0.50  &  wide Kroupa  &    0.4445  &    0.9707  &   0.2827  &    0.9882  \\\hline
1.00  &  wide Kroupa  &    0.4244  &    0.9754  &   0.6725  &    0.9515  \\\hline
2.00  &  wide Kroupa  &    0.2409  &    0.9973  &   0.1958  &    1.0024  \\\hline
3.00  &  wide Kroupa  &    0.1042  &    1.0136  &   0.0840  &    1.0151  \\\hline
%0.50  &  th salp  &    3.2595  &    0.6052  &    6.9659  &    0.2669 \\\hline
%1.00  &  th salp  &    3.2868  &    0.6259  &    5.7522  &    0.4079 \\\hline
%2.00  &  th salp  &    3.4986  &    0.6190  &    3.2089  &    0.6543 \\\hline
%3.00  &  th salp  &    2.8414  &    0.7105  &    2.1081  &    0.7443 \\\hline
0.50  &  wide $Z_{\odot}$  &    0.8464  &    0.9292  &   -0.9575  &    1.0978 \\\hline
1.00  &  wide $Z_{\odot}$  &    -0.0264  &    1.0077  &   -0.5233  &    1.0531 \\\hline
2.00  &  wide $Z_{\odot}$  &    0.7532  &    0.9225  &    0.5159  &    0.9452 \\\hline
3.00  &  wide $Z_{\odot}$  &    0.6843  &    0.9314  &    0.4773  &    0.9514 \\\hline
0.50  &  only-$\tau$  &    1.0651  &    0.9057  &   -2.8838  &    1.2630 \\\hline
1.00  &  only-$\tau$  &    -0.1275  &    1.0171  &   -0.8565  &    1.0844 \\\hline
2.00  &  only-$\tau$  &    0.5824  &    0.9383  &    0.4459  &    0.9510 \\\hline
3.00  &  only-$\tau$  &    0.8008  &    0.9197  &    0.4467  &    0.9554 \\\hline
%0.50  &  SSP &    -0.6057  &    1.0807  &   -1.4678  &    1.1741 \\\hline
%1.00  &  SSP &    -0.0842  &    1.0330  &   -0.7707  &    1.1354 \\\hline
%2.00  &  SSP &    -0.3447  &    1.0536  &   -3.4109  &    1.4226 \\\hline
%3.00  &  SSP &    -1.0393  &    1.1474  &   -3.9928  &    1.4932 \\\hline
0.50  &  wide BC03  &    -0.0349  &    0.9887  &    0.0385  &    0.9611 \\\hline
1.00  &  wide BC03  &    0.0887  &    0.9675  &   -0.2077  &    0.9939 \\\hline
2.00  &  wide BC03  &    0.9366  &    0.8855  &    0.8272  &    0.8947 \\\hline
3.00  &  wide BC03  &    1.0761  &    0.8785  &    0.7205  &    0.9101 \\\hline
0.50  &  BC03 only-$\tau$   &    0.2524  &    0.9617  &   -2.4432  &    1.1980 \\\hline
1.00  &  BC03 only-$\tau$   &    0.2369  &    0.9487  &   -0.1809  &    0.9906 \\\hline
2.00  &  BC03 only-$\tau$   &    1.2843  &    0.8480  &    1.1366  &    0.8613 \\\hline
3.00  &  BC03 only-$\tau$   &    1.3341  &    0.8523  &    1.0544  &    0.8759 \\\hline
\end{tabular}
\end{center}
\label{scale}
\end{table}%
\begin{table}%[htdp]
\caption{Scaling relations for passive galaxies in the form of $logM^*_{wide Salpeter}=a_1+a_2*x$ where $x$ stands for $logM^*$ of the various fitting setups. The coefficients $a_1$ and $a_2$ represent the unreddened case, $b_1$ and $b_2$ represent the reddened case.}
\begin{center}
\begin{tabular}{@{}clcccc}\hline
redshift & x & $a_1$ & $a_2$ & $b_1$ & $b_2$\\\hline
0.50  &  wide Chabrier   &    0.0243  &    1.0095  &    0.3167  &    0.9821 \\\hline
1.00  &  wide Chabrier   &    0.0913  &    1.0042  &    0.1660  &    0.9974 \\\hline
2.00  &  wide Chabrier   &    0.1477  &    1.0007  &    0.2017  &    0.9960 \\\hline
3.00  &  wide Chabrier   &    0.1612  &    1.0001  &    0.1615  &    1.0001 \\\hline
0.50  &  wide Kroupa   &   -0.0188  &    1.0124  &    0.0795  &    1.0022 \\\hline
1.00  &  wide Kroupa   &    0.0785  &    1.0035  &   -0.7200  &    1.0731 \\\hline
2.00  &  wide Kroupa   &    0.1289  &    0.9996  &    0.1785  &    0.9952 \\\hline
3.00  &  wide Kroupa   &    0.1217  &    1.0004  &    0.1214  &    1.0005 \\\hline
%0.50  &  th   &   -0.6635  &    0.9631  &    3.7492  &    0.6070 \\\hline
%1.00  &  th   &   -1.0417  &    0.9969  &    5.2470  &    0.4930 \\\hline
%2.00  &  th   &   -0.9025  &    0.9930  &   -0.8167  &    0.9862 \\\hline
%3.00  &  th   &   -1.1013  &    1.0139  &   -1.0828  &    1.0125 \\\hline
0.50  &  wide $Z_{\odot}$  &    0.1568  &    0.9865  &   -0.1059  &    1.0129 \\\hline
1.00  &  wide $Z_{\odot}$  &    0.0227  &    0.9983  &   -0.0418  &    1.0051 \\\hline
2.00  &  wide $Z_{\odot}$  &    0.0773  &    0.9941  &   -0.0145  &    1.0022 \\\hline
3.00  &  wide $Z_{\odot}$  &    0.0006  &    0.9999  &    0.0003  &    1.0000 \\\hline
0.50  &  only-$\tau$   &    -0.0161  &    1.0012  &   -0.2548  &    1.0261 \\\hline
1.00  &  only-$\tau$   &    0.0442  &    0.9966  &   -1.0724  &    1.0948 \\\hline
2.00  &  only-$\tau$   &    0.0592  &    0.9963  &    0.0013  &    1.0005 \\\hline
3.00  &  only-$\tau$   &    -0.0512  &    1.0060  &   -0.0336  &    1.0046 \\\hline
0.50  &  SSPs   &    -0.0401  &    1.0043  &    0.0877  &    0.9914 \\\hline
1.00  &  SSPs   &    -0.0498  &    1.0039  &    0.3950  &    0.9628 \\\hline
2.00  &  SSPs   &    0.2261  &    0.9792  &    0.2829  &    0.9744 \\\hline
3.00  &  SSPs   &    0.0685  &    0.9956  &    0.0715  &    0.9952 \\\hline
0.50  &  wide BC03    &    0.7390  &    0.9240  &    0.8017  &    0.9245 \\\hline
1.00  &  wide BC03    &    1.3682  &    0.8783  &   -0.6348  &    1.0528 \\\hline
2.00  &  wide BC03    &    7.7711  &    0.4620  &    0.0007  &    0.9854 \\\hline
3.00  &  wide BC03    &    8.8104  &    0.6450  &    0.0241  &    0.9793 \\\hline
\end{tabular}
\end{center}
\label{scalepass}
\end{table}%

\section{Summary and conclusions}\label{summ}
We analyse the problem of recovering galaxy stellar population properties - such as ages, star formation rates, stellar masses -  through broad-band spectral energy distribution fitting as a function of the various parameters of the fit, namely: star formation history, metallicity, age grid, the Initial Mass Function (IMF), reddening law, wavelength coverage and filter setup used in the fitting. 
This is a crucial question in galaxy formation and evolution as most conclusions that are derived from data are fully based on the galaxy properties that are obtained through this method \citep[see e.g. the recent review by][]{Shapley2011}.
As test particles we use mock galaxies whose properties are known and well defined. We consider both star-forming and passive galaxies and a large redshift range, from 0.5 to 3. Mock star-forming galaxies are taken from a semi-analytical galaxy formation model (GalICS) which uses as input the same stellar population model (by Maraston 2005) used for the fitting templates, but we also experiment with other stellar population models for both the fitting and the semi-analytic galaxies (results for this are fully discussed in the Appendix). Mock passive galaxies are simulated using solar metallicity simple stellar populations. We then treat the mocks as real observed galaxies, i.e. we obtain their observed-frame magnitudes, apply SED-fitting with various templates and compare the properties derived from the fitting to their true values.\\
This work complements an earlier paper (Maraston et al. 2010) where we discuss the problem of deriving star formation rates and masses for real $z\sim2$ star-forming galaxies right at the peak of the cosmic star formation rate. There we found that the adopted star formation history affects the results, as well as the priors that are set on the galaxy formation ages. This happens because the age parameter becomes very poorly constrained for galaxies with on-going star formation, due to the overshining of older generations by the youngest stars, even when these make up just a few percent of the total stellar mass. As a main result of that paper we put forward a star formation history that allows for a very good determination of star formation rates and masses of star-forming galaxies. This is an exponentially-increasing star formation with high formation redshift and a star formation timescale $\tau\sim 0.4$ Gyr. We also find that normally adopted $\tau$-models underestimate the stellar mass and overestimate the star formation rates. In this paper we quantify these various offsets using simulated galaxies (Appendix \ref{overres}).\\

\noindent Our results for mock star-forming galaxies can be summarised as follows:\\
- The match between template and real SFH is crucial for a {\it simultaneous} determination of the galaxy physical parameters. For example, at high redshift - when galaxy ages span a confined range - stellar masses are very reasonably determined - within 0.1 dex - with a variety of template choices. However, this comes as a compensating effect of other properties being underestimated/overestimated.\\ 
- When there is a SFH mismatch, the direction is typically at underestimating ages and stellar masses and overestimating reddening and SFRs, both due to overshining from the youngest stellar generations and the age-dust degeneracy. We quantify these effects as functions of adopted fitting setup. \\
- Inverted-$\tau$ models with the exact same prescriptions as in M10 allow for a virtually perfect mass and SFR determination of $z\sim 2-3$ galaxies.\\
- When galaxies are older and have residual star formation, as is the case for the semi-analytic galaxies at lower redshifts, the larger age span and the age-dust degeneracy cause the mass (and the age) to be severely underestimated (by up to 0.6 dex) using standard templates composed of a variety of star formation histories. \\
- An efficient trick to reduce this effect is to switch off reddening in the fit as this helps avoiding unrealistically young and dusty solutions. Reddening and star formation rates should then be obtained through a separate fit.\\
- A similar effect is obtained by constraining the minimum age in the fitting as already done in the literature.\\
- Ages derived from the best fit are best compared to mass-weighted ages. They are generally underestimated.\\
- Metallicity effects play a secondary role.\\
- The exact type of reddening law can be correctly identified when dust reddening is present in the galaxies even if at high redshift differences in stellar mass arising from the use of various reddening laws are negligible. \\
- Photometric uncertainties alone produce an uncertainty of on average 0.08 dex in stellar mass and SFR and of 0.07 dex in age.\\
- At all redshifts, a wavelength coverage from the rest-frame UV to the rest-frame near-IR allows to obtain the most robust results.\\
- Uncertainties stemming from different stellar population models amount to $\sim0.2$ dex in stellar mass ($\sim0.4$ dex in the reddened case at z=0.5) and $\sim0.3$ dex in age ($\sim1.2$ dex in the reddened case at z=0.5).\\
 %This depends only very little on the wavelength coverage.\\
 
\noindent For mock passive galaxies our main results are: \\
- Stellar masses can be recovered very well. The dispersion from the true value amounts on average to less than 0.05 dex.\\
- Metallicity effects become important for aged passive galaxies. The use of various metallicities in the fitting appears to be the most secure option.\\
- The effect from photometric errors alone on the ages and stellar masses is $\sim0.2$ dex.\\
- The wavelength dependence of the best fit ages for old passive galaxies is smaller than in the case of star-forming galaxies, but the scatter is substantially reduced by including the rest-frame near-IR in the fit.\\

\noindent We conclude that the stellar population parameters of star-forming and passive galaxies can be reasonably well determined provided one uses the right setup and wavelength coverage. Although degeneracies between age, dust and metallicity can have a significant effect, the mismatch between template and real SFH dominates at low redshift. Because of this and of the larger spread in stellar ages the stellar population properties of galaxies at high redshift are better recovered than those at low redshift. Based on our extensive study we conclude that at $z=0.5$ and $1$ stellar population parameters are best recovered with a wide fitting setup, setting a minimum age of 0.1 Gyr in the fitting and excluding reddening from the fitting parameters. This solution is acceptable, though not optimal, and future work will be devoted in envisaging a good analytical description for this type of galaxies.\\
For $z\geq2-3$ inverted-$\tau$ models give the best results for both mass and star formation rates.\\
In all cases, a wavelength coverage from UV to near-IR rest-frame wavelengths in the fitting ensures the best results, though in some cases it is possible to find alternative filter band setups that can give reasonable results. We quantify these effects which will be useful for the planning of surveys and observational proposals.\\
Finally, due to the variety of assumptions made in the literature regarding fitting methods, stellar population models and adopted filters, caution is required when comparing SED-fitting results between different studies (\citet{Lee2009}). In order to ease these comparisons we provide scaling relations for the stellar mass that allow transformation from one set of fitting parameters and population models to another.

\section*{Acknowledgments}
The authors would like to thank the anonymous referee for a prompt and useful report. For relevant discussion and suggestions we thank E. Daddi, D. Thomas, M. Bolzonella, A. Renzini, A. Aragon-Salamanca, C. Conselice, S. Ellison. We would like to thank A. Cimatti for suggesting the exercise in section \ref{scalerel}. We would also like to thank M. Bolzonella for providing \hs \space and support with the fitting code. JP, CT and CM have been supported by the Marie Curie Excellence Team Grant "UniMass", MEXT-CT-2006-042754, P.I. Claudia Maraston, throughout the project Numerical computations were carried out on the Sciama High Performance Compute (HPC) cluster which is supported by the ICG, SEPNet and the University of Portsmouth.
% The calculations for this [paper/thesis/report] were performed on the COSMOS Consortium supercomputer within the DiRAC Facility jointly funded by STFC, the Large Facilities Capital Fund of BIS and the University of Portsmouth."
%\addcontentsline{toc}{chapter}{Literaturverzeichnis} 
\bibliographystyle{mn2e}
\bibliography{biblioJaninePhDmnras.bib}
%\begin{thebibliography}{}
%\bibitem[]{×}
%\end{thebibliography}

\appendix
\section{Overview over fitting setups}
We summarise the fitting setups explored throughout this work in Table \ref{overfits}.
\begin{table*}
\begin{minipage}{186mm}
\caption{Overview of all fits carried out with various template and filter setups for simulated galaxies at redshift 0.5, 1, 2 and 3. Fits that involve constraining or rebinning the age grid were only carried out for mock star-forming galaxies. Fits with inverted-$\tau$ models as a function of wavelength dependence were only carried out for mock star-forming galaxies at z=2.\label{overfits}}
\centering{
\begin{tabular}{@{}lccccccccccccc}
%\begin{tabular}{|l|c|c|c|c|c|c|c|c|c|c|c|c|c|}
\hline
Template Setup & \begin{sideways}UBVRI JHK IRAC\end{sideways} & \begin{sideways}UBVRI IRAC\end{sideways} & \begin{sideways}UBVRI JHK\end{sideways} & \begin{sideways}UBVRI JH\end{sideways} & \begin{sideways}UBVRI J\end{sideways} & \begin{sideways}UBVRI\end{sideways} & \begin{sideways}BVRI JHK IRAC\end{sideways} & \begin{sideways}VRI JHK IRAC\end{sideways} & \begin{sideways}RI JHK IRAC\end{sideways} & \begin{sideways}u'g'r'i'z'\end{sideways} & \begin{sideways}UBVRI Y JHK IRAC    \end{sideways}& \begin{sideways}BRIK    \end{sideways}\\\hline\hline
wide setup                             & x & x & x & x & x & x & x & x & x & x & x & x\\\hline
wide, age rebin                      & x & - & x & - & - & - & - & - & - & - & - & - \\\hline
wide, age $\geq$ 0.1 Gyr      & x & - & - & - & - & - & - & - & - & - & - & - \\\hline
wide, $Z=0.004$		           & x & - & - & - & - & - & - & - & - & - & - & - \\\hline
wide, $Z=0.01$ 		           & x & - & - & - & - & - & - & - & - & - & - & - \\\hline
wide, $Z=0.02$ 		           & x & - & - & - & - & - & - & - & - & - & - & - \\\hline
wide, $Z=0.04$ 			    & x & - & - & - & - & - & - & - & - & - & - & - \\\hline
wide, Kroupa			           & x & - & - & - & - & - & - & - & - & - & - & - \\\hline
wide, Chabrier			    & x & - & - & - & - & - & - & - & - & - & - & - \\\hline
wide, top-heavy	                  & x & - & - & - & - & - & - & - & - & - & - & - \\\hline
only-$\tau$                            & x & - & x & - & - & - & - & - & - & - & - & - \\\hline
inverted-$\tau$                      & x & x & x & x & x & x & x & x & x & x & x & x \\\hline
only SSPs                              & x & - & - & - & - & - & - & - & - & - & - & - \\\hline
solar SSP                              & x & - & - & - & - & - & - & - & - & - & - & - \\\hline
BC03                                     & x & - & x & - & - & x & - & - & - & - & - & - \\\hline
\end{tabular}}
\end{minipage}
\end{table*}%

\section{Overview over fitting results}\label{overres}
We summarise median offsets and 68\% confidence ranges for age, E(B-V), stellar mass and SFR in Tables \ref{sfoverres} - \ref{poverres2} for a selected number of fitting setups.
\begin{table*}%[h!tbp]
\begin{minipage}{174mm}
\caption{Median offsets and ranges including 68\% of the solutions for $\Delta$ log age, $\Delta$ E(B-V), $\Delta$ log M* and $\Delta$ log SFR of mock star-forming galaxies at $z=0.5$ derived with various fitting setups. SFRs of zero are set to 0.001 in order to compute the logarithm often causing large negative values for $\Delta$ log SFR.}
\centering{%\small{%\begin{sideways}
\begin{tabular}{@{}lllllllll}
\hline
setup & $\Delta$ log age & & $\Delta$ E(B-V)&  & $\Delta$ log M* &  & $\Delta$ log SFR & \\\hline
z=0.5, no reddening &median &range&median&range&median&range&median&range\\\hline
wide Salpeter 			& -0.35   &    -0.91   to     0.07   &     0.00    &    0.00   to     0.00   &    -0.31   &    -0.54  to     -0.04   &    -0.02   &    -2.31   to     0.10\\\hline
only-$\tau$            		& -0.55   &    -0.93   to     0.03   &     0.00    &    0.00   to     0.00   &    -0.35   &    -0.70  to     -0.04   &     0.00   &    -0.62   to     0.13\\\hline
wide Kroupa           		& -0.39   &    -0.97   to     0.05   &     0.00    &    0.00   to     0.00   &    -0.46   &    -0.77  to     -0.18   &    -0.20   &    -2.31   to    -0.07\\\hline
wide Chabier          		& -0.38   &    -0.89   to     0.00   &     0.00    &    0.00   to     0.00   &    -0.45   &    -0.71  to     -0.19   &    -0.39   &    -2.31   to    -0.11\\\hline
wide top-heavy              	& -0.22   &    -0.84   to     0.34   &     0.00    &    0.00   to     0.00   &     0.90   &     1.63  to      0.14    &   -0.68    &   -3.05    to   -0.51\\\hline
Z=0.004          			& -0.34   &    -0.80   to     0.16   &     0.00    &    0.00   to     0.00   &    -0.32   &    -0.57  to     -0.05   &    -0.14   &    -2.69   to     0.06\\\hline
Z=0.04            			& -0.91   &    -1.06   to    -0.62   &     0.00    &    0.00   to     0.00   &    -0.56   &    -0.82  to     -0.37   &    -0.15   &    -2.95   to     0.03\\\hline
UBVRI+IRAC           		& -0.35   &    -0.92   to     0.06   &     0.00    &    0.00   to     0.00   &    -0.28   &    -0.57  to     -0.04   &    -0.01   &    -2.31   to     0.10\\\hline
UBVRIJHK             		& -0.26   &    -0.80   to     0.10   &     0.00    &    0.00   to     0.00   &    -0.21   &    -0.53  to      0.03   &    -0.04   &    -1.51   to     0.10\\\hline
ugriz           				& -0.31   &    -0.74   to     0.03   &     0.00    &    0.00   to     0.00   &    -0.23   &    -0.47  to     -0.03   &     0.02   &    -0.34   to     0.15\\\hline
RIJHK+IRAC            		& -0.23   &    -1.00   to     0.19   &     0.00    &    0.00   to     0.00   &    -0.20   &    -0.62  to      0.07   &    -0.73   &    -3.29   to     0.30\\\hline
BRIK            				& -0.20   &    -0.87   to     0.19   &     0.00    &    0.00   to     0.00   &    -0.19   &    -0.56  to      0.02   &    -0.03   &    -2.10   to     0.12\\\hline
wide, age $\geq0.1$ Gyr   & -0.33   &    -0.89   to     0.07   &     0.00    &    0.00   to     0.00   &    -0.30   &    -0.54  to     -0.04   &    -0.01   &    -2.24   to     0.10\\\hline
wide BC03     			& -0.18   &    -0.50  to      0.01   &     0.00   &     0.00  to      0.00   &    -0.09   &    -0.29  to      0.07   &    -0.07   &    -0.28  to       0.04\\\hline
z=0.5, with reddening &&&&&&&&\\\hline
wide Salpeter   	        	& -1.78   &    -2.83   to    -0.37   &     0.29    &        0.00  to  0.59   &    -0.63   &    -1.38   to    -0.25   &    -2.44   &    -3.39   to     0.42\\\hline
only-$\tau$             		& -1.27   &    -2.52   to    -0.19   &     0.29    &        0.00  to  0.54   &    -0.47   &    -0.71   to    -0.17   &     0.37   &    -0.17    to    2.08\\\hline
wide Kroupa             		& -2.05   &    -2.81   to    -0.71   &     0.29    &        0.00  to  0.59   &    -0.88   &    -1.59   to    -0.45   &    -2.74   &    -3.48    to    0.15\\\hline
wide Chabrier          		& -2.07   &    -2.76   to    -0.80   &     0.30    &        0.05  to  0.54   &    -0.96   &    -1.59   to    -0.49   &    -2.74   &    -3.39    to    0.12\\\hline
wide top-heavy               	& -2.10   &    -2.89   to    -0.35   &     0.26    &        0.10  to  0.64   &    -1.06   &    -1.88   to     0.41   &    -3.02   &    -3.52    to   -0.28\\\hline
Z=0.004            			& -1.55   &    -2.80   to    -0.16   &     0.35    &        0.05  to  0.59   &    -0.60   &    -1.24   to    -0.19   &    -2.27   &    -3.45    to    0.61\\\hline
Z=0.04             			& -2.49   &    -2.93   to    -1.02   &     0.40    &        0.10  to  0.64   &    -0.68   &    -1.19   to    -0.40   &    -2.70   &    -3.41    to    0.39\\\hline
UBVRI+IRAC            		& -2.14   &    -2.86   to    -0.60   &     0.32    &        0.05  to  0.64   &    -0.70   &    -1.41   to    -0.30   &    -2.59   &    -3.39    to    0.42\\\hline
UBVRIJHK              		& -2.15   &    -2.78   to    -0.55   &     0.29    &        0.00  to  0.54   &    -0.69   &    -1.46   to    -0.29   &    -2.68   &    -3.41    to    0.19\\\hline
ugriz             				& -1.95   &    -2.76   to    -0.28   &     0.34    &        0.06  to  0.66   &    -0.51   &    -1.44   to    -0.13   &    -1.56   &    -3.41    to    0.59\\\hline
RIJHK+IRAC             		& -2.21   &    -2.83   to    -0.30   &     0.29    &        0.06  to  0.74   &    -0.69   &    -1.42   to    -0.19   &    -1.69   &    -3.41    to    2.21\\\hline
BRIK             				& -2.15   &    -2.89   to    -1.05   &     0.39    &        0.07  to  0.84   &    -0.62   &    -1.11   to    -0.37   &    -2.74   &    -3.41    to    1.38\\\hline
wide, age $\geq0.1$ Gyr   & -0.87   &    -1.36   to    -0.03   &     0.14    &        0.00  to  0.35   &    -0.38   &    -0.59   to    -0.08   &     0.12   &    -2.38    to    0.71\\\hline
wide BC03      			& -0.59   &     -2.71  to     0.06   &      0.12   &        0.00  to  0.54   &    -0.24   &     -0.93  to      0.02   &    0.00   &    -2.87  to    0.45\\\hline
\end{tabular}
%\end{sideways}
\label{sfoverres}}%}
\end{minipage}
\end{table*}%

\begin{table*}%[h!tbp]
\begin{minipage}{174mm}
\caption{Same as Table \ref{sfoverres} for mock star-forming galaxies at $z=2.0$.}
\centering{%\small{%\begin{sideways}
\begin{tabular}{@{}lllllllll}
\hline
setup & $\Delta$ log age & & $\Delta$ E(B-V)&  & $\Delta$ log M* &  & $\Delta$ log SFR & \\\hline
z=2, no reddening &median &range&median&range&median&range&median&range\\\hline
wide Salpeter    			&       -0.08   &       -0.72  to        0.15   &        0.00   &       0.00  to        0.00   &       -0.10   &       -0.29  to        0.00   &        0.01   &       -1.52  to        0.10\\\hline
only-$\tau$      			&       -0.06   &       -0.44  to        0.15   &        0.00   &       0.00  to        0.00   &       -0.05   &       -0.43  to        0.01   &        0.09   &        0.00  to        0.19\\\hline
inverted-$\tau$			&       -0.47   &       -0.56  to       -0.31   &        0.00   &       0.00  to        0.00   &        0.02   &       -0.06  to        0.09   &        0.10   &        -0.03  to       0.18\\\hline
wide Kroupa      			&       -0.19   &       -0.96  to        0.06   &        0.00   &       0.00  to        0.00   &       -0.37   &       -0.63  to       -0.20   &       -0.20   &       -3.33  to       -0.08\\\hline
wide Chabrier      			&       -0.13   &       -0.96  to        0.04   &        0.00   &       0.00  to        0.00   &       -0.29   &       -0.66  to       -0.19   &       -0.25   &       -3.26  to       -0.12\\\hline
wide top-heavy         		&       -0.17   &       -1.14  to        0.24   &        0.00   &       0.00  to        0.00   &        0.39   &       -0.68  to        0.73   &       -3.74   &       -4.17  to       -0.69\\\hline
Z=0.004      				&       -0.03   &       -0.70  to        0.15   &        0.00   &       0.00  to        0.00   &       -0.11   &       -0.24  to       -0.04   &       -0.02   &       -1.52  to        0.08\\\hline
Z=0.04       				&       -0.39   &       -1.29  to       -0.19   &        0.00   &       0.00  to        0.00   &       -0.16   &       -0.79  to       -0.06   &        0.04   &       -3.33  to        0.15\\\hline
UBVRI+IRAC      			&       -0.14   &       -0.96  to        0.15   &        0.00   &       0.00  to        0.00   &       -0.12   &       -0.62  to       -0.01   &        0.03   &       -3.26  to        0.14\\\hline
UBVRIJHK        			&       -0.16   &       -1.03  to        0.15   &        0.00   &       0.00  to        0.00   &       -0.11   &       -0.62  to        0.01   &        0.00   &       -3.02  to        0.09\\\hline
ugriz       				&       -0.21   &       -1.53  to        0.44   &        0.00   &       0.00  to        0.00   &       -0.18   &       -1.00  to        0.43   &        0.00   &       -3.88  to        0.13\\\hline
RIJHK+IRAC       			&       -0.11   &       -0.88  to        0.12   &        0.00   &       0.00  to        0.00   &       -0.12   &       -0.35  to       -0.01   &       -0.02   &       -3.33  to        0.10\\\hline
BRIK       				&       -0.09   &       -0.83  to        0.23   &        0.00   &       0.00  to        0.00   &       -0.10   &       -0.49  to        0.04   &        0.05   &       -3.02  to        0.15\\\hline
wide, age $\geq0.1$ Gyr 	&        0.00   &       -0.22  to        0.24   &        0.00   &       0.00  to        0.00   &       -0.10   &       -0.21  to        0.00   &        0.04   &       -0.05  to        0.10\\\hline
wide BC03			       &        0.20   &       -0.19  to        0.44   &        0.00   &       0.00  to        0.00   &        0.12   &       -0.18  to        0.24   &        0.01   &       -0.08  to        0.08\\\hline
z=2, with reddening &&&&&&&&\\\hline
wide Salpeter      			&       -0.15   &       -0.89  to        0.14   &        0.00   &       -0.01  to        0.06   &       -0.12   &       -0.51  to       -0.03   &        0.06   &       -0.35  to        0.44\\\hline
only-$\tau$       			&       -0.07   &       -0.96  to        0.14   &        0.01   &       -0.01  to        0.06   &       -0.06   &       -0.51  to        0.01   &        0.12   &        -0.09  to       0.61\\\hline
inverted-$\tau$ 			&       -0.47   &       -0.56  to       -0.31   &        0.01   &        0.00  to        0.10   &        0.01   &        0.06  to       -0.05   &        0.17   &        -0.04  to       0.40\\\hline 
wide Kroupa       			&       -0.22   &       -1.07  to        0.11   &        0.02   &       -0.02  to        0.07   &       -0.38   &       -0.66  to       -0.21   &       -0.09   &       -3.49  to        0.35\\\hline
wide Chabrier     			&       -0.20   &       -1.04  to        0.02   &        0.03   &       -0.02  to        0.06   &       -0.33   &       -0.74  to       -0.22   &       -0.17   &       -3.33  to        0.25\\\hline
wide top-heavy         		&       -0.08   &       -1.54  to        0.33   &        0.09   &        0.00  to        0.19   &        0.39   &       -1.05  to        0.67   &       -0.73   &       -4.12  to       -0.08\\\hline
Z=0.004       				&       -0.09   &       -0.82  to        0.16   &        0.02   &       -0.01  to        0.10   &       -0.12   &       -0.47  to       -0.04   &        0.09   &       -0.24  to        0.60\\\hline
Z=0.04        				&       -0.39   &       -1.28  to       -0.18   &        0.04   &       -0.03  to        0.13   &       -0.16   &       -0.64  to       -0.07   &        0.17   &       -0.21  to        0.83\\\hline
UBVRI+IRAC       			&       -0.29   &       -1.22  to        0.12   &        0.04   &       -0.02  to        0.10   &       -0.19   &       -0.73  to       -0.05   &        0.11   &       -3.26  to        0.55\\\hline
UBVRIJHK         			&       -0.22   &       -1.55  to        0.16   &        0.03   &        0.00  to        0.06   &       -0.15   &       -0.99  to        0.00   &        0.00   &       -3.76  to        0.22\\\hline
ugriz        				&       -0.92   &       -2.04  to        0.22   &        0.02   &       -0.03  to        0.10   &       -0.61   &       -1.27  to        0.23   &       -0.22   &       -3.98  to        0.39\\\hline
RIJHK+IRAC        		&       -0.26   &       -1.07  to        0.14   &        0.00   &       -0.04  to        0.13   &       -0.14   &       -0.46  to       -0.03   &       -0.05   &       -3.65  to        0.56\\\hline
BRIK        				&       -0.13   &       -1.26  to        0.28   &        0.00   &       -0.02  to        0.06   &       -0.10   &       -0.72  to        0.08   &        0.09   &       -2.69  to        0.49\\\hline
wide, age $\geq0.1$ Gyr   &       -0.03   &       -0.31  to        0.19   &        0.00   &       -0.01  to        0.05   &       -0.11   &       -0.22  to       -0.03   &        0.06   &       -0.10  to        0.21\\\hline
wide BC03			       &        0.18   &       -0.77  to        0.42   &        0.00   &       -0.03  to        0.06   &        0.12   &       -0.43  to        0.25   &        0.03   &       -0.20  to        0.40\\\hline
\end{tabular}
%\end{sideways}
\label{sfoverres2}}%}
\end{minipage}
\end{table*}%

\begin{table*}%[h!tbp]
\begin{minipage}{174mm}
\caption{Same as Table \ref{sfoverres} for mock passive galaxies at $z=0.5$.}
\centering{%\small{%\begin{sideways}
\begin{tabular}{@{}lllllllll}
\hline
setup & $\Delta$ log age & & $\Delta$ E(B-V)&  & $\Delta$ log M* &  & $\Delta$ log SFR & \\\hline
z=0.5, no reddening &median &range&median&range&median&range&median&range\\\hline
wide Salpeter     	&        0.03   &       -0.02  to        0.15   &        0.00   &        0.00  to        0.00   &        0.01   &       -0.04  to        0.05   &        0.00   &        0.00  to        0.00\\\hline
only-$\tau$       	&        0.03   &        0.00  to        0.10   &        0.00   &        0.00  to        0.00   &        0.01   &       -0.04  to        0.05   &        0.00   &        0.00  to        0.00\\\hline
only SSPs        	&        0.01   &       -0.03  to        0.06   &        0.00   &        0.00  to        0.00   &        0.01   &       -0.05  to        0.04   &        0.00   &        0.00  to        0.00\\\hline
only solar SSP     &        0.01   &       -0.03  to        0.03   &        0.00   &        0.00  to        0.00   &        0.00   &       -0.05  to       -0.03   &        0.00   &        0.00  to        0.00\\\hline
wide Kroupa       	&        0.05   &        0.00  to        0.15   &        0.00   &        0.00  to        0.00   &       -0.10   &       -0.16  to       -0.06   &        0.00   &        0.00  to        0.00\\\hline
wide Chabrier      &        0.05   &        0.00  to        0.15   &        0.00   &        0.00  to        0.00   &       -0.12   &       -0.18  to       -0.06   &        0.00   &        0.00  to        0.00\\\hline
wide top-heavy    &        0.08   &        0.00  to        0.22   &        0.00   &        0.00  to        0.00   &        1.14   &        1.02  to        1.26   &        0.00   &        0.00  to        0.00\\\hline
Z=0.004      		&        0.39   &        0.15  to        0.75   &        0.00   &        0.00  to        0.00   &        0.03   &       -0.32  to        0.50   &        0.00   &        0.00  to        0.00\\\hline
Z=0.04       		&       -0.30   &       -0.39  to       -0.24   &       0.00   &        0.00  to        0.00   &       -0.27   &       -0.44  to       -0.18   &        0.00   &        0.00  to        0.00\\\hline
UBVRI+IRAC      	&        0.05   &        0.00  to        0.23   &        0.00   &        0.00  to        0.00   &        0.03   &       -0.03  to        0.08   &        0.00   &        0.00  to        0.00\\\hline
UBVRIJHK        	&        0.04   &       -0.02  to        0.19   &        0.00   &        0.00  to        0.00   &        0.02   &       -0.03  to        0.10   &        0.00   &        0.00  to        0.00\\\hline
ugriz       		&        0.02   &       -0.18  to        0.31   &        0.00   &        0.00  to        0.00   &       -0.01   &       -0.12  to        0.23   &        0.00   &        0.00  to        0.00\\\hline
RIJHK+IRAC       	&        0.00   &       -0.18  to        0.26   &        0.00   &        0.00  to        0.00   &       -0.04   &       -0.24  to        0.09   &        0.00   &        0.00  to        3.00\\\hline
BRIK       		&        0.06   &       -0.12  to        0.28   &        0.00   &        0.00  to        0.00   &        0.02   &       -0.18  to        0.13   &        0.00   &        0.00  to        2.30\\\hline
wide BC03      	&        0.05   &       -0.03  to        0.44   &        0.00   &        0.00  to        0.00   &        0.07   &        0.00  to        0.32   &        0.00   &        0.00  to        2.30\\\hline
z=0.5, with reddening &&&&&&&&\\\hline
wide Salpeter     	&        0.02   &       -0.09  to        0.12   &        0.00   &        0.00  to        0.07   &        0.01   &       -0.07  to        0.08   &        0.00   &        0.00  to        0.00\\\hline
only-$\tau$        	&        0.00   &       -0.09  to        0.07   &        0.00   &        0.00  to        0.06   &        0.00   &       -0.08  to        0.05   &        0.00   &        0.00  to        0.00\\\hline
only SSPs         	&        0.01   &       -0.09  to        0.10   &        0.00   &        0.00  to        0.07   &        0.02   &       -0.05  to        0.09   &        0.00   &        0.00  to        0.00\\\hline
only solar SSP     &        0.00   &       -0.09  to        0.03   &        0.00   &        0.00  to        0.06   &        0.00   &       -0.07  to        0.03   &        0.00   &        0.00  to        0.00\\\hline
wide Kroupa        &        0.04   &       -0.08  to        0.20   &        0.00   &        0.00  to        0.07   &       -0.11   &       -0.18  to        0.00   &        0.00   &        0.00  to        0.00\\\hline
wide Chabrier      &        0.04   &       -0.15  to        0.18   &        0.00   &        0.00  to        0.07   &       -0.12   &       -0.22  to       -0.03   &        0.00   &        0.00  to        0.00\\\hline
wide top-heavy    &        0.08   &       -0.12  to        0.30   &        0.00   &        0.00  to        0.13   &        1.20   &        1.02  to        1.46   &        0.00   &        0.00  to        0.00\\\hline
Z=0.004       		&        0.31   &        0.08  to        0.56   &        0.19   &        0.13  to        0.26   &        0.31   &        0.17  to        0.47   &        0.00   &        0.00  to        0.00\\\hline
Z=0.04        		&       -0.34   &       -0.55  to       -0.26  &        0.00   &        0.00  to        0.00   &       -0.27   &       -0.44  to       -0.18   &        0.00   &        0.00  to        0.00\\\hline
UBVRI+IRAC       	&        0.04   &       -0.12  to        0.20   &        0.00   &        0.00  to        0.07   &        0.03   &       -0.06  to        0.17   &        0.00   &        0.00  to        0.00\\\hline
UBVRIJHK         	&        0.03   &       -0.07  to        0.23   &        0.00   &        0.00  to        0.07   &        0.02   &       -0.06  to        0.15   &        0.00   &        0.00  to        0.00\\\hline
ugriz        		&       -0.04   &       -1.50  to        0.10   &        0.00   &        0.00  to        0.84   &       -0.01   &       -0.14  to        0.14   &        0.00   &        0.00  to        0.00\\\hline
RIJHK+IRAC        &       -0.13   &       -2.36  to        0.11   &        0.07   &        0.00  to        0.71   &       -0.14   &       -1.16  to        0.07   &        0.00   &        0.00  to        2.60\\\hline
BRIK        		&       -0.07   &       -1.39  to        0.18   &        0.07   &        0.00  to        0.59   &       -0.04   &       -0.46  to        0.12   &        0.00   &        0.00  to        0.00\\\hline
wide BC03      	&        0.00   &       -0.08  to        0.21   &        0.00   &        0.00  to        0.13   &        0.07   &       -0.01  to        0.22   &        0.00   &        0.00  to        0.00\\\hline
\end{tabular}
%\end{sideways}
\label{poverres}}%}
\end{minipage}
\end{table*}%

\begin{table*}%[h!tbp]
\begin{minipage}{174mm}
\caption{Same as Table \ref{poverres} for mock passive galaxies at $z=2.0$.}
\centering{%\small{%\begin{sideways}
\begin{tabular}{@{}lllllllll}
\hline
setup & $\Delta$ log age & & $\Delta$ E(B-V)&  & $\Delta$ log M* &  & $\Delta$ log SFR & \\\hline
z=2, no reddening &median &range&median&range&median&range&median&range\\\hline
wide Salpeter     	&        0.01   &        0.00  to        0.19   &        0.00   &        0.00  to        0.00   &       -0.01   &       -0.02  to        0.00   &        0.00   &        0.00  to        0.00\\\hline
only-$\tau$       	&        0.01   &        0.00  to        0.03   &        0.00   &        0.00  to        0.00   &       -0.02   &       -0.05  to        0.00   &        0.00   &        0.00  to        0.00\\\hline
only SSPs       	&        0.01   &        0.00  to        0.17   &        0.00   &        0.00  to        0.00   &        0.00   &       -0.03  to        0.03   &        0.00   &        0.00  to        0.00\\\hline
only solar SSP     &        0.00   &       -0.02  to        0.01   &        0.00   &        0.00  to        0.00   &       -0.02   &       -0.07  to        0.00   &        0.00   &        0.00  to        0.00\\\hline
wide Kroupa       	&        0.01   &        0.00  to        0.17   &        0.00   &        0.00  to        0.00   &       -0.14   &       -0.15  to       -0.12   &        0.00   &        0.00  to        0.00\\\hline
wide Chabrier      &        0.01   &        0.00  to        0.17   &        0.00   &        0.00  to        0.00   &       -0.17   &       -0.18  to       -0.14   &        0.00   &        0.00  to        0.00\\\hline
wide top-heavy    &        0.03   &        0.01  to        0.17   &        0.00   &        0.00  to        0.00   &        1.00   &        0.87  to        1.07   &        0.00   &        0.00  to        0.00\\\hline
Z=0.004      		&        0.30   &        0.18  to        0.48   &        0.00   &        0.00  to        0.00   &       -1.51   &       -2.19  to       -0.48   &        0.00   &        0.00  to        0.00\\\hline
Z=0.04       		&       -0.27   &       -0.29  to       -0.24   &       0.00   &        0.00  to        0.00   &       -0.25   &       -0.31  to       -0.19   &        0.00   &        0.00  to        0.00\\\hline
UBVRI+IRAC      	&        0.01   &        0.00  to        0.15   &        0.00   &        0.00  to        0.00   &       -0.01   &       -0.05  to        0.04   &        0.00   &        0.00  to        0.00\\\hline
UBVRIJHK        	&        0.01   &        0.00  to        0.17   &        0.00   &        0.00  to        0.00   &        0.00   &       -0.04  to        0.03   &        0.00   &        0.00  to        0.00\\\hline
ugriz       		&        0.01   &       -0.09  to        0.17   &        0.00   &        0.00  to        0.00   &        0.01   &       -0.06  to        0.07   &        0.00   &        0.00  to        0.00\\\hline
RIJHK+IRAC      	&        0.14   &        0.00  to        0.30   &        0.00   &        0.00  to        0.00   &        0.05   &        -0.07  to        0.16   &        0.00   &        0.00  to        2.60\\\hline
BRIK       		&        0.00   &       -0.27  to        0.01   &        0.00   &        0.00  to        0.00   &       -0.04   &       -0.15  to        0.01   &        0.00   &        0.00  to        0.00\\\hline
wide BC03      	&       -0.42   &       -0.54  to       -0.24   &        0.00   &        0.00  to        0.00   &       -3.83   &       -4.49  to       -2.89   &        0.00   &        0.00  to        0.00\\\hline
z=2, with reddening &&&&&&&&\\\hline
wide Salpeter      	&        0.01   &        0.00  to        0.19   &        0.00   &        0.00  to        0.00   &       -0.01   &       -0.02  to        0.02   &        0.00   &        0.00  to        0.00\\\hline
only-$\tau$        	&        0.01   &       -0.02  to        0.03   &        0.00   &        0.00  to        0.05   &       -0.01   &       -0.02  to        0.01   &        0.00   &        0.00  to        0.00\\\hline
only SSPs         	&        0.01   &        0.00  to        0.17   &        0.00   &        0.00  to        0.00   &        0.00   &       -0.02  to        0.03   &        0.00   &        0.00  to        0.00\\\hline
only solar SSP     &        0.00   &       -0.02  to        0.01   &        0.00   &        0.00  to        0.00   &       -0.02   &       -0.07  to        0.00   &        0.00   &        0.00  to        0.00\\\hline
wide Kroupa        &        0.01   &        0.00  to        0.05   &        0.00   &        0.00  to        0.00   &       -0.14   &       -0.15  to       -0.11   &        0.00   &        0.00  to        0.00\\\hline
wide Chabrier      &        0.01   &        0.00  to        0.17   &        0.00   &        0.00  to        0.00   &       -0.17   &       -0.18  to       -0.14   &        0.00   &        0.00  to        0.00\\\hline
wide top-heavy    &        0.01   &        0.00  to        0.17   &        0.00   &        0.00  to        0.00   &        1.00   &        0.87  to        1.06   &        0.00   &        0.00  to        0.00\\\hline
Z=0.004       		&        0.30   &        0.18  to        0.48   &        0.37   &        0.22  to        0.44   &        0.42   &        0.32  to        0.53   &        0.00   &        0.00  to        0.00\\\hline
Z=0.04        		&       -0.27   &       -0.29  to       -0.24  &        0.00   &        0.00  to        0.00   &       -0.25   &      -0.30  to       -0.19   &        0.00   &        0.00  to        0.00\\\hline
UBVRI+IRAC       &        0.01   &        0.00  to        0.15   &        0.00   &        0.00  to        0.00   &       -0.01   &       -0.05  to        0.04   &        0.00   &        0.00  to        0.00\\\hline
UBVRIJHK         	&        0.01   &       -0.02  to        0.15   &        0.00   &        0.00  to        0.05   &        0.02   &       -0.01  to        0.07   &        0.00   &        0.00  to        0.00\\\hline
ugriz        		&        0.00   &       -0.24  to        0.12   &        0.00   &        0.00  to        0.06   &        0.02   &       -0.06  to        0.12   &        0.00   &        0.00  to        0.00\\\hline
RIJHK+IRAC        &        0.18   &       -0.04  to        0.30   &        0.00   &        0.00  to        0.26   &        0.07   &       -0.07  to        0.21   &        0.00   &        0.00  to        2.30\\\hline
BRIK        		&        0.00   &       -0.27  to        0.11   &        0.00   &        0.00  to        0.05   &       -0.03   &       -0.14  to        0.02   &        0.00   &        0.00  to        0.00\\\hline
wide BC03      	&       -0.42   &       -0.49  to       -0.19   &        0.59   &        0.44  to        0.66   &        0.13   &        0.09  to        0.29   &        0.00   &        0.00  to        0.00\\\hline
\end{tabular}
%\end{sideways}
\label{poverres2}}%}
\end{minipage}
\end{table*}%

\section{Fitting BC03 templates to M05 galaxies}\label{bc03results}
\begin{figure}
\includegraphics[width=84mm]{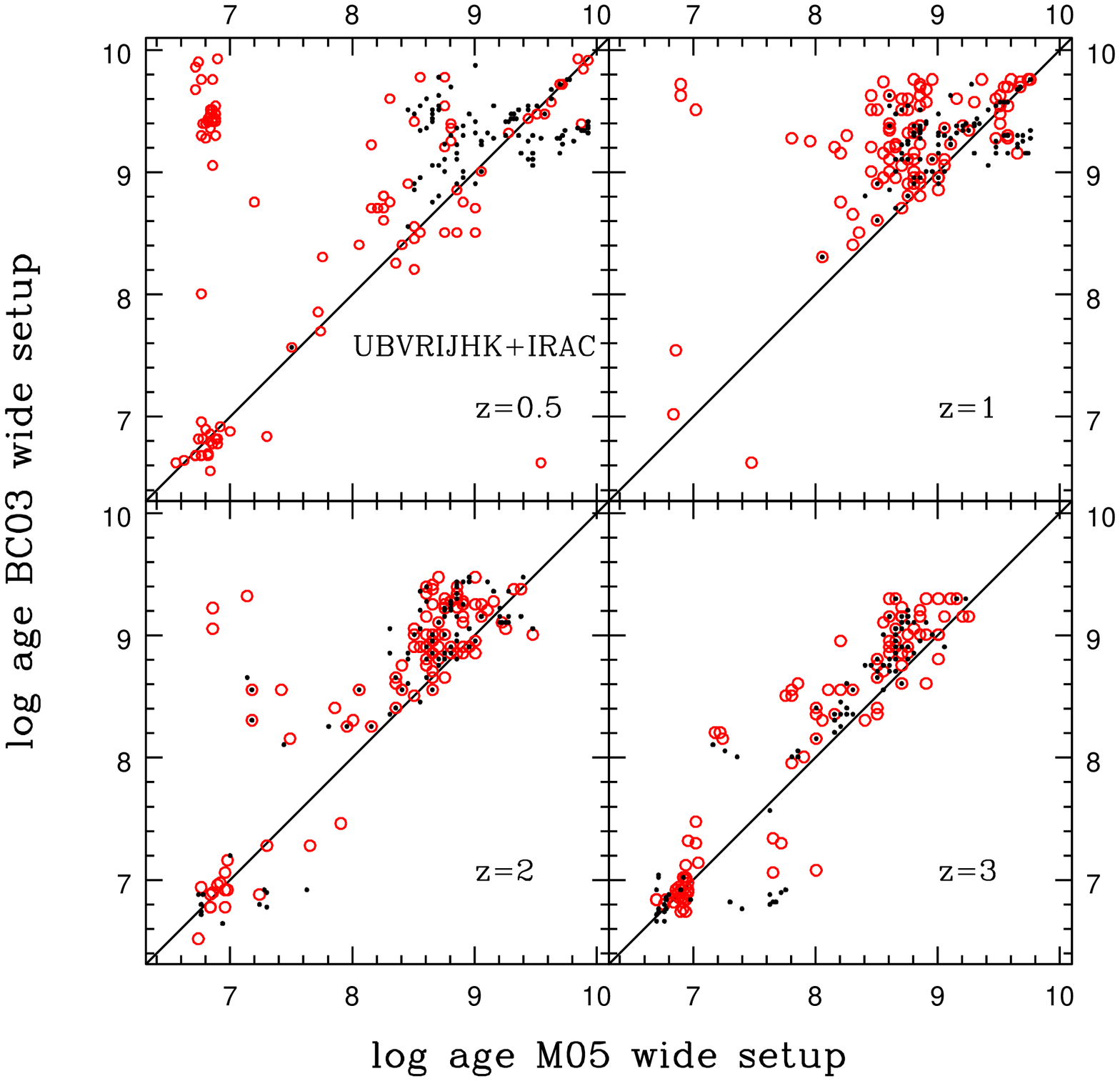}                                                                   
\caption{\label{Bage} Ages derived from SED-fitting with different stellar population synthesis models (M05 and BC03) for mock star-forming galaxies based on M05 photometry. A wide setup and wavelength coverage was used. Black refers to the unreddened case, red to the case including reddening.}
\end{figure}

\begin{figure}
\includegraphics[width=84mm]{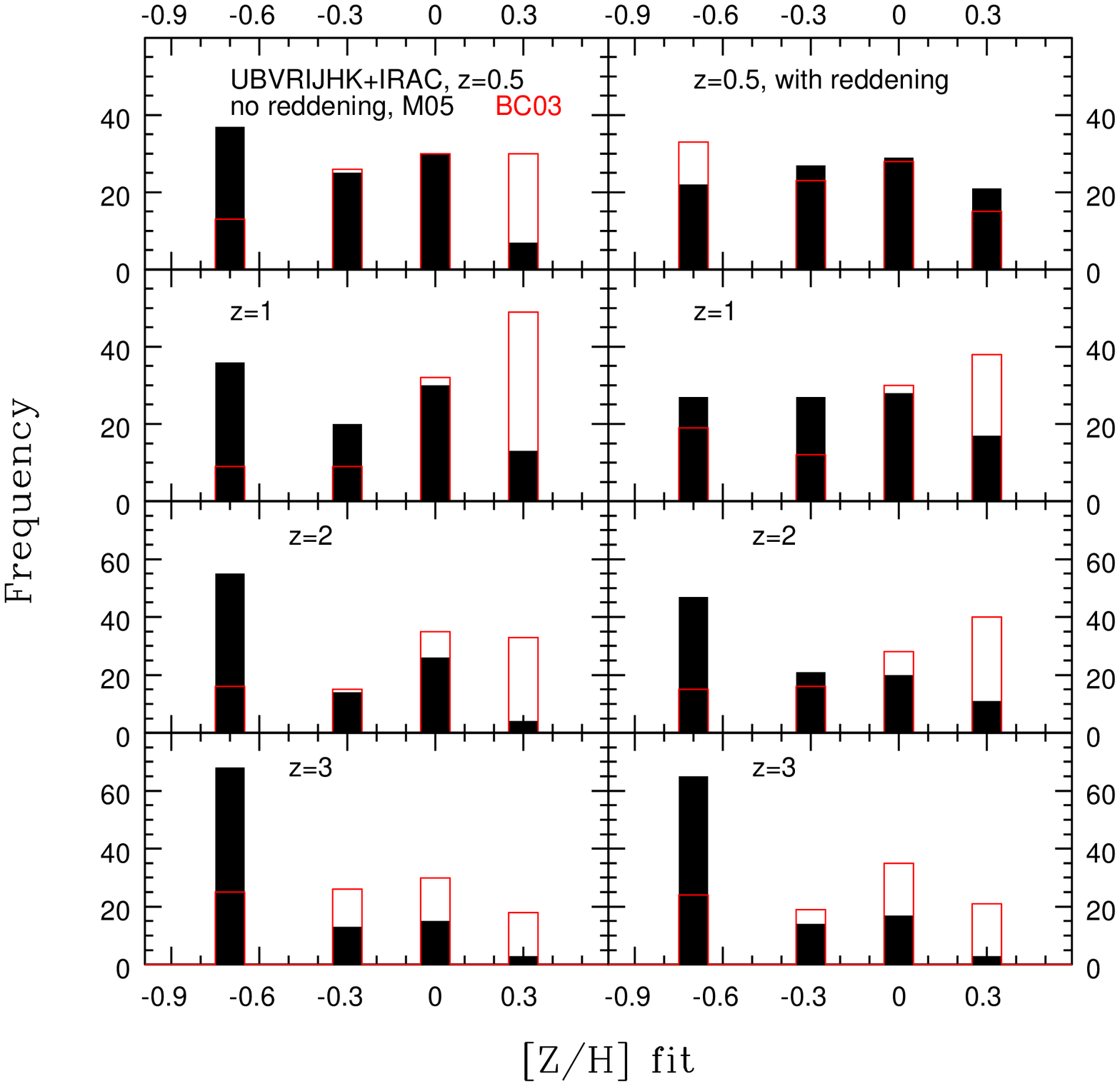}                  
\caption{\label{Bmetal} Same as Fig. \ref{Bage} for the derived metallicities. Red histograms stand for BC03-based results, black shaded histograms for M05-based results.}
\end{figure}

\begin{figure}
\includegraphics[width=84mm]{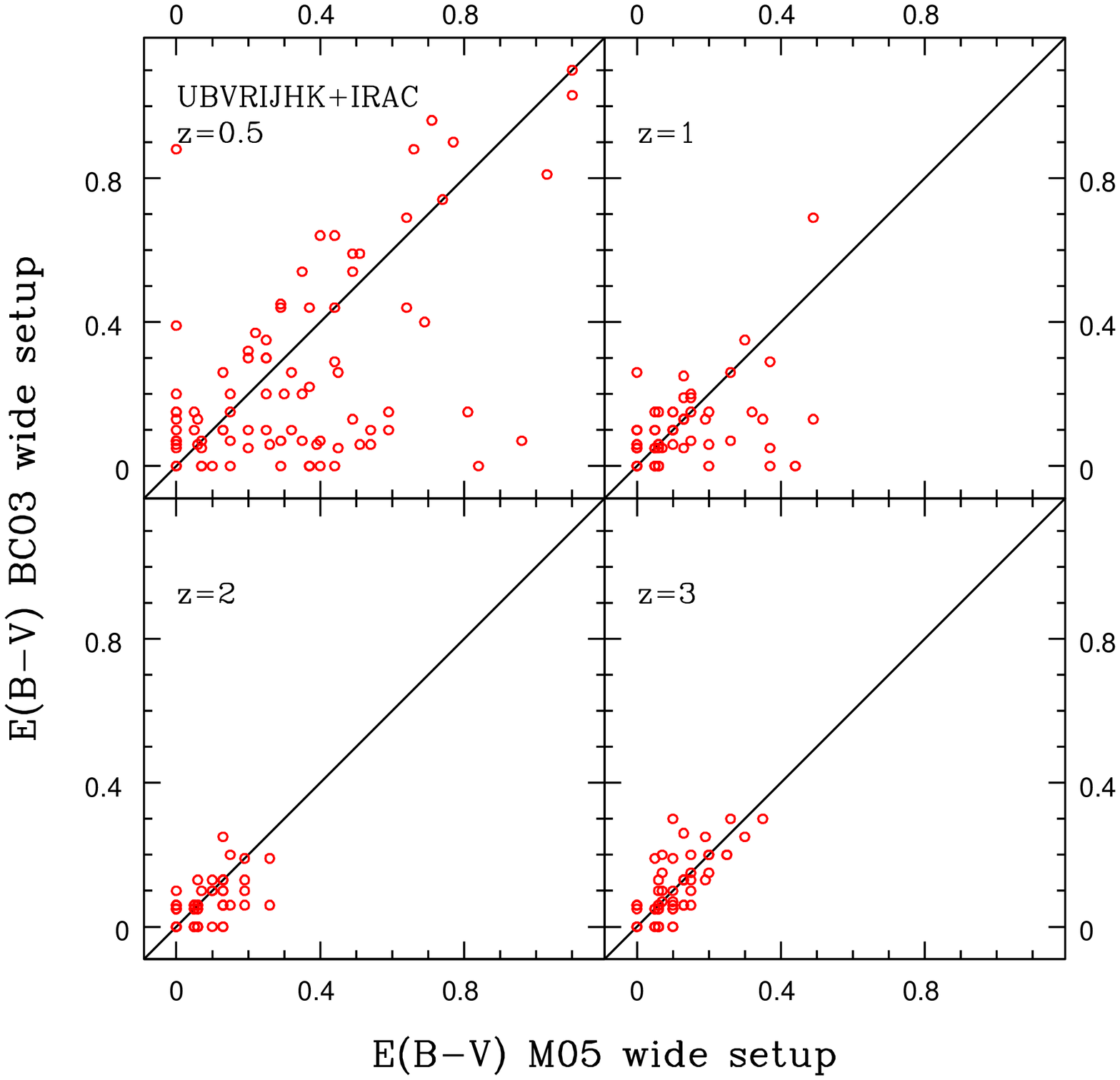}                                                                                                                                  
\caption{\label{Bebv} Same as Fig. \ref{Bage} for the comparison of derived dust attenuation.}
\end{figure}

\begin{figure}
\includegraphics[width=84mm]{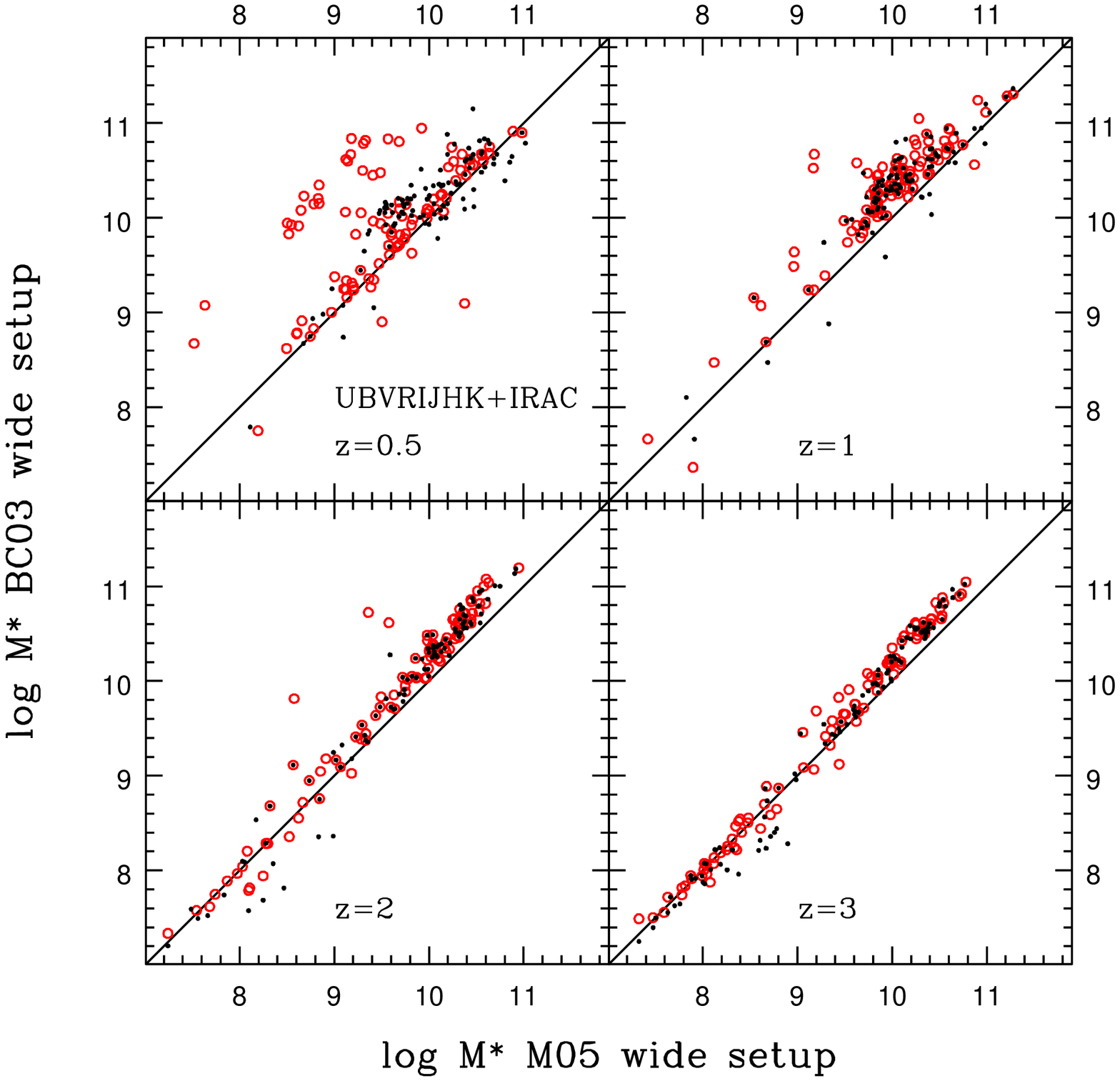}                                                                                                                              
\caption{\label{Bmass} Same as Fig. \ref{Bage} for the comparison of derived stellar masses. Black refers to the unreddened case, red to the case including reddening.}
\end{figure}

\begin{figure}
\includegraphics[width=84mm]{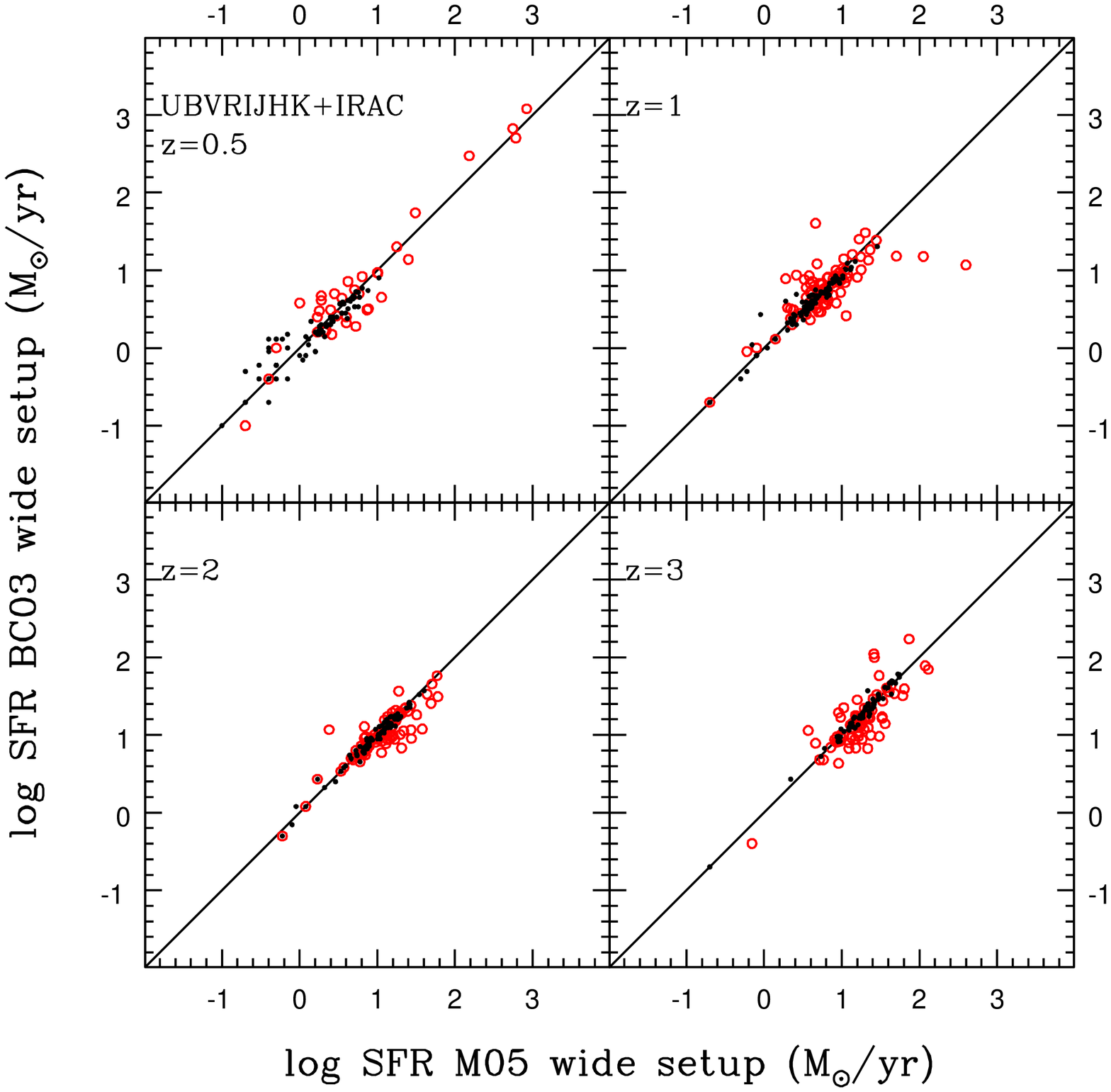}                                                                                      
\caption{\label{Bsfr} Same as Fig. \ref{Bage} for the comparison of derived star formation rates. Black refers to the unreddened case, red to the case including reddening.}
\end{figure}
It is interesting to see the effect of fitting galaxies adopting a certain stellar evolution with templates adopting another prescription. To this aim we fitted the semi-analytic star-forming galaxies whose spectra adopt the M05 stellar population models with templates based on the BC03 stellar population models. Results are shown in Figs. \ref{Bage} - \ref{Bsfr}. Median offsets and 68\% ranges are given in Tables \ref{sfoverres} and \ref{sfoverres2}. Ages derived from BC03 templates are on average older than those from the M05-based templates (Fig. \ref{Bage}). Median offsets differ by at least $\sim0.2$ dex and can differ by up to $\sim1.2$ dex in the reddened case at low redshift. This is clearly a TP-AGB effect for ages between $\sim0.2$ and 2 Gyr (M05, M06). M05 templates contain more flux in the rest-frame near-IR caused by the TP-AGB. BC03 templates have a lower contribution from this phase, hence an older age of these templates is required to match the near-IR flux. At ages $\geq 2$ Gyr (and metallicities $\geq$ Z$_{\odot}$) M05 models are bluer than BC03 models because of the warmer Red Giant Branch\footnote{most likely caused by a different treatment of the mixing length in these tracks in comparison to the Padova tracks used in BC03, see M05 Fig. 9} (RGB) in the Frascati stellar tracks used in M05. The optical/near-IR ratio is most affected by these differences in the stellar tracks, for example the V-K colour is 0.08 mag bluer in the Frascati tracks (see M05 for more details). This is compensated by younger ages obtained with BC03 templates. \\
For the youngest ages (at low redshift) differences between M05 and BC03-based models stem from the different input stellar tracks for young populations, namely Geneva for M05 and Padova for BC03. Geneva tracks in particular are redder than Padova tracks around 10 Myr, an effect due to the assumed mass-loss and post-MS evolution. This implies a redder M05 model at very young ages which may drive BC03 templates to fit with older ages.\\ 
% \textbf{Are M05 bluer for ages $< 10^8$ yrs???} %BC03 stellar population models use Padova stellar tracks (which include overshooting), causing a later onset of the Red Giant Branch (RGB). 
%A detailed comparison between the two stellar population models can be found in M05 and \citet{M06}.\\
Fig. \ref{Bmetal} shows that the best fit solution obtained from BC03-based templates is on average more metal-rich than that from M05-based templates. This in combination (and degeneracy) with age compensates the lower fuel in TP-AGB and red supergiant stars in the model.\\
At low redshift the reddening is better determined with BC03-based templates which on average provide less reddening (Fig. \ref{Bebv}, median offset of $\Delta E(B-V)=0.12$). Obviously, this is because ages are older and metallicities are higher. However, as the model galaxies evolve $\rmn{\grave{a}}$ la M05, this better agreement is contrived as it originates from the compensation of wrong determination of parameters. At high redshift, the reddening values from both models agree very well.\\ 
Masses derived with templates based on BC03 stellar population models are on average higher than M05-based masses. Median offsets differ by $\sim0.2$ to $\sim0.4$ dex. The mass for $M^*>10^{9.5}\,M_{\odot}$ at $z\geq1$ is higher by up to 0.6 dex (Fig. \ref{Bmass}). This is in agreement with previous findings of M06 and \citet{Pozzetti2007}. In the unreddened case this clearly points out the different TP-AGB treatment, particularly at high redshift where a significant amount of stars is found in this evolutionary phase. In the reddened case the age-dust degeneracy dominates. At $z\leq1$ the lower TP-AGB in the BC03 models causes just the right amount of overestimation in age to correctly predict stellar masses. Given that the physical effect is to underestimate the mass, the larger masses derived with BC03 templates are closer to the correct mass but for the wrong reason. At $z\geq2$ BC03 templates overestimate the masses at the high-mass end for this reason.\\
The SFRs derived with both models agree very well at low redshift (Fig. \ref{Bsfr}), at high redshift BC03-based template setups give lower SFRs on average in the reddened case.\\

\section{Fitting M05 and BC03 templates to Pegase-based semi-analytic galaxies}\label{pegase}
\begin{figure*}\includegraphics[width=144mm]{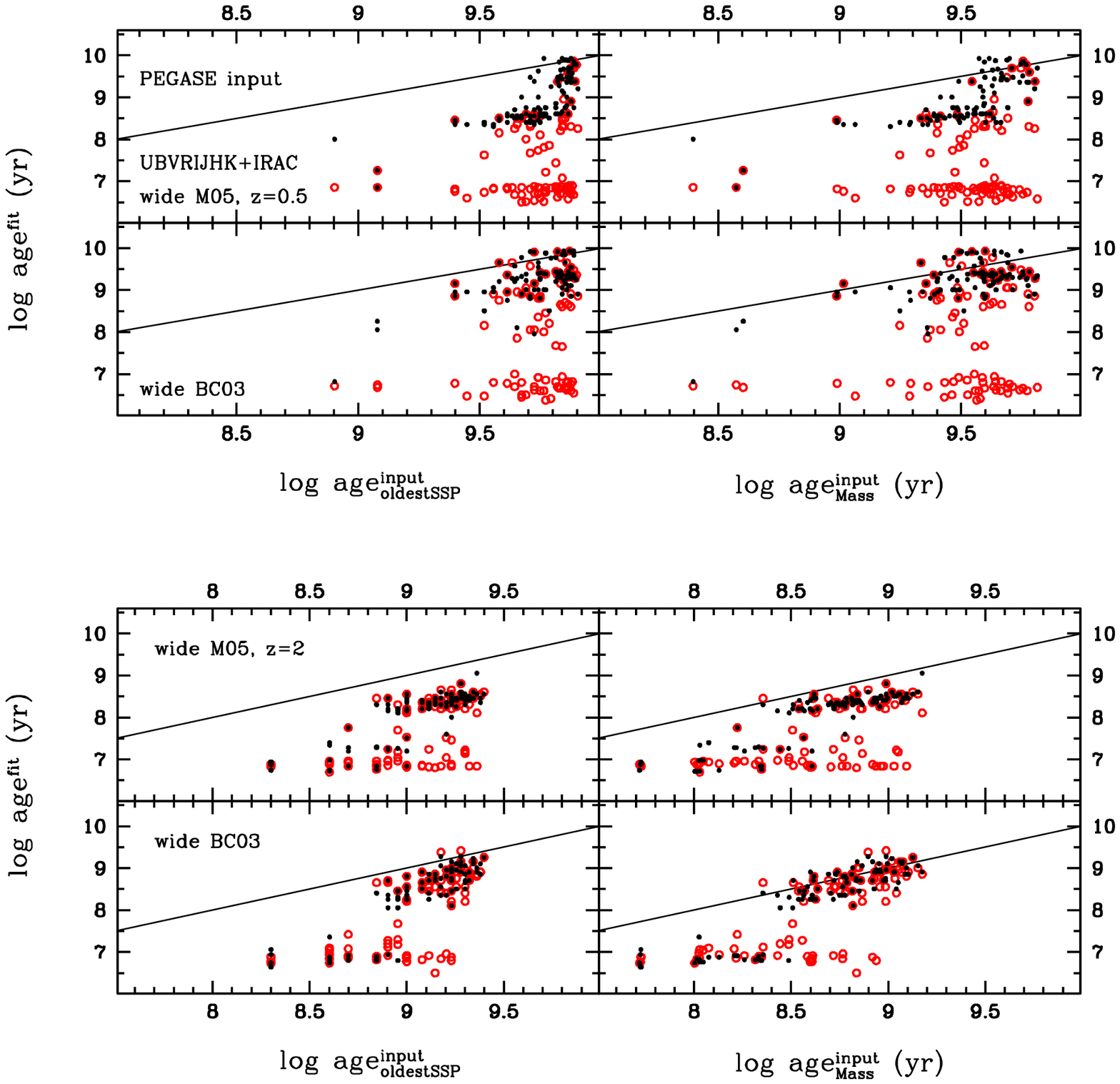}
\caption{\label{Page} Ages derived from SED-fitting with different stellar population synthesis models (M05 and BC03) for mock star-forming galaxies based on Pegase stellar evolution as a function of the age of oldest SSP in the galaxy (left panels) and mass-weighted ages (right panels). Shown are the cases for z=0.5 (upper part) and z=2 (lower part). A wide setup and wavelength coverage was used for both, M05 and BC03. Black refers to the unreddened case, red to the case including reddening.}
\end{figure*}

\begin{figure*}\includegraphics[width=144mm]{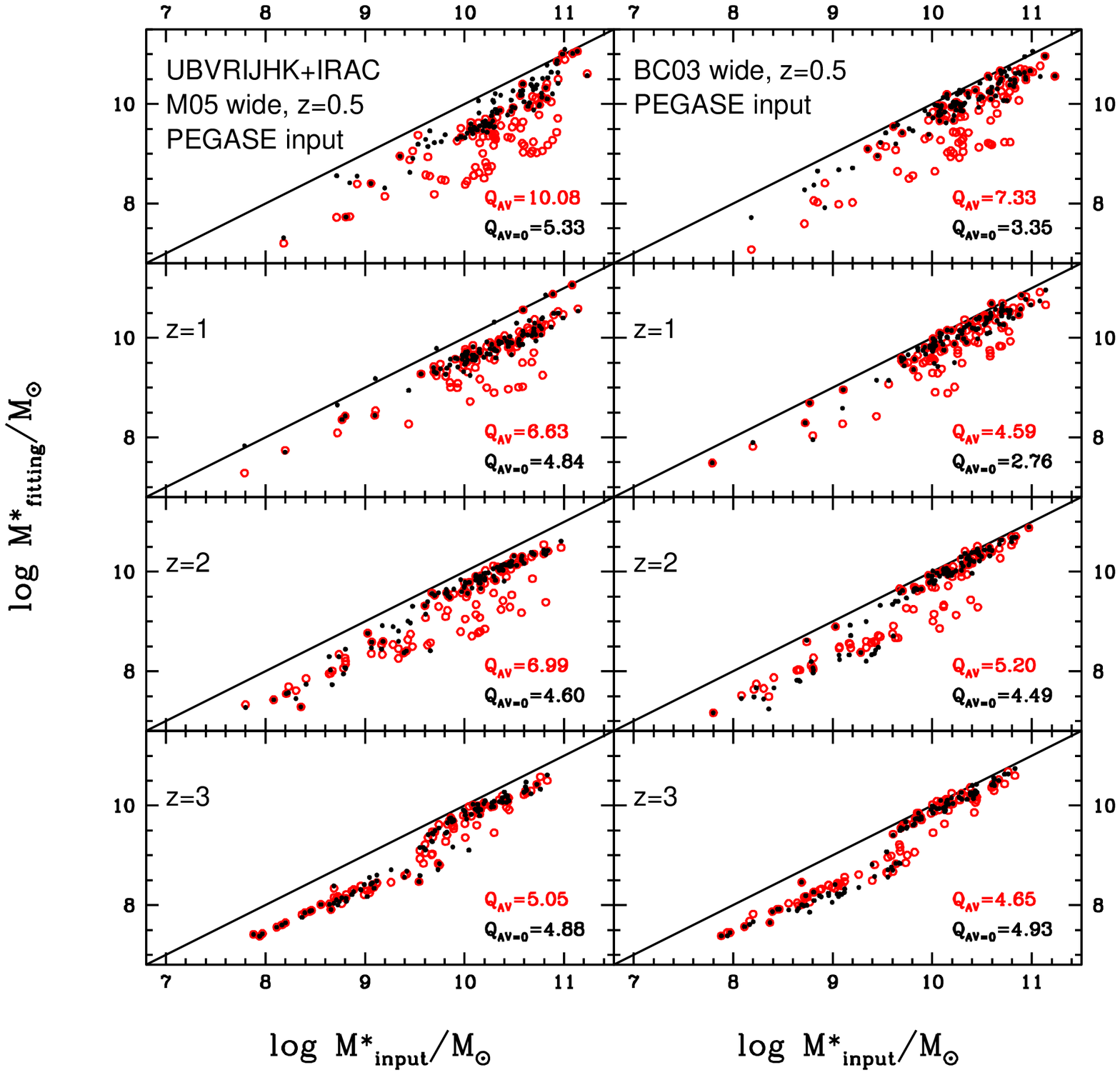}
\caption{\label{Pmass} Same as in the previous figure for stellar masses. Quality factors are given for the entire mass range.}
\end{figure*}

\begin{figure}\includegraphics[width=84mm]{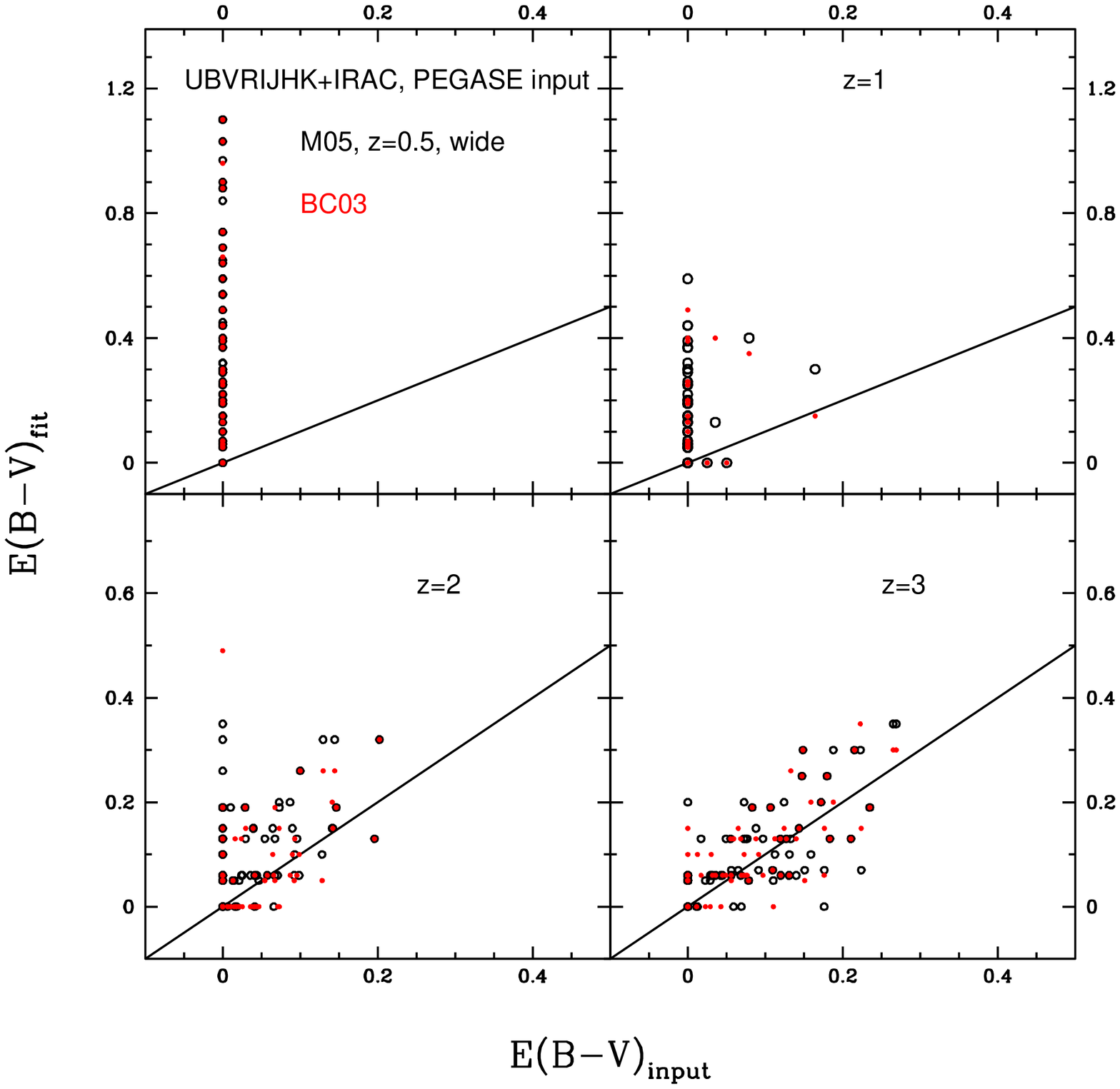}
\caption{\label{Pebv} Same as Fig. \ref{Bebv} for the comparison of derived dust attenuation when mock star-forming galaxies are based on Pegase models. Red dots show results derived with BC03-based templates in the fitting, black those with M05-based templates.}
\end{figure}

\begin{figure*}\includegraphics[width=144mm]{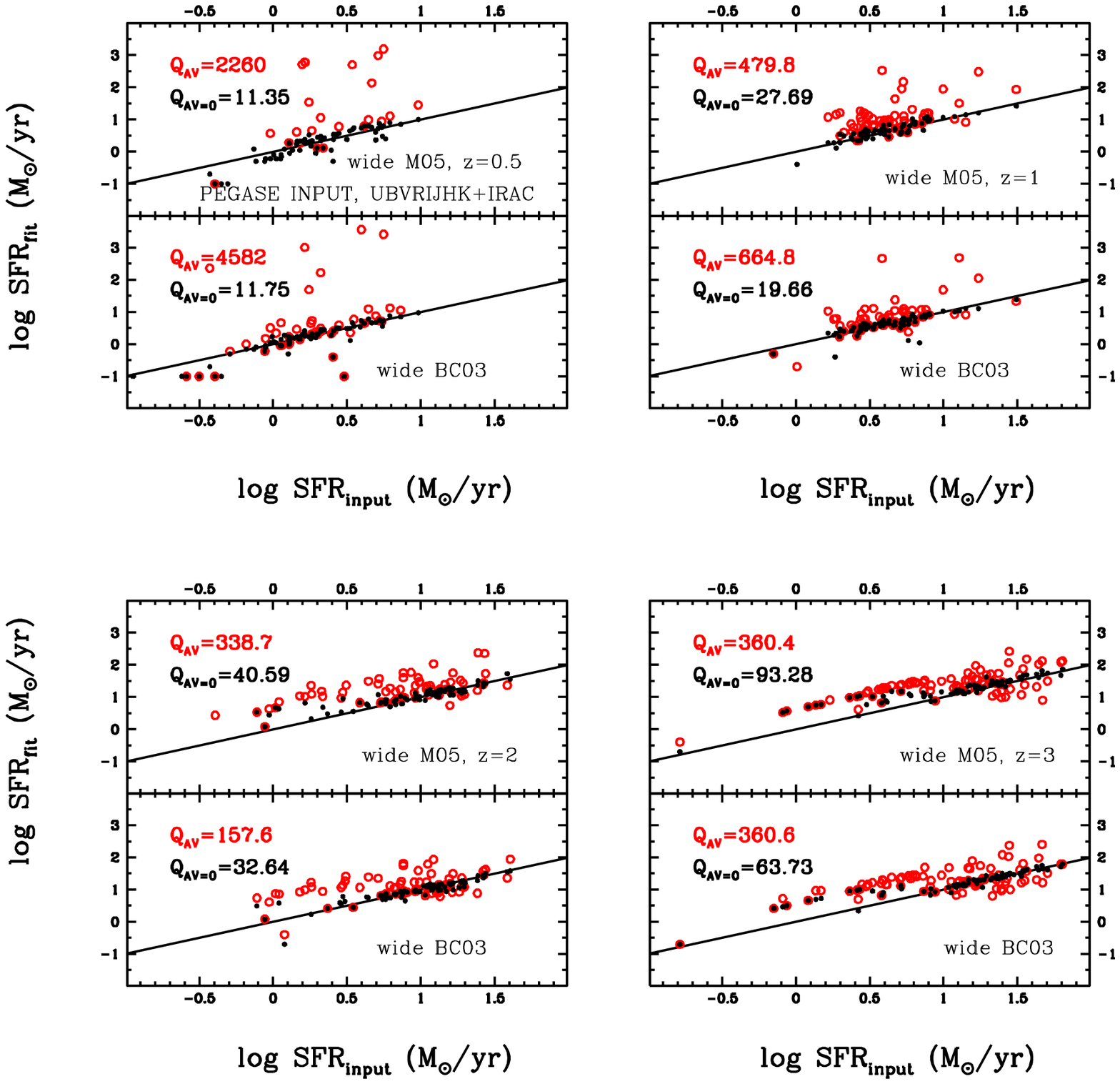}
\caption{\label{Psfr} Same as Fig. \ref{Pmass} for the comparison of derived star formation rates.}
\end{figure*}

\begin{table*}%[h!tbp]
\begin{minipage}{174mm}
\caption{Same as Table \ref{sfoverres} for mock star-forming galaxies at $z=0.5$ and $z=2.0$ based on Pegase photometry.}
\centering{%\small{%\begin{sideways}
\begin{tabular}{@{}lllllllll}
\hline
setup & $\Delta$ log age & & $\Delta$ E(B-V)&  & $\Delta$ log M* &  & $\Delta$ log SFR & \\\hline
z=0.5, no reddening &median &range&median&range&median&range&median&range\\\hline
wide M05     &       -0.73   &       -0.97  to       -0.11   &        0.00   &        0.00  to        0.00   &       -0.53   &       -0.68  to       -0.21   &       -0.16   &       -2.76  to        0.13\\\hline
wide BC03      &       -0.30   &       -0.54  to        0.16   &        0.00   &        0.00  to        0.00   &       -0.28   &       -0.45  to       -0.06   &       -0.02   &       -2.00  to        0.08\\\hline
z=0.5, with reddening &&&&&&&&\\\hline
wide M05      &       -2.49   &       -2.91  to       -0.88   &        0.29   &        0.05  to        0.64   &       -0.85   &       -1.51  to       -0.49   &       -2.88   &    -3.41 to 0.39\\\hline
wide BC03       &       -1.00   &       -2.96  to       -0.20   &        0.15   &        0.00  to        0.64   &       -0.49   &       -1.13  to       -0.20   &       -0.70   &    -3.26 to 0.30\\\hline
z=2, no reddening &&&&&&&&\\\hline
wide M05     &       -0.52   &       -1.05  to       -0.33   &        0.00   &       0.00  to        0.00   &       -0.35   &       -0.63  to       -0.24   &       -0.01   &       -3.45  to        0.14\\\hline
wide BC03      &       -0.22   &       -1.21  to        0.13   &        0.00   &       0.00  to        0.00   &       -0.20   &       -0.78  to       -0.08   &       -0.03   &       -3.26  to        0.08\\\hline
z=2, with reddening &&&&&&&&\\\hline
wide M05      &       -0.89   &       -1.77  to       -0.41   &        0.06   &        0.00  to        0.13   &       -0.52   &       -1.02  to       -0.28   &        0.09   &       -3.89  to        0.72\\\hline
wide BC03       &       -0.36   &       -1.46  to        0.03   &        0.05   &     -0.02  to       0.13   &       -0.25   &       -0.81  to       -0.09   &        0.01   &       -3.64  to        0.64\\\hline
\end{tabular}
%\end{sideways}
\label{overresP}}%}
\end{minipage}
\end{table*}%
Here we investigate the reverse exercise as in Section \ref{bc03results}, namely we fit M05 or BC03 templates to semi-analytic galaxies based on Pegase templates \citep{Pegase}. Results are shown in Figs. \ref{Page} - \ref{Psfr}. Median offsets and 68\% ranges are listed in Table \ref{overresP}.\\
Overall, we find the same trends as in section \ref{results}, ages and stellar masses are underestimated, reddening and SFRs are overestimated.\\
At each redshift, M05-based templates underestimate the age more than BC03-based templates which is evident in a clear offset at z=2 in Fig. \ref{Page}. Median offsets between recovered and mass-weighted age differ by $\sim0.3$ dex  in the unreddened case and by $\sim0.5$ dex in the reddened case. BC03 templates recover the age of the Pegase galaxies similarly well as M05 templates recovered the age of M05 galaxies in Fig. \ref{agedt05}. For M05 galaxies BC03 templates overestimate the age.\\
This effect is carried through to the derived stellar masses. While fitting with M05 templates leads to underestimated masses (with a clear offset), BC03 templates recover masses at high redshift very well (Fig. \ref{Pmass}). At low redshift masses are underestimated. The median offsets in stellar mass differ by $\sim0.3$ dex between the two different stellar population models.\\ 
Importantly here BC03 templates show the same behaviour as M05 templates for M05 galaxies (compare to Fig. \ref{massdt05}), namely the mass is underestimated. This shows again that when fitting M05-type galaxies with BC03 templates the correct determination of the mass is just due to a good concatenation of compensating effects, namely the determination of an older age due to TP-AGB effects, which is by good chance similar to the age of the underlying older population.\\
Reddening and SFRs are comparable for M05 and BC03 templates, independently of the flavour of the input galaxies.\\

\section{Comparison with Wuyts et al. (2009)}
\begin{figure*}\includegraphics[width=144mm]{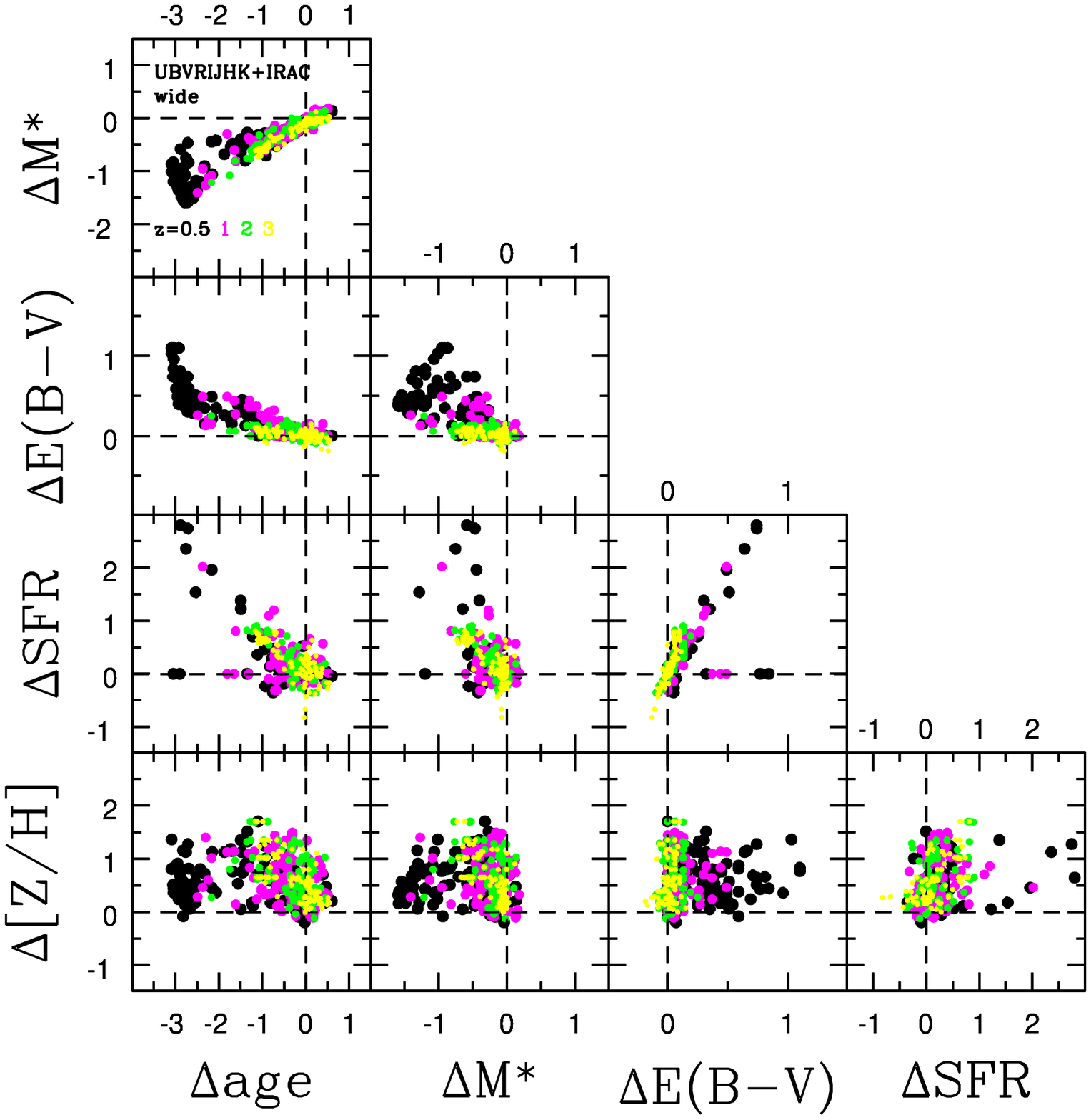}
\caption{\label{trends} Basic correlations between age, dust reddening, stellar mass and SFR derived from SED-fitting using a wide setup and a wavelength coverage of UBVRIJHK+IRAC bands in the reddened case. $\Delta$ is the difference between estimated and true quantity. For stellar mass, age and SFR we used the logarithmic values. The true age for the mock galaxies is represented by the mass-weighted age. Negative values of $\Delta$ mean underestimation, positive values stand for overestimation. We show objects at $z=0.5$ in black, at $z=1$ in magenta, $z=2$ in green and at $z=3$ in yellow. -- Reproduced from \citet{Wuyts2009} with added metallicity row and omitted $A_V$ row.}
\end{figure*}

For a direct comparison with Wuyts et al. (2009) we show here our version of their Fig. 8 in which we swapped the $A_V$ row for a row showing the metallicity (Fig. \ref{trends}). We also list median values with 68\% confidence levels of the difference between derived and true age, mass, SFR and E(B-V), together with those of Wuyts et al. (2009) in Table \ref{wuytscomp}. 
\begin{table*}%[h!tbp]
\begin{minipage}{174mm}
\caption{Comparison of our SED-fitting performance for age, reddening, stellar mass and SFR with \citet{Wuyts2009}. We list the median ($\Delta_m$) values of $\Delta$ with their 68\% intervals, respectively, for mock star-forming galaxies for 4 setups: 1) a wide setup with UBVRIJHK+IRAC filter bands in the reddening case, 2) same as 1) with a minimum age of 0.1 Gyr, 3)  inverted-$\tau$ at z=2 (age fixed to 2 Gyr) and z=3 (age fixed to 1.1 Gyr). For mock passive galaxies we list only the cases for the wide setup and no reddening in the fitting. Values of Wuyts et al. are medians with 68\% intervals. $\Delta$ is defined as the difference between recovered and true value. For stellar mass, age and SFR the logarithmic values are used. The true age for the mock star-forming galaxies is represented by the mass-weighted age. For inverted-$\tau$ models we compare to the oldest age. Ages from Wuyts et. al. refer to mass-weighted ages. Negative values of $\Delta$ mean underestimation of the according property, positive values stand for overestimation. In cases in which either true or best-fit SFR equal zero, we adopt a dummy value of 0.001 $M_{\odot}$/yr to be able to express them in logarithm.}
\centering{%\small{%\begin{sideways}
\begin{tabular}{@{}lrrrr}
\hline
Setup/literature & $\Delta_{m}log\, age$ & $\Delta_{m} E(B-V)$ & $\Delta_{m} log\, M^*$ & $\Delta_{m} log\, SFR$ \\\hline\hline
Wuyts+ disk            & $0.03^{+0.19}_{-0.42}$ & $-0.02^{+0.13}_{-0.07}$ & $-0.06^{+0.06}_{-0.14}$ & $-0.22^{+0.23}_{-0.28}$\\\hline
Wuyts+ merger      & $-0.12^{+0.40}_{-0.26}$ & $-0.02^{+0.08}_{-0.08}$ & $-0.13^{+0.10}_{-0.14}$ & $-0.44^{+0.32}_{-0.31}$\\\hline
Wuyts+ spheroid   & $-0.03^{+0.12}_{-0.14}$ & $-0.03^{+0.11}_{-0.07}$ & $-0.02^{+0.06}_{-0.11}$ & $-0.23^{+0.62}_{-0.47}$ \\\hline
mock star-forming  &    &   &      &\\
with reddening  &    &   &      &\\\hline
wide setup             &     &   &      &\\
z=1              &  $-0.35^{+0.45}_{-0.64}$ & $0.05^{+0.14}_{-0.05}$ & $-0.22^{+0.21}_{-0.22}$ & $0.11^{+0.29}_{-0.43}$\\
z=2              & $-0.15^{+0.29}_{-0.73}$ & $0.00^{+0.06}_{-0.01}$ & $-0.12^{+0.09}_{-0.39}$ & $0.06^{+0.39}_{-0.41}$\\
z=3              &  $-0.12^{+0.34}_{-0.83}$  & $0.02^{+0.04}_{-0.05}$ & $-0.15^{+0.12}_{-0.43}$ & $0.11^{+0.53}_{-0.78}$\\\hline
wide, age $\geq0.1$ Gyr             &    &   &      &\\
z=1              & $-0.33^{+0.43}_{-0.49}$ & $0.05^{+0.10}_{-0.05}$ & $-0.22^{+0.21}_{-0.20}$ & $0.15^{+0.27}_{-0.32}$\\
z=2              & $-0.03^{+0.22}_{-0.28}$ & $0.00^{+0.05}_{-0.01}$ & $-0.11^{+0.08}_{-0.10}$ & $0.06^{+0.14}_{-0.16}$\\
z=3              & $0.09^{+0.27}_{-0.21}$ & $0.00^{+0.04}_{-0.04}$ & $-0.11^{+0.08}_{-0.11}$ & $-0.02^{+0.16}_{-0.25}$\\\hline
inverted-$\tau$ &    &    &    &\\
z=2                 &  $0.19^{+0.41}_{-0.19}$ & $0.01^{+0.09}_{-0.01}$  & $0.01^{+0.05}_{-0.06}$  & $0.17^{+0.23}_{-0.20}$\\
z=3                 &  $0.26^{+0.30}_{-0.26}$ & $0.00^{+0.05}_{-0.04}$  & $0.02^{+0.05}_{-0.06}$  & $-0.04^{+0.33}_{-0.29}$\\\hline
mock passive  &    &    &    & \\
no reddening   &    &    &    & \\\hline
wide   &    &    &    & \\
z=0.5             &   $0.03^{+0.12}_{-0.05}$ & - & $0.01^{+0.04}_{-0.05}$  & $0.00^{+0.00}_{-0.00}$    \\
z=1             &   $0.02^{+0.04}_{-0.02}$  & - & $0.00^{+0.02}_{-0.01}$  &  $0.00^{+0.00}_{-0.00}$ \\\hline
with reddening   &    &    &    & \\\hline
wide   &    &    &    & \\
z=0.5             &   $0.02^{+0.10}_{-0.11}$ & $0.00^{+0.07}_{-0.00}$ & $0.01^{+0.07}_{-0.08}$  & $0.00^{+0.00}_{-0.00}$    \\
z=1             &   $0.02^{+0.03}_{-0.02}$  & $0.00^{+0.00}_{-0.00}$ & $0.01^{+0.02}_{-0.02}$  &  $0.00^{+0.00}_{-0.00}$ \\\hline
\end{tabular}
%\end{sideways}
\label{wuytscomp}}%}
\end{minipage}
\end{table*}%

\clearpage

\end{document}